\documentclass[12pt,a4paper]{article}

\title{Branes wrapped on orbifolds and their gravitational blocks}
\author{Federico Faedo, Alessio Fontanarossa, Dario Martelli}

\usepackage[hmargin=2.5cm,vmargin=3cm]{geometry}
\usepackage[english]{babel}
\usepackage[utf8]{inputenc}
\usepackage{amsmath,amssymb,amsfonts}
\usepackage{mathrsfs}

\usepackage{caption}
\usepackage{subcaption}

\usepackage[bbgreekl]{mathbbol}
\DeclareSymbolFontAlphabet{\mathbb}{AMSb}
\DeclareSymbolFontAlphabet{\mathbbl}{bbold}

\numberwithin{equation}{section}
\usepackage[bottom]{footmisc}
\usepackage{cite}

\usepackage{xcolor}
\usepackage[
  bookmarksnumbered,
  linktocpage=true,
  colorlinks=true,
  citecolor={green!50!black},
	pdfusetitle
]{hyperref}

\usepackage{tikz}
\usetikzlibrary{calc}

\DeclareMathOperator{\rank}{rank}

\newcommand{\ie}{\textit{i.e.}}
\newcommand{\eg}{\textit{e.g.}}

\newcommand{\cf}{\textit{cf.}}
\newcommand{\pvec}[1]{\vec{#1}\mkern2mu\vphantom{#1}}

\newcommand{\NN}{\mathbb{N}}
\newcommand{\ZZ}{\mathbb{Z}}
\newcommand{\QQ}{\mathbb{Q}}
\newcommand{\RR}{\mathbb{R}}
\newcommand{\CC}{\mathbb{C}}
\newcommand{\dd}{\mathrm{d}}
\newcommand{\ee}{\mathrm{e}}
\newcommand{\ii}{\mathrm{i}}

\newcommand{\AdS}{\mathrm{AdS}}
\newcommand{\vol}[1]{\mathrm{vol}(#1)}
\newcommand{\spindle}{\mathbbl{\Sigma}}
\newcommand{\riemann}{{\Sigma_\mathrm{g}}}
\newcommand{\hemi}{{\mathbbl{S}^4}}
\newcommand{\Morb}{\mathbb{M}}

\newcommand{\mc}[1]{\mathcal{#1}}

\newcommand{\XA}{x_N}
\newcommand{\XB}{x_S}
\newcommand{\XAB}{x_{N,S}}
\newcommand{\YA}{y_N}
\newcommand{\YB}{y_S}
\newcommand{\YAB}{y_{N,S}}
\newcommand{\x}{\mathtt{x}}

\newcommand{\spang}[1]{\nu_{#1}}

\newcommand{\fls}{\mathfrak{s}}
\newcommand{\flt}{\mathfrak{t}}
\newcommand{\flp}{\mathfrak{p}}
\renewcommand{\flq}{\mathfrak{q}}

\newcommand{\Ispindle}{F}
\newcommand{\fp}{v}
\newcommand{\fpN}{n}
\newcommand{\vv}{v}

\begin{document}

\begin{titlepage}
\vskip 2cm

\begin{center}

\vspace*{2cm}

{\Large \bf Branes wrapped on orbifolds\\[3mm] and their gravitational blocks}

\vskip 1.5cm

{Federico Faedo, Alessio Fontanarossa and Dario Martelli}

\vskip 0.7cm

\textit{Dipartimento di Matematica ``Giuseppe Peano'', Universit\`a di Torino,\\
Via Carlo Alberto 10, 10123 Torino, Italy}

\vskip 0.2cm

\textit{INFN, Sezione di Torino,\\Via Pietro Giuria 1, 10125 Torino, Italy}

\end{center}

\vskip 3.5cm

\begin{abstract}

\noindent
We construct new supersymmetric $\AdS_2\times\Morb_4$ solutions of $D=6$ gauged supergravity, where $\Morb_4$ are certain four-dimensional orbifolds.
After uplifting to massive type~IIA supergravity these correspond to the near-horizon limit of a system of $N$ D4-branes and $N_f$ D8-branes wrapped on $\Morb_4$.
In one class of solutions $\Morb_4 = \riemann\ltimes\spindle$ is a spindle fibred over a smooth Riemann surface of genus $\mathrm{g}>1$, while in another class $\Morb_4 = \spindle\ltimes\spindle$ is a spindle fibred over another spindle. Both classes can be thought of as orbifold generalizations of Hirzebruch surfaces and, in the second case, we describe the solutions in terms of toric geometry.
We show that the entropy associated with these solutions is reproduced by extremizing an entropy function obtained by gluing gravitational blocks, using a general recipe for orbifolds that we propose.
We also discuss how our prescription can be used to define an off-shell central charge whose extremization reproduces the gravitational central charge of analogous $\AdS_3\times\Morb_4$ solutions of $D=7$ gauged supergravity, arising from wrapping M5-branes on $\Morb_4$.

\end{abstract}

\end{titlepage}

\tableofcontents

\section{Introduction}

For two decades several examples of the AdS/CFT correspondence have been engineered considering setups involving branes wrapped on compact cycles, preserving supersymmetry via a topological twist on the brane world-volumes \cite{Maldacena:2000mw}.
On the supergravity side, these give rise to supersymmetric solutions involving a warped product of an anti-de Sitter (AdS) spacetime, together with an internal compact manifold. The corresponding dual field theories are obtained from suitably twisted compactifications of the parent theories on smooth Riemann surfaces embedded holomorphically in Calabi--Yau manifolds, or different higher-dimensional calibrated submanifolds in special holonomy manifolds.

Solutions with an $\AdS_2$ factor are particularly interesting because on general grounds these are expected to correspond to the near-horizon limit of supersymmetric black holes arising in string or M-theory. In this case, the AdS/CFT correspondence gives access to the microscopic entropy of these black holes, obtained via counting the microstates of the dual supersymmetric quantum mechanics, which becomes particularly useful when the latter arises from a twisted compactification of a higher-dimensional gauge theory.
The paradigmatic example of this successful strategy is that of the four-dimensional static black holes of \cite{Cacciatori:2009iz}, for which a microscopic account of the entropy has been obtained in \cite{Benini:2015eyy}.

The construction presented in  \cite{Ferrero:2020laf} changed somewhat the rules of the game, showing that one can obtain examples of holographic dual pairs
with genuine novel features by considering D3-branes wrapped on the orbifold $\spindle  = \mathbb{WCP}^1_{[n_-,n_+]}$, where $n_-,n_+$ are  co-prime integers, sometimes referred to as \emph{spindle}.
Perhaps the most striking aspect of this construction is that supersymmetry is preserved via a novel mechanism different from the topological twist, that was later called anti-twist. Further examples of various branes wrapping the spindle have been discussed in \cite{Ferrero:2020twa,Hosseini:2021fge,Boido:2021szx,Ferrero:2021wvk,Ferrero:2021ovq,Couzens:2021rlk,Faedo:2021nub,Ferrero:2021etw,Giri:2021xta,Couzens:2021cpk,Arav:2022lzo,Couzens:2022yiv}
and closely related constructions appeared in \cite{Bah:2021mzw,Bah:2021hei,Couzens:2021tnv,Suh:2021ifj,Suh:2021aik,Suh:2021hef,Karndumri:2022wpu,Couzens:2022yjl}.
In general, the only two possible ways to preserve supersymmetry on a spindle are by the anti-twist of \cite{Ferrero:2020laf} or the twist introduced in \cite{Ferrero:2021ovq}, which is a global version of the standard topological twist \cite{Ferrero:2021etw}.

Once it has been established that it makes sense to wrap branes on spaces with orbifold singularities, it is natural to extend the exploration of the landscape of holographic dualities to branes wrapping orbifolds of dimension higher than two. Simple examples of branes wrapped on four-dimensional orbifolds of the type $\Morb_4 = \riemann\times\spindle$ were discussed in \cite{Boido:2021szx} for M5-branes and in \cite{Faedo:2021nub,Giri:2021xta} for D4-branes (see also \cite{Suh:2022olh}).
In this paper we will focus mainly on D4-branes, extending the results of \cite{Faedo:2021nub} in various ways. In particular, we will present two classes of new supersymmetric solutions of gauged $D=6$ supergravity of the type $\AdS_2\times \Morb_4$, where $\Morb_4$ are four-dimensional orbifolds. After uplifting to massive type~IIA supergravity these
describe a system of $N$ D4-branes and $N_f$ D8-branes wrapped on $\Morb_4$, and are expected to correspond to the near-horizon limit of novel supersymmetric $\AdS_6$ black holes.
Previously, supersymmetric $\AdS_2\times M_4$ solutions of the same theory, where $M_4$ is either a negatively curved K\"ahler--Einstein manifold or the product of two Riemann surfaces $\Sigma_{\mathrm{g}_1}\times \Sigma_{\mathrm{g}_2}$ (with $\mathrm{g}_1,\mathrm{g}_2>1$), were discussed in \cite{Hosseini:2018usu} and \cite{Suh:2018szn}, with the latter reference also presenting numerical solutions for supersymmetric $\AdS_6$ black holes with $\Sigma_{\mathrm{g}_1}\times \Sigma_{\mathrm{g}_2}$ horizons.
Below we will construct two new classes of solutions. In one class $\Morb_4 = \riemann\ltimes\spindle$ is a spindle fibred over a smooth Riemann surface of genus $\mathrm{g}>1$, while in the other class $\Morb_4 =\spindle \ltimes \spindle $ is a spindle fibred over another spindle. Both classes can be thought of as generalizations of Hirzebruch surfaces and analogous $\AdS_3\times\Morb_4$ solutions of $D=7$ gauged supergravity, arising from wrapping M5-branes on $\Morb_4$, appeared in \cite{Cheung:2022ilc}.

One of the motivations for this paper was to begin investigating entropy functions whose extremization reproduces the entropy associated with the corresponding supergravity solutions, comprising orbifolds of dimension higher than two. In the two-dimensional case,  entropy functions associated with compactifications on the spindle were discussed in \cite{Hosseini:2021fge,Cassani:2021dwa,Ferrero:2021ovq} and a conjectural extension to more general off-shell free energies was put forward in \cite{Faedo:2021nub}. In particular, for all cases, precise formulas for the off-shell free energies and for the associated constraints obeyed by the magnetic fluxes and fugacities were given in terms of data at the north and south poles of the spindle.
These are the fixed points of the Killing vector generating its $U(1)$ azimuthal symmetry and the contributions to the off-shell free energies arising at these fixed points are referred to as gravitational blocks \cite{Hosseini:2019iad}. Here we will extend the conjecture of \cite{Faedo:2021nub} presenting an \emph{orbifold entropy function} whose contributions (blocks) arise from fixed points of a $U(1)^2$ action on $\Morb_4$,
together with a recipe for determining the appropriate constraints.

The key for formulating this proposal is to provide a toric description of our solutions. Despite the fact that our orbifolds are not symplectic and therefore there are no moment maps with associated polytope, assuming a $U(1)^2$ action, we can still define a set of \emph{toric data} from the co-dimension two loci where one $U(1)\subset U(1)^2$ degenerates. In particular, these can be determined by studying the Killing vectors of the explicit metrics on $\Morb_4$. We can then associate a compact polytope defined using these data as normal vectors to the edges and identify the vertices as fixed points of the torus action, as in symplectic toric geometry.
Our proposal for the entropy function reproduces the results of the explicit supergravity solutions presented here, but we formulate it in a way that is in principle applicable  to more generic toric orbifolds, including toric manifolds, for which explicit solutions have not yet been found.

Finally, we discuss how our prescription can be extended to describe analogous supersymmetric $\AdS_3\times\Morb_4$ solutions of $D=7$ gauged supergravity arising from wrapping M5-branes on $\Morb_4$.
In these cases one can employ the method of integrating the M5-branes anomaly polynomial  on $\Morb_4$ to obtain an off-shell central charge of the corresponding $d = 2$, $\mathcal{N}=(0,2)$ SCFTs and our proposal reproduces the results available in the literature. In particular, when $\Morb_4$ is actually a toric manifold, our recipe reduces to the construction of \cite{Hosseini:2020vgl}, whereas for the specific case of $\Morb_4 = \spindle\ltimes\spindle$ we recover the results of \cite{Cheung:2022ilc}.

The rest of the paper is organized as follows. In section~\ref{sec:6dsugra} we present the new solutions in $D=6$ and demonstrate their supersymmetry by solving the relevant Killing spinor equations. In section~\ref{sec:uplift} we uplift our solutions to massive type~IIA supergravity and calculate their associated geometric entropies. In section~\ref{sec:toric_geometry} we work out the toric data of the relevant solutions. In section~\ref{sec:entropy-function} we present our conjectural entropy functions and use these to recover the entropies of the above solutions, as well as the central charge associated with the solutions in \cite{Cheung:2022ilc}.
Appendix~\ref{app:quantization.exe} contains a discussion about the different quantization conditions on the fluxes and their resolution. In appendix~\ref{app:spindle-2-riemann} we present a special limit in which a Riemann surface can be retrieved from a spindle. In appendix~\ref{app:toric-theory} we derive toric data from different metrics, including the solutions of~\cite{Cheung:2022ilc}.

\vskip 5mm

\noindent
{\bf Note added}: as this work was being finalized we became aware that there would be
significant overlap with the results of~\cite{Couzens:2022lvg}, which appeared on the arXiv on the same day.

\section{Solutions in $D=6$ gauged supergravity}
\label{sec:6dsugra}

In this paper we discuss solutions of a $D=6$ gauged supergravity with gauge group $U(1)^2$, comprising two gauge fields $A_1$, $A_2$, a two-form $B$ and two real scalar fields $\vec{\varphi}=(\varphi_1,\varphi_2)$. This model can also be obtained as a sub-sector of an extension of Romans $F(4)$ gauged supergravity~\cite{Romans:1985tw}, coupled to three vector multiplets~\cite{DAuria:2000afl,Andrianopoli:2001rs}. The bosonic part of the action reads\footnote{Here and in what follows we define, for any $p$-form $\omega$, $|\omega|^2 = \frac{1}{p!} \, \omega_{\mu_1\ldots\mu_p} \omega^{\mu_1\ldots\mu_p}$.}
\begin{equation} \label{6d_action}
	\begin{split}
		S_\text{6D} &= \frac{1}{16\pi G_{(6)}} \int \dd^6x \, \sqrt{-g} \biggl[ R - V - \frac12 |\dd\vec{\varphi}|^2 - \frac12 \sum_{i=1}^2 X_i^{-2} |F_i|^2 - \frac18 (X_1 X_2)^2 |H|^2 \\
		& - \frac{m^2}{4} (X_1 X_2)^{-1} |B|^2 - \frac{1}{16} \frac{\varepsilon^{\mu\nu\rho\sigma\tau\lambda}}{\sqrt{-g}} B_{\mu\nu} \Bigl( F_{1\,\rho\sigma} F_{2\,\tau\lambda} + \frac{m^2}{12} B_{\rho\sigma} B_{\tau\lambda} \Bigr) \biggr] \,,
	\end{split}
\end{equation}
where $F_i=\dd A_i$, $H=\dd B$ and the scalar fields $\vec{\varphi}$ are parameterized as
\begin{equation}
	X_i = \ee^{-\frac12 \vec{a}_i\cdot\vec{\varphi}}  \qquad  \text{with}  \qquad  \vec{a}_1 = \bigl(2^{1/2}, 2^{-1/2}\bigr) \,,  \qquad  \vec{a}_2 = \bigl(-2^{1/2}, 2^{-1/2}\bigr) \,.
\end{equation}
The scalar potential is
\begin{equation}
	V = m^2 X_0^2 - 4g^2 X_1 X_2 - 4mg \, X_0 (X_1 + X_2) \,,
\end{equation}
with $g$ the gauge coupling and $m$ the mass parameter, and where for later convenience we defined $X_0=(X_1 X_2)^{-3/2}$.

A solution to the equations of motion of the model is supersymmetric if and only if it satisfies also the following set of Killing spinor equations~\cite{DAuria:2000afl}:
\begin{align}
	\begin{split} \label{KSE_grav}
		\mathcal{D}_\mu \epsilon^A + \frac18 \bigl[ g (X_1 + X_2) + m X_0 \bigr] \Gamma_\mu \epsilon^A & \\
		+ \frac{1}{32} \bigl[ m (X_1 X_2)^{-1/2} B_{\nu\lambda} \Gamma_7 \delta^A_B + \ii \bigl( X_1^{-1} F_1 + X_2^{-1} F_2 \bigr)_{\nu\lambda} (\sigma^3)^A_{\,\ B} \bigr] \bigl(\Gamma_\mu^{\ \nu\lambda} - 6\delta_\mu^\nu \, \Gamma^\lambda \bigr) \epsilon^B & \\
		- \frac{1}{96} (X_1 X_2) H_{\nu\lambda\rho} \Gamma_7 \bigl(\Gamma_\mu^{\ \nu\lambda\rho} - 3\delta_\mu^\nu \, \Gamma^{\lambda\rho} \bigr) \epsilon^A &= 0 \,,
	\end{split} \\[0.5em]
	\begin{split} \label{KSE_dila}
		\frac14 \bigl( X_1^{-1} \partial_\mu X_1 + X_2^{-1} \partial_\mu X_2 \bigr) \Gamma^\mu \epsilon^A - \frac18 \bigl[ g (X_1 + X_2) - 3m X_0 \bigr] \epsilon^A & \\
		+ \frac{1}{32} \bigl[ m (X_1 X_2)^{-1/2} B_{\mu\nu} \Gamma_7 \delta^A_B - \ii \bigl( X_1^{-1} F_1 + X_2^{-1} F_2 \bigr)_{\mu\nu} (\sigma^3)^A_{\,\ B} \bigr] \Gamma^{\mu\nu} \epsilon^B & \\
		+ \frac{1}{96} (X_1 X_2) H_{\mu\nu\lambda} \Gamma_7 \Gamma^{\mu\nu\lambda} \epsilon^A &= 0 \,,
	\end{split} \\[0.5em]
	\begin{split} \label{KSE_gauge}
		\frac12 \bigl( X_1^{-1} \partial_\mu X_1 - X_2^{-1} \partial_\mu X_2 \bigr) \Gamma^\mu (\sigma^3)^A_{\,\ B} \epsilon^B - g (X_1 - X_2) (\sigma^3)^A_{\,\ B} \epsilon^B & \\
		- \frac{\ii}{4} \bigl( X_1^{-1} F_1 - X_2^{-1} F_2 \bigr)_{\mu\nu} \Gamma^{\mu\nu} \epsilon^A &= 0 \,,
	\end{split}
\end{align}
where
\begin{equation} \label{cov-D}
	\mathcal{D}_\mu \epsilon^A \equiv \partial_\mu \epsilon^A + \frac14 \, \omega_\mu^{\ ab} \Gamma_{ab} \epsilon^A - \frac{\ii}{2} \, g (A_1 + A_2)_\mu (\sigma^3)^A_{\,\ B} \epsilon^B \,.
\end{equation}
These follow from setting to zero the supersymmetry variations of the fermionic fields of the theory with three vector multiplets \cite{DAuria:2000afl}, that do not vanish automatically in the sub-truncation that we are considering. Here $(\sigma^3)^A_{\,\ B}$ is the usual third Pauli matrix, $\{\Gamma_a,\Gamma_b\}=2\eta_{ab}$ and $\Gamma_7 \equiv \Gamma^0 \Gamma^1 \Gamma^2 \Gamma^3 \Gamma^4 \Gamma^5$. The $SU(2)$ indices $A,B$ are raised and lowered as $\epsilon^A=\varepsilon^{AB}\epsilon_B$ and $\epsilon_A=\epsilon^B\varepsilon_{BA}$, where $\varepsilon_{AB}=-\varepsilon_{BA}$ and its inverse matrix $\varepsilon^{AB}$ is defined such that $\varepsilon^{AB}\varepsilon_{AC}=\delta^B_C$. The supersymmetry parameter $\epsilon^A$ is an eight-component symplectic-Majorana spinor, hence it satisfies the condition
\begin{equation} \label{symp-Maj}
	\varepsilon^{AB} \epsilon_B^* = \mc{B}_6 \epsilon_A \,,
\end{equation}
where $\mc{B}_6$ is related to the six-dimensional charge conjugation matrix $\mc{C}_6$ by $\mc{B}_6=-\ii\,\mc{C}_6\Gamma^0$.

\subsection{AdS$_2\times \riemann \ltimes \spindle_2$ solutions}
\label{subsec:6d_riem}

The first local solutions to the supergravity action~\eqref{6d_action} that we consider are of the type  $\AdS_2 \times \riemann \ltimes \spindle_2$, which consist of a two-dimensional spindle~$\spindle_2$ non-trivially fibred over a Riemann surface~$\riemann$, with genus $\mathrm{g}>1$. When $\mathrm{g}=0$ a solution to the equations of motion still exists, but it is not supersymmetric. Specifically, the backgrounds are described by the following set of fields solving the equations of motion
\begin{equation} \label{riem_solution}
\begin{aligned}
\dd s^2 &= (y^2 h_1 h_2)^{1/4} \biggl[ \frac14 \, \dd s_{\AdS_2}^2 + \frac12 \, \dd s_{\riemann}^2 + \frac{y^2}{F} \, \dd y^2 + \frac{F}{h_1 h_2} \Bigl( \dd z - \frac{1}{2m} \, \omega_\riemann \Bigr)^2 \biggr] \,, \\
A_i &= -\frac{y^3}{h_i} \Bigl( \dd z - \frac{1}{2m} \, \omega_\riemann \Bigr) \,,  \qquad  X_i = (y^2 h_1 h_2)^{3/8} h_i^{-1} \,, \\[0.5em]
B &= \frac{y}{2m} \, \vol{\AdS_2} \, ,
\end{aligned}
\end{equation}
where $\dd s_{\AdS_2}^2$ denotes the unit radius metric on~$\AdS_2$, $\dd s_{\riemann}^2$ the unit radius metric on~$\riemann$ and~$\omega_\riemann$ is such that $\dd \omega_\riemann = \vol{\riemann}$. The functions $h_i$ and $F$ are given by
\begin{equation} \label{func-h+F}
h_i(y) = \frac{2g}{3m} \, y^3 + q_i \,,  \qquad  F(y) = m^2 h_1 h_2 - y^4 \,,
\end{equation}
with $q_1$, $q_2$ two real parameters. A curvature singularity lies at $y=0$, hence without loss of generality, in what follows we will restrict to $y>0$.

The first step we take in the analysis of these solutions is to prove that they are supersymmetric by constructing the local form of the Killing spinors solving
the equations~\eqref{KSE_grav}--\eqref{KSE_gauge}. We employ the following orthonormal frame
\begin{equation}
\begin{aligned}
e^{\hat{a}} &= \frac{y^{1/4} (h_1 h_2)^{1/8}}{2} \, \hat{e}^{\hat{a}} \,,  \qquad  & e^{\check{a}} &= \frac{y^{1/4} (h_1 h_2)^{1/8}}{\sqrt2} \, \check{e}^{\check{a}} \,, \\
e^4 &= \frac{y^{5/4} (h_1 h_2)^{1/8}}{F^{1/2}} \, \dd y \,,  \qquad  & e^5 &= \frac{y^{1/4} F^{1/2}}{(h_1 h_2)^{3/8}} \Bigl( \dd z - \frac{1}{2m} \, \omega_\riemann \Bigr) \,,
\end{aligned}
\end{equation}
where $\hat{e}^{\hat{a}}$, $\hat{a}=0,1$, is the zweibein on $\AdS_2$, whose coordinates are denoted as $x^{\hat{\mu}}$, and $\check{e}^{\check{a}}$, $\check{a}=2,3$, is the zweibein on $\riemann$, whose coordinates are denoted as $x^{\check{\mu}}$. Equation~\eqref{KSE_grav} then splits into the following system
\begin{equation} \label{riem_KSE_grav-group}
\begin{split}
\biggl( \partial_{\hat{\mu}} + \frac14 \omega_{\hat{\mu}}^{\ \hat{a}\hat{b}} \Gamma_{\hat{a}\hat{b}} \biggr) \epsilon^A + \hat{e}^{\hat{a}}_{\hat{\mu}} \biggl[ -\frac{\ii}{4} \, \Gamma_{\hat{a}}^{\ 45} (\sigma^3)^A_{\,\ B} \epsilon^B + \frac14 \, \Gamma_{\hat{a}}^{\ 2345} \epsilon^A \biggr] &= 0 \,, \\
\partial_{\check{\mu}} \epsilon^A - \frac12 \biggl[ (\omega_\riemann)_{\check{\mu}} + \frac{1}{\sqrt2} \, \Gamma_{\check{\mu}}^{\ 45} \biggr] \bigl( \Gamma^{23} \epsilon^A + \ii (\sigma^3)^A_{\,\ B} \epsilon^B \bigr) &= 0 \,, \\
\partial_y \epsilon^A - \frac{1}{16 y} \biggl[ (2 + y \tilde h') \epsilon^A - \ii\,\frac{4y^2}{F^{1/2}} (4 - y \tilde h') \Gamma^5 (\sigma^3)^A_{\,\ B} \epsilon^B \biggr] &= 0 \,, \\
\partial_z \epsilon^A - \ii\,\frac{g}{2} \biggl( \alpha_1+\alpha_2 - \frac{2m}{g} \biggr) (\sigma^3)^A_{\,\ B} \epsilon^B &= 0 \,,
\end{split}
\end{equation}
where we defined $\tilde h \equiv \log ( h_1h_2)$ and we applied the gauge transformation $A_i \mapsto A_i + \alpha_i \, \dd z$. Equations~\eqref{KSE_dila} and~\eqref{KSE_gauge} yield the same constraint
\begin{equation} \label{riem_KSE_other}
F^{1/2} \Gamma^4 \epsilon^A + m (h_1 h_2)^{1/2} \epsilon^A + \ii\,y^2 \Gamma^{45} (\sigma^3)^A_{\,\ B} \epsilon^B = 0 \,.
\end{equation}
We note that the equations along~$y$ and~$z$ in~\eqref{riem_KSE_grav-group} and equation~\eqref{riem_KSE_other} are the same as in the $\AdS_4\times\spindle$ system (\cf\ equations~(3.3) and~(3.4) of~\cite{Faedo:2021nub}), a fact that will play an important role later on. From the fourth equation in~\eqref{riem_KSE_grav-group} we see immediately that setting
\begin{equation}
\alpha_1 + \alpha_2 = \frac{2m}{g}
\end{equation}
leads to Killing spinors independent of~$z$ and  we will adopt this choice in the reminder of this section.

Before tackling the Killing spinor equations, we note the following fact about the $\AdS_4\times\spindle$ solution of~\cite{Faedo:2021nub}: choosing a suitable decomposition of the six-dimensional gamma matrices, it is possible to write the six-dimensional Killing spinors as tensor products of a single $\AdS_4$ spinor and a single spinor on~$\spindle$. Specifically, we consider the decomposition
\begin{equation} \label{gamma-deco_6to4}
\Gamma^{\tilde{a}} = \gamma^{\tilde{a}} \otimes \rho_* \,,  \qquad\quad  \Gamma^{4,5} = I_4 \otimes \rho^{1,2} \,,
\end{equation}
where $\gamma^{\tilde{a}}$, $\tilde{a}=0,\ldots,3$, are the (Lorentzian) gamma matrices in $D=4$, $\rho^{\hat{\imath}}$, $\hat{\imath}=1,2$, are the (Euclidean) gamma matrices in $D=2$ and $\rho_* = -\ii\,\rho^1\rho^2$ is the related chiral matrix. For $\rho^{\hat{\imath}}$ we choose the following representation
\begin{equation} \label{2d-gamma_rep}
\rho^{\hat{\imath}} = \sigma^{\hat{\imath}} \,,  \qquad\quad  \rho_* = \sigma^3 \,,
\end{equation}
and we take $\mc{B}_2^\rho=-\sigma^2$. For consistency, the six-dimensional matrix~$\mc{B}_6$ decomposes as
\begin{equation}
\mc{B}_6 = (\mc{B}_4 \gamma_5) \otimes (\mc{B}_2^\rho \rho_*) \,,
\end{equation}
where $\gamma_5 = \ii\,\gamma^0\gamma^1\gamma^2\gamma^3$ is the four-dimensional chiral matrix. Adopting this decomposition, the symplectic-Majorana condition~\eqref{symp-Maj} and the Killing spinor equations~(3.3)--(3.4) of~\cite{Faedo:2021nub} can be solved to give
\begin{equation}
\epsilon^1 = (\gamma_5 \vartheta_{(4)}) \otimes \eta^1 \,,  \qquad  \epsilon^2 = \vartheta_{(4)} \otimes \eta^2 \,,
\end{equation}
where $\vartheta_{(4)} = \vartheta_{(4)}(x^{\tilde{\mu}})$ is a Majorana Killing spinor on~$\AdS_4$ and
\begin{equation} \label{eta}
\eta^1 = \xi_\eta \, y^{1/8} (h_1 h_2)^{-3/16} \begin{pmatrix} f_1^{1/2} \\ -f_2^{1/2} \end{pmatrix} \,,  \qquad
\eta^2 = -\ii\,\xi_\eta^* \, y^{1/8} (h_1 h_2)^{-3/16} \begin{pmatrix} f_2^{1/2} \\ -f_1^{1/2} \end{pmatrix} \,.
\end{equation}
Here, $\xi_\eta$ is a complex constant and we have defined
\begin{equation}
f_1(y) \equiv m (h_1 h_2)^{1/2} + y^2 \,,  \qquad  f_2(y) \equiv m (h_1 h_2)^{1/2} - y^2 \,,
\end{equation}
which satisfy $F(y)=f_1(y)\,f_2(y)$. Notice that $\eta^1$ and $\eta^2$ are related by $\eta^2 = \ii\,\sigma^1(\eta^1)^*$, which descends from the symplectic-Majorana condition~\eqref{symp-Maj}.

Going back to the $\AdS_2 \times \riemann \ltimes \spindle_2$ background, in order to solve its Killing spinor equations we adopt the same decomposition of the six-dimensional gamma matrices as in~\eqref{gamma-deco_6to4}. Moreover, we decompose the four-dimensional gamma matrices~$\gamma^{\tilde{a}}$ as
\begin{equation} \label{gamma-deco_4to2}
\gamma^{\hat{a}} = \beta^{\hat{a}} \otimes \tau_* \,,  \qquad\quad  \gamma^{2,3} = I_2 \otimes \tau^{1,2} \,,
\end{equation}
where $\beta^{\hat{a}}$ are the (Lorentzian) gamma matrices in $D=2$, $\tau^{\hat{\imath}}$, $\hat{\imath}=1,2$, are the (Euclidean) gamma matrices in $D=2$ and $\tau_* = -\ii\,\tau^1\tau^2$ is the related chiral matrix. Explicitly, we adopt the same representation as in~\eqref{2d-gamma_rep}
\begin{equation}
\tau^i = \sigma^i \,,  \qquad\quad  \tau_* = \sigma^3 \,,
\end{equation}
and we take $\mc{B}_2^\tau=\sigma^1$. The four-dimensional matrices~$\mc{B}_4$ and~$\gamma_5$ decompose as
\begin{equation}
\mc{B}_4 = (\mc{B}_2^\beta \beta_*) \otimes \mc{B}_2^\tau \,,  \qquad\quad  \gamma_5 = \beta_* \otimes \tau_* \,,
\end{equation}
where $\beta_* = -\beta^0\beta^1$ is the (Lorentzian) chiral matrix in $D=2$. The ansatz for the six-dimensional Killing spinors is
\begin{equation} \label{spinor-deco_6to2}
\epsilon^A = \vartheta \otimes \chi^A \otimes \eta^A \,,
\end{equation}
where $\vartheta=\vartheta(x^{\hat{\mu}})$ is a Majorana Killing spinor on~$\AdS_2$, hence $\hat{\nabla}_{\hat{\mu}} \vartheta = \frac12 \beta_{\hat{\mu}} \vartheta$ and $\vartheta^* = \mc{B}_2^\beta \vartheta$. $\chi^A = \chi^A(x^{\check{\mu}})$ are two two-component spinors defined on the Riemann surface~$\riemann$, while $\eta^1$ and $\eta^2$ are given in~\eqref{eta}. Putting all the ingredients together, the symplectic-Majorana condition~\eqref{symp-Maj} and the Killing spinor equations~\eqref{riem_KSE_grav-group}--\eqref{riem_KSE_other} can be solved to give
\begin{equation}
\chi^1 = \begin{pmatrix} 0 \\ \xi_\chi \end{pmatrix} \,,  \qquad  \chi^2 = \begin{pmatrix} -\xi_\chi^* \\ 0 \end{pmatrix} \,,
\end{equation}
where $\xi_\eta$ is a complex constant. Notice that we can express $\chi^2$ as $\chi^2 = -\ii\,\sigma^2(\chi^1)^*$.

We can now count the number of supersymmetries preserved by our $\AdS_2 \times \riemann \ltimes \spindle_2$ solution. $\vartheta$ is a Majorana spinor, hence it has two real degrees of freedom, while the tensor product $\chi^A \otimes \eta^A$ is fully determined by the product $\xi_\chi \xi_\eta$, which has two real degrees of freedom. Therefore, there are four real independent Killing spinors, that is one quarter of the number of supersymmetries of the six-dimensional $\mc{N}=(1,1)$ theory, hence the solution is $1/4$-BPS.

We now proceed to the global analysis of the local solution~\eqref{riem_solution}, for which we set $m=2g/3$ without loss of generality\footnote{Taking $g>0$ and $m>0$, we can apply the rescaling~(2.4) of~\cite{Faedo:2021nub}, together with $B \mapsto (m/l g)^{-1/2} B$, in order to absorb $m$ in the coupling constant.}. Fixed a generic point on the Riemann surface~$\riemann$, the metric on the spindle~$\spindle_2$ reads
\begin{equation} \label{spin2_metric}
\dd s_{\spindle_2}^2 = \frac{y^2}{F} \, \dd y^2 + \frac{F}{h_1 h_2} \, \dd z^2 \,.
\end{equation}
In order to have a well-defined metric and positive scalars $X_i$ we need to take $F>0$, $h_1>0$, $h_2>0$ in a closed interval not containing the curvature singularity in $y=0$,
thus without loss of generality we restrict to $y>0$. Moreover, denoting $[\YA,\YB]$ the range of the coordinate~$y$, $\spindle_2$ is a proper spindle given that~\cite{Faedo:2021nub}
\begin{equation}
\frac{g F'(\YA)}{3\YA^3} \, \Delta z = \frac{2\pi}{n_-} \,,  \qquad\quad  \frac{g F'(\YB)}{3\YB^3} \, \Delta z = -\frac{2\pi}{n_+} \,,
\end{equation}
where $n_\pm$ are two co-prime integers and $\Delta z$ is the periodicity of the $z$ coordinate. These two relations ensure that at the poles $y=\YAB$ there are $\ZZ_{n_\mp}$ orbifold singularities, respectively. An additional constraint comes from the quantization of the magnetic fluxes across~$\spindle_2$, which arises from the requirement that $g A_i$ be well-defined connection one-forms on $U(1)$ bundles over~$\spindle_2$. Given any point of~$\riemann$, we must have
\begin{equation} \label{quant_F-spin2}
\flt_i = \frac{g}{2\pi} \int_{\spindle_2} F_i = \frac{p_i}{n_+ n_-} \,,  \qquad  p_i \in \ZZ \,.
\end{equation}
It can be shown that~\cite{Faedo:2021nub}
\begin{equation}
p_1 + p_2 = n_+ + n_- \,,
\end{equation}
meaning that the $R$-symmetry gauge field $A_R \equiv g(A_1+A_2)$ is a connection one-form on the line bundle~$\mc{O}(n_++n_-)$ over~$\spindle_2$, and therefore the integers~$p_i$ can be conveniently parameterized as
\begin{equation}
p_1 = \frac{n_+ + n_-}{2} (1 + \mathtt{z}) \,,  \qquad  p_2 = \frac{n_+ + n_-}{2} (1 - \mathtt{z}) \,,
\end{equation}
where $\mathtt{z}$ is an appropriate rational number. The value of the sum $p_1+p_2$ implies that $A_R$ realizes on the bundle the type of twist that was dubbed ``twist'', as for the D4~\cite{Faedo:2021nub} and M5-branes~\cite{Ferrero:2021wvk} wrapped on a spindle.

All the conditions we presented were studied in~\cite{Faedo:2021nub}, to which we refer for a detailed analysis, and it was proven that they can be satisfied if $n_-<n_+$, $p_1<0$, $p_2>0$, the expressions of $q_i$ and $\YAB$ are determined as in equations~(3.39) and~(3.31) of~\cite{Faedo:2021nub} and $z$ has periodicity
\begin{equation} \label{delta-z}
\Delta z = \chi_2 \, \frac{3\pi (\x^2 + 3) (\mu - \x)}{8g \x^2} \,.
\end{equation}
Here, $\chi_2$, the Euler characteristic of~$\spindle_2$, and $\mu$ are defined as
\begin{equation}\label{euler_and_mu_spin2}
\chi_2 = \frac{n_+ + n_-}{n_+ n_-} \,,  \qquad  \mu \equiv \frac{n_+ - n_-}{n_+ + n_-} \,,
\end{equation}
and $\x$ is the only solution of the quartic equation
\begin{equation} \label{quartic}
\x^4 + \bigl(8\mathtt{z}^2 - 3 - 9\mu^2\bigr) \x^2 + 12\mu \x - 9\mu^2 = 0
\end{equation}
lying inside the range $0<\x<1$.

As we already mentioned, the internal space has the structure of a spindle~$\spindle_2$ fibred over a Riemann surface~$\riemann$. This fibration is well-defined in the orbifold sense if the one-form~$\eta$ describing the fibration is globally defined, \ie
\begin{equation} \label{riem_quant-t}
	\frac{1}{2\pi} \int_\riemann \dd\eta = t \,,  \qquad  t \in \ZZ \,,
\end{equation}
where
\begin{equation}
	\eta \equiv \frac{2\pi}{\Delta z} \Bigl( \dd z - \frac{3}{4g} \, \omega_\riemann \Bigr) \,.
\end{equation}
This requirement yields a quantization condition relating the spindle data $(n_\pm, \mathtt{z})$ and the genus~$\mathrm{g}$, namely
\begin{equation} \label{riem_t}
	t = -\frac{8\x^2 (\mathrm{g} - 1)}{\chi_2 (\x^2 + 3) (\mu - \x)} \in \ZZ \,.
\end{equation}
We refer to appendix~\ref{app:quantization.exe} for the analysis of this constraint.

The last conditions follow from the quantization of the fluxes~$F_i$ through two-cycles in~$\riemann\ltimes\spindle_2$. Their integration through~$\spindle_2$ were computed in~\eqref{quant_F-spin2}, hence they are already appropriately quantized. Next, we define the two-cycles $S_- \equiv \{y=\YA\}$ and $S_+ \equiv \{y=\YB\}$, corresponding to the two poles of~$\spindle_2$, and compute, \eg,
\begin{equation} \label{riem_quantN}
	\begin{split}
		\fls_1^- &= \frac{g}{2\pi} \int_{S_-} F_1 = \frac32 \frac{\YA^3}{h_1(\YA)} \, (\mathrm{g} - 1) \\
&= t \chi_2 \, \frac{\x^3 - (2 + 2\mathtt{z} + 3\mu)\x^2 + (3 - 2\mathtt{z}) \x - 3\mu}{8 \x^2} \,,
	\end{split}
\end{equation}
where we substituted~$t$ as in~\eqref{riem_t}. Similarly to what happened in~\eqref{riem_quant-t}, the quantization condition requires~\eqref{riem_quantN} to be an integer. In appendix~\ref{app:quantization.exe} we present some examples in which this constraint is satisfied. The quantization through~$S_+$ can be addressed easily since an explicit computation shows that
\begin{equation} \label{riem_quantS}
	\fls_i^+ = \frac{g}{2\pi} \int_{S_+} F_i = \fls_i^- + t \, \frac{p_i}{n_+ n_-} \,,
\end{equation}
therefore the related condition is automatically satisfied due to the fact that $n_\pm$ divide~$t$, as explained in appendix~\ref{app:quantization.exe}. We also notice that~\eqref{riem_quantS} agrees with the homology relation $S_+ - S_- = t\,\spindle_2$. The fluxes of~$F_2$ can be obtained flipping the sign of~$\mathtt{z}$ in the previous formulas and, as a consequence, the quantization of its fluxes automatically holds exchanging $p_1$ with~$p_2$.

Focussing on the $R$-symmetry gauge field, its fluxes read
\begin{equation}
	\frac{1}{2\pi} \int_{S_-} F_R = 2(\mathrm{g} - 1) - \frac{t}{n_-} \,,  \qquad
	\frac{1}{2\pi} \int_{S_+} F_R = 2(\mathrm{g} - 1) + \frac{t}{n_+} \,,
\end{equation}
and, as expected, they are correctly quantized because $n_\pm$ divide~$t$.

\subsection{AdS$_2\times \spindle_1 \ltimes \spindle_2$ solutions}
\label{subsec:6d_2spin}

A second class of solutions to the model described by action~\eqref{6d_action} is the $\AdS_2 \times \spindle_1 \ltimes \spindle_2$ background, where a two-dimensional spindle~$\spindle_2$ is non-trivially fibred over a different two-dimensional spindle~$\spindle_1$. Remarkably, there exists a specific limit of the coordinates and the parameters in which the $\AdS_2 \times \riemann \ltimes \spindle_2$ solutions are retrieved, both their bosonic content and the Killing spinors. All the details and computations are left to appendix~\ref{app:spindle-2-riemann}. Going back to the systems under exam, they read
\begin{equation} \label{2spin_solution}
\begin{split}
\dd s^2 &= (y^2 h_1 h_2)^{1/4} \biggl[ \frac{x^2}{4} \, \dd s_{\AdS_2}^2 + \frac{x^2}{q} \, \dd x^2 + \frac{q}{4x^2} \, \dd\psi^2 + \frac{y^2}{F} \, \dd y^2 \\
& + \frac{F}{h_1 h_2} \Bigl( \dd z - \frac{1}{2m} \Bigl(1 - \frac{\mathtt{a}}{x}\Bigr) \, \dd\psi \Bigr)^2 \biggr] \,, \\
A_i &= -\frac{y^3}{h_i} \Bigl( \dd z - \frac{1}{2m} \Bigl(1 - \frac{\mathtt{a}}{x}\Bigr) \, \dd\psi \Bigr) \,,  \qquad  X_i = (y^2 h_1 h_2)^{3/8} h_i^{-1} \,, \\
B &= \frac{\mathtt{a} y}{2m} \, \vol{\AdS_2} \,,
\end{split}
\end{equation}
where $\dd s_{\AdS_2}^2$ denotes the unit radius metric on~$\AdS_2$ and~$\mathtt{a}$ is a real parameter. The functions $h_i$ and $F$ are the same as in~\eqref{func-h+F}, while $q$ is given by
\begin{equation}
q(x) = x^4 - 4x^2 + 4\mathtt{a} x - \mathtt{a}^2 \,.
\end{equation}
Notice that formally taking $\mathtt{a}=0$ we retrieve the $\AdS_4\times\spindle$ background studied in~\cite{Faedo:2021nub}.

In order to construct the local form of the Killing spinors and to demonstrate that this solution is supersymmetric, we must first specialize the Killing spinor equations~\eqref{KSE_grav}--\eqref{KSE_gauge} to our system employing the orthonormal frame
\begin{gather}
	e^{\hat{a}} = \frac{x \, y^{1/4} (h_1 h_2)^{1/8}}{2} \, \hat{e}^{\hat{a}} \,,  \qquad  e^2 = \frac{x \, y^{1/4} (h_1 h_2)^{1/8}}{q^{1/2}} \, \dd x \,,  \qquad  e^3 = \frac{q^{1/2} y^{1/4} (h_1 h_2)^{1/8}}{2x} \, \dd\psi \,, \nonumber \\
	e^4 = \frac{y^{5/4} (h_1 h_2)^{1/8}}{F^{1/2}} \, \dd y \,,  \qquad  e^5 = \frac{y^{1/4} F^{1/2}}{(h_1 h_2)^{3/8}} \Bigl( \dd z - \frac{1}{2m} \Bigl(1 - \frac{\mathtt{a}}{x}\Bigr) \, \dd\psi \Bigr) \,,
\end{gather}
where $\hat{e}^{\hat{a}}$, $\hat{a}=0,1$, is the zweibein on $\AdS_2$, whose coordinates are denoted as $x^{\hat{\mu}}$. Equation~\eqref{KSE_grav} then splits into the following system
\begin{equation} \label{2spin_KSE_grav-group}
\begin{split}
\biggl( \partial_{\hat{\mu}} + \frac14 \omega_{\hat{\mu}}^{\ \hat{a}\hat{b}} \Gamma_{\hat{a}\hat{b}} \biggr) \epsilon^A + \hat{e}^{\hat{a}}_{\hat{\mu}} \biggl[ \frac{q^{1/2}}{4x} \, \Gamma_{\hat{a}}^{\ 2} \epsilon^A - \ii \, \frac{x}{4} \, \Gamma_{\hat{a}}^{\ 45} (\sigma^3)^A_{\,\ B} \epsilon^B + \frac{\mathtt{a}}{4x} \, \Gamma_{\hat{a}}^{\ 2345} \epsilon^A \biggr] &= 0 \,, \\
\partial_x \epsilon^A - \frac{\mathtt{a}}{2x q^{1/2}} \, \Gamma^{345} \epsilon^A - \ii \, \frac{x}{2q^{1/2}} \, \Gamma^{245} (\sigma^3)^A_{\,\ B} \epsilon^B &= 0 \,, \\
\partial_\psi \epsilon^A - \frac{\ii}{2} \Bigl(1 - \frac{\mathtt{a}}{x}\Bigr) (\sigma^3)^A_{\,\ B} \epsilon^B + \frac{2q - x q'}{8x^3} \, \Gamma^{23} \epsilon^A + \frac{\mathtt{a} q^{1/2}}{4x^3} \, \Gamma^{245} \epsilon^A & \\
- \ii \, \frac{q^{1/2}}{4x} \, \Gamma^{345} (\sigma^3)^A_{\,\ B} \epsilon^B &= 0 \,, \\
\partial_y \epsilon^A - \frac{1}{16 y} \biggl[ (2 + y \tilde h') \epsilon^A - \ii\,\frac{4y^2}{F^{1/2}} (4 - y \tilde h') \Gamma^5 (\sigma^3)^A_{\,\ B} \epsilon^B \biggr] &= 0 \,, \\
\partial_z \epsilon^A - \ii\,\frac{g}{2} \biggl( \alpha_1+\alpha_2 - \frac{2m}{g} \biggr) (\sigma^3)^A_{\,\ B} \epsilon^B &= 0 \,,
\end{split}
\end{equation}
where $\tilde h \equiv \log ( h_1h_2)$ is defined as before and we employed the gauge transformation $A_i \mapsto A_i + \alpha_i \, \dd z$. Equations~\eqref{KSE_dila} and~\eqref{KSE_gauge} yield, again, the same constraint
\begin{equation} \label{2spin_KSE_other}
F^{1/2} \Gamma^4 \epsilon^A + m (h_1 h_2)^{1/2} \epsilon^A + \ii\,y^2 \Gamma^{45} (\sigma^3)^A_{\,\ B} \epsilon^B = 0 \,.
\end{equation}
Also in this case, the Killing spinor equations along~$y$ and~$z$ and the last constraint are identical to those obtained in the $\AdS_4\times\spindle$ background. Once again we fix the gauge such that $\alpha_1+\alpha_2 = 2m/g$, in order to have a $z$-independent Killing spinors.

The explicit construction of the Killing spinors proceeds in a way similar to the $\AdS_2 \times \riemann \ltimes \spindle_2$ case. First of all, we employ the following decomposition of the six-dimensional gamma matrices (\cf\ equations~\eqref{gamma-deco_6to4} and~\eqref{gamma-deco_4to2}):
\begin{equation}
\Gamma^{\hat{a}} = \beta^{\hat{a}} \otimes \tau_* \otimes \rho_* \,,  \qquad  \Gamma^{2,3} = I_2 \otimes \tau^{1,2} \otimes \rho_* \,,  \qquad  \Gamma^{4,5} = I_2 \otimes I_2 \otimes \rho^{1,2} \,,
\end{equation}
which implies,
\begin{equation}
\mc{B}_6 = \mc{B}_2^\beta \otimes (\mc{B}_2^\tau \tau_*) \otimes (\mc{B}_2^\rho \rho_*) \,,
\end{equation}
where $\beta^{\hat{a}}$ are the (Lorentzian) gamma matrices in $D=2$, $\rho^{\hat{\imath}}$ and $\tau^{\hat{\imath}}$, $\hat{\imath}=1,2$, are the (Euclidean) gamma matrices in $D=2$ and $\rho_*$ and $\tau_*$ are the related chiral matrices. In the Euclidean sectors we adopt the representation~\eqref{2d-gamma_rep}
\begin{equation}
\rho^{\hat{\imath}} = \tau^{\hat{\imath}} = \sigma^{\hat{\imath}} \,,  \qquad\quad  \rho_* = \tau_* = \sigma^3 \,,
\end{equation}
and we take $\mc{B}_2^\rho=-\sigma^2$ and $\mc{B}_2^\tau=\sigma^1$. The six-dimensional Killing spinors are assumed to consist of the tensor product of spinors living on~$\AdS_2$, $\spindle_1$ and~$\spindle_2$. Namely,
\begin{equation}
\epsilon^A = \vartheta \otimes \chi^A \otimes \eta^A \,,
\end{equation}
with $\vartheta=\vartheta(x^{\hat{\mu}})$ Majorana Killing spinor on~$\AdS_2$, $\chi^A=\chi^A(x,\psi)$ two two-component spinors defined on the spindle~$\spindle_1$ and $\eta^A$ given in~\eqref{eta}. The symplectic-Majorana condition~\eqref{symp-Maj} and the Killing spinor equations \eqref{2spin_KSE_grav-group}--\eqref{2spin_KSE_other} are satisfied given that
\begin{equation}
\chi^1 = \xi_\chi \, x^{-1/2} \begin{pmatrix} Q_1^{1/2} \\ -Q_2^{1/2} \end{pmatrix} \,,  \qquad
\chi^2 = \xi_\chi^* \, x^{-1/2} \begin{pmatrix} Q_2^{1/2} \\ Q_1^{1/2} \end{pmatrix} \,,
\end{equation}
where $\xi_\chi$ is a complex constant and
\begin{equation}
Q_1(x) \equiv x^2 - (2x - \mathtt{a}) \,,  \qquad  Q_2(x) \equiv x^2 + (2x - \mathtt{a}) \,,
\end{equation}
which satisfy $q(x)=Q_1(x)\,Q_2(x)$. Notice that the relation $\chi^2 = -\ii\,\sigma^2(\chi^1)^*$ holds also in this case. The count of the real degrees of freedom is identical to the $\AdS_2 \times \riemann \ltimes \spindle_2$ case and shows that the solution is $1/4$-BPS, since it preserves four real supercharges.

The first part of the global analysis of the $\AdS_2 \times \spindle_1 \ltimes \spindle_2$ solution~\eqref{2spin_solution} proceeds along the path traced in the previous section. Chosen a generic point on the base spindle~$\spindle_1$, the metric of~$\spindle_2$ is the same as in~\eqref{spin2_metric}, therefore are identical also the conditions for it to describe a proper spindle with $\ZZ_{n_\mp}$ orbifold singularities. In particular, $\Delta z$ must be as in~\eqref{delta-z}. Moreover, at a generic point of~$\spindle_1$ (different from the north and south poles) the magnetic fluxes are quantized as in~\eqref{quant_F-spin2}:
\begin{equation}
\label{nonstackyintegral}
\flt_i = \frac{g}{2\pi} \int_{\spindle_2} F_i = \frac{p_i}{n_+ n_-} \,,  \qquad  p_i \in \ZZ \,,
\end{equation}
with $p_1 + p_2 = n_+ + n_-$.

The base spindle~$\spindle_1$, described by the metric
\begin{equation} \label{spin1_metric}
\dd s_{\spindle_1}^2 = \frac{x^2}{q} \, \dd x^2 + \frac{q}{4x^2} \, \dd\psi^2 \,,
\end{equation}
was thoroughly studied in~\cite{Ferrero:2020twa}, thus we will take advantage of the results presented therein, setting $\mathtt{j}=0$. $\spindle_1$ is indeed a spindle, characterized by the two co-prime integers~$m_\pm$, with $m_-<m_+$\footnote{Notice that we exchanged $m_-$ and $m_+$ with respect to~\cite{Ferrero:2020twa}.}, if the parameter~$\mathtt{a}$ and the periodicity of the coordinate~$\psi$ are
\begin{equation} \label{spin1_a+psi}
\mathtt{a} = \frac{m_+^2 - m_-^2}{m_+^2 + m_-^2} \,,  \qquad  \Delta\psi = \frac{\sqrt{m_+^2 + m_-^2}}{\sqrt2 \, m_+ m_-} \, 2\pi \,.
\end{equation}
The coordinate~$x$ is restricted between the two middle roots of the quartic polynomial~$q(x)$, namely $x\in[\XA,\XB]$ with
\begin{equation}
\XA = -1 + \frac{\sqrt2 \, m_+}{\sqrt{m_+^2 + m_-^2}} \,,  \qquad  \XB = 1 - \frac{\sqrt2 \, m_-}{\sqrt{m_+^2 + m_-^2}} \,.
\end{equation}
At the north and south poles of~$\spindle_1$, namely $x=\XAB$, are present $\ZZ_{m_\mp}$ orbifold singularities, respectively. Moreover, defined the vector field
\begin{equation}
A_{4\mathrm{d}} = \frac12 \Bigl( 1 - \frac{\mathtt{a}}{x} \Bigr) \, \dd\psi \,,
\end{equation}
its magnetic flux through the spindle~$\spindle_1$ is
\begin{equation} \label{4d_flux}
\frac{1}{2\pi} \int_{\spindle_1} F_{4\mathrm{d}} = \frac{m_+ - m_-}{2 m_+ m_-} \,.
\end{equation}
In the four-dimensional theory this implies that $2A_{4\mathrm{d}}$ is a connection one-form on the line bundle~$\mc{O}(m_+-m_-)$ over~$\spindle_1$ and, as a consequence, that we have anti-twist.

Similarly to the $\riemann \ltimes \spindle_2$ system, the fibration of $\spindle_2$ over~$\spindle_1$ is well-defined given that
\begin{equation}
\frac{1}{2\pi} \int_{\spindle_1} \dd\eta = \frac{t}{m_+ m_-} \,,  \qquad  t \in \ZZ \,,
\label{pippo}
\end{equation}
where, in this case,
\begin{equation}
\eta \equiv \frac{2\pi}{\Delta z} \Bigl( \dd z - \frac{3}{4g} \Bigl(1 - \frac{\mathtt{a}}{x}\Bigr) \, \dd\psi \Bigr) \,.
\end{equation}
We derive the following constraint on the quantum numbers~$n_\pm$, $m_\pm$ and~$\mathtt{z}$
\begin{equation} \label{2spin_t}
t =  -\frac{4\x^2 (m_+ - m_-)}{\chi_2 (\x^2 + 3) (\mu - \x)} \in \ZZ \,,
\end{equation}
which is solved in appendix~\ref{app:quantization.exe}.

Like we did in the previous section, we study the quantization of the fluxes of~$F_i$, now through two-cycles in~$\spindle_1\ltimes\spindle_2$. The magnetic fluxes across~$\spindle_2$ were computed in~\eqref{quant_F-spin2} and are, thus, already quantized. Again we define the two-cycles $S_- \equiv \{y=\YA\}$ and $S_+ \equiv \{y=\YB\}$, which corresponds to two copies of the base spindle~$\spindle_1$, and compute
\begin{equation} \label{2spin_quantN}
	\begin{split}
		\fls_1^- &= \frac{g}{2\pi} \int_{S_-} F_1 = \frac{3}{4} \frac{\YA^3}{h_1(\YA)} \frac{m_+ - m_-}{m_+ m_-} \\
&= \frac{t \chi_2}{m_+ m_-} \, \frac{\x^3 - (2+2\mathtt{z}+3\mu)\x^2 + (3-2\mathtt{z})\x - 3\mu}{8\x^2} \,.
	\end{split}
\end{equation}
In analogy to~\eqref{riem_quantN}, the flux~\eqref{2spin_quantN} must be an integer divided by $(m_+ m_-)$, and we refer to appendix~\ref{app:quantization.exe} for the analysis of this constraint. The same quantization condition, but with $F_1$ integrated across~$S_+$, is automatically satisfied due to the relation
\begin{equation} \label{2spin_quantS}
	\fls_i^+ = \frac{g}{2\pi} \int_{S_+} F_i = \fls_i^- + \frac{t}{m_+ m_-} \, \frac{p_i}{n_+ n_-} \,,
\end{equation}
which can be both computed explicitly and obtained from the homology relation $S_+ - S_- = \frac{t}{m_+ m_-}\,\spindle_2$. On the other hand, the fluxes of~$F_2$ can be obtained switching the sign of~$\mathtt{z}$, \ie\ exchanging $p_1$ with~$p_2$, in the previous formulas and, as a consequence, the quantization follows directly.
The fluxes of the $R$-symmetry gauge potential read
\begin{equation}
	\frac{1}{2\pi} \int_{S_-} F_R = \frac{m_+ - m_-}{m_+ m_-} - \frac{t}{m_+ m_- n_-} \,,  \qquad
	\frac{1}{2\pi} \int_{S_+} F_R = \frac{m_+ - m_-}{m_+ m_-} + \frac{t}{m_+ m_- n_+} \,,
\end{equation}
and are well quantized, \ie\ they are integers divided by $m_+ m_-$, because $n_\pm$ divide~$t$.

\subsection{Truncation to $D=4$ minimal gauged supergravity}

An accurate inspection of the $\AdS_2 \times \riemann \ltimes \spindle_2$ and $\AdS_2 \times \spindle_1 \ltimes \spindle_2$ solutions given in~\eqref{riem_solution} and~\eqref{2spin_solution}, respectively, hints at the existence of an underlying consistent truncation of matter-coupled $D=6$, $U(1)^2$ gauged supergravity on a spindle~$\spindle_2$, down to $D=4$, $\mathcal{N}=2$ minimal gauged supergravity. Even though we do not demonstrate the consistency of this truncation\footnote{This truncation has been proved to be consistent and to preserve supersymmetry in~\cite{Couzens:2022lvg}.}, we conjecture its validity in analogy with the results of~\cite{Cheung:2022ilc} and we will provide supporting evidence.
For completeness, we report here the action describing the four-dimensional theory
\begin{equation}
\begin{split}
S_\text{4D} &= \frac{1}{16\pi G_{(4)}} \int \dd^4x \, \sqrt{-g} \, \bigl( {}^{(4)}\!R + 6 - \mathscr{F}_{\mu\nu} \mathscr{F}^{\mu\nu} \bigr) \,,
\end{split}
\end{equation}
along with the corresponding Killing spinor equations
\begin{equation}
\Bigl( \nabla_\mu - \ii \, \mathscr{A}_\mu + \frac12 \, \gamma_\mu + \frac{\ii}{4} \, \mathscr{F}_{\rho\sigma} \gamma^{\rho\sigma} \gamma_\mu \Bigr) \epsilon = 0 \,,
\end{equation}
where $\mathscr{F} = \dd\mathscr{A}$, $\{\gamma_a,\gamma_b\}=2\eta_{ab}$ and the $\AdS_4$ radius has been set to 1.

Given a solution of $D=4$ minimal gauged supergravity comprising a metric~$\dd s_{(4)}^2$ and a gauge potential~$\mathscr{A}$, we propose the following ansatz for the six-dimensional solution:
\begin{equation}
\begin{aligned} \label{truncation}
\dd s_{(6)}^2 &= (y^2 h_1 h_2)^{1/4} \biggl[ \dd s_{(4)}^2 + \frac{y^2}{F} \, \dd y^2 + \frac{F}{h_1 h_2} \Bigl( \dd z - \frac{1}{m} \, \mathscr{A} \Bigr)^2 \biggr] \,, \\
A_i &= -\frac{y^3}{h_i} \Bigl( \dd z - \frac{1}{m} \, \mathscr{A} \Bigr) \,,  \qquad  X_i = (y^2 h_1 h_2)^{3/8} h_i^{-1} \,, \\[0.25em]
B &= \frac{2y}{m} \star_4\!\mathscr{F} \,.
\end{aligned}
\end{equation}
Here, $\star_4$ is the Hodge dual with respect to the metric~$\dd s_{(4)}^2$ and (\cf~\eqref{func-h+F})
\begin{equation}
h_i(y) = \frac{2g}{3m} \, y^3 + q_i \,,  \qquad  F(y) = m^2 h_1 h_2 - y^4 \,.
\end{equation}

This ansatz may be justified comparing the two six-dimensional solutions already mentioned with the following $\AdS_2 \times \riemann$~\cite{Caldarelli:1998hg} and $\AdS_2 \times \spindle_1$~\cite{Ferrero:2020twa} backgrounds
\begin{align}
\label{4d_riemann}
\dd s_{(4)}^2 &= \frac14 \, \dd s_{\AdS_2}^2 + \frac12 \, \dd s_{\riemann}^2 \,,  \qquad
& \mathscr{A} &= \frac12 \, \omega_\riemann \,, \\
\label{4d_spindle}
\dd s_{(4)}^2 &= \frac{x^2}{4} \, \dd s_{\AdS_2}^2 + \frac{x^2}{q} \, \dd x^2 + \frac{q}{4x^2} \, \dd\psi^2 \,,  \qquad
& \mathscr{A} &= \frac12 \Bigl(1 - \frac{\mathtt{a}}{x}\Bigr) \, \dd\psi \,,
\end{align}
with $q(x) = x^4 - 4x^2 + 4\mathtt{a} x - \mathtt{a}^2$. In both cases, plugging the four-dimensional solutions into the ansatz~\eqref{truncation} gives the corresponding ones in $D=6$. An additional  example is given by the $\AdS_4 \times \spindle_2$ solution of~\cite{Faedo:2021nub}, which can be obtained starting from the $\AdS_4$ vacuum.

In addition to the truncation of the bosonic sector, we also propose an ansatz connecting the Killing spinors of the two supergravity theories in $D=4$ and $D=6$. This conjecture is driven both by the construction presented in~\cite{Cheung:2022ilc} and by the fact, already observed, that some of the Killing spinor equations are the same for the two background considered in this paper, as well as  for the $\AdS_4 \times \spindle_2$ solution.

The truncation ansatz for the Killing spinors heavily depends on the choice of the decomposition of the six-dimensional gamma matrices. In what follows we will consider~\eqref{gamma-deco_6to4}
\begin{equation}
\Gamma^{\tilde{a}} = \gamma^{\tilde{a}} \otimes \sigma^3 \,,  \qquad\quad  \Gamma^{4,5} = I_4 \otimes \sigma^{1,2} \,,
\end{equation}
where $\gamma^{\tilde{a}}$, $\tilde{a}=0,\ldots,3$, are the (Lorentzian) gamma matrices in $D=4$.  For consistency, $\mc{B}_6$ decomposes as
\begin{equation}
\mc{B}_6 = (\mc{B}_4 \gamma_5) \otimes (\mc{B}_2 \sigma^3) \,,
\end{equation}
where $\gamma_5 = \ii\,\gamma^0\gamma^1\gamma^2\gamma^3$ is the four-dimensional chiral matrix and $\mc{B}_2=-\sigma^2$. We conjecture that the first Killing spinor~$\epsilon^1$ takes the expression
\begin{equation}
\epsilon^1 = \zeta \otimes \eta \,,
\end{equation}
where $\zeta$ is a Killing spinor of $D=4$ minimal gauged supergravity and $\eta$ is given by\footnote{Here we relabelled~$\eta^1$ with~$\eta$ and $\xi_\eta$ with~$\xi$ with respect to~\eqref{eta}.}
\begin{equation}
\eta = \xi \, y^{1/8} (h_1 h_2)^{-3/16} \begin{pmatrix} f_1^{1/2} \\ -f_2^{1/2} \end{pmatrix} \,,
\end{equation}
where $\xi$ is a complex constant and
\begin{equation}
f_1(y) \equiv m (h_1 h_2)^{1/2} + y^2 \,,  \qquad  f_2(y) \equiv m (h_1 h_2)^{1/2} - y^2 \,,
\end{equation}
satisfying $F(y)=f_1(y)\,f_2(y)$. The second Killing spinor $\epsilon^2$ is determined by the symplectic-Majorana condition~\eqref{symp-Maj}, explicitly
\begin{equation}
\epsilon^2 = \ii \, (\gamma_5 \zeta^c) \otimes (\sigma^1 \eta^*) \,,
\end{equation}
where we defined the charge conjugate spinor $\zeta^c = \mc{B}_4^{-1} \zeta^*$.

As for the bosonic part of the truncation, we can justify this ansatz making contact with the Killing spinors of $D=4$ minimal gauged supergravity. Assuming, additionally, the decomposition of the four-dimensional gamma matrices in~\eqref{gamma-deco_4to2}, the Killing spinors of the $\AdS_2 \times \riemann$~\cite{Caldarelli:1998hg} and $\AdS_2 \times \spindle_1$~\cite{Ferrero:2021ovq} (see also~\cite{Ferrero:2020twa}) backgrounds are respectively\footnote{Here $\zeta$ is a superposition of the two Killing spinors~$\widetilde{\epsilon}_1$ and~$\widetilde{\epsilon}_2$ of~\cite{Ferrero:2021ovq}.}
\begin{align}
	\label{4d_riemann-killing}
	\zeta &= \vartheta \otimes \begin{pmatrix} 0 \\ 1 \end{pmatrix} \,, \\
	\label{4d_spindle-killing}
	\zeta &= \vartheta \otimes x^{-1/2} \begin{pmatrix} Q_1^{1/2} \\ -Q_2^{1/2} \end{pmatrix} \,,
\end{align}
where $\vartheta$ is a Majorana Killing spinor on $\AdS_2$ and we defined $Q_1(x) \equiv x^2 - (2x - \mathtt{a})$ and $Q_2(x) \equiv x^2 + (2x - \mathtt{a})$. Uplifting these two spinors by means of the proposed recipe we obtain exactly the Killing spinors of the corresponding six-dimensional solution. For what concerns the $\AdS_4 \times \spindle_2$ background, we must take $\zeta = \gamma_5\vartheta_{(4)}$, with $\vartheta_{(4)}$ a Killing spinor on $\AdS_4$, due to a slightly different convention on the four-dimensional Killing spinor equations.

The existence and the consistency of the truncation for the Killing spinors would prove that any supersymmetric solution of $D=4$ minimal gauged supergravity gives rise to a supersymmetric solution of $D=6$, $U(1)^2$ gauged supergravity after uplifting on $\spindle_2$ via~\eqref{truncation}.

\section{Uplift to massive type IIA supergravity}
\label{sec:uplift}

The local solutions presented in section~\ref{sec:6dsugra} can be embedded in massive type~IIA supergravity by means of an appropriate ansatz. A consistent truncation for the field content of interest is unfortunately not available. However, if we set $m=2g/3$ we can consider an enlarged version of the ansatz presented in~\cite{Faedo:2021nub}.
This proposal\footnote{A more general truncation ansatz, which does not require $m = 2g/3$, has been proposed in~\cite{Couzens:2022lvg}.} comes as a superposition of the formulas presented in~\cite{Cvetic:1999un} and~\cite{Cvetic:1999xx}, which we recover in the respective limits. In the former only one scalar is considered, but the six-dimensional two-form $B$ is non-zero, while the latter encompasses four scalar fields, but vanishing $B$ field. The ansatz for the metric in the string frame and the dilaton is the same as in~\cite{Faedo:2021nub}
\begin{align}
	\begin{split} \label{uplift_metric}
		\dd s_\mathrm{s.f.}^2 &= \mu_0^{-1/3} (X_1 X_2)^{-1/4} \bigl\{ \Delta^{1/2} \dd s_6^2 \\
		& + g^{-2} \Delta^{-1/2} \bigl[ X_0^{-1} \dd\mu_0^2 + X_1^{-1} \bigl(\dd\mu_1^2 + \mu_1^2 \sigma_1^2\bigr) + X_2^{-1} \bigl(\dd\mu_2^2 + \mu_2^2 \sigma_2^2\bigr) \bigr] \bigr\} \,,
	\end{split} \\
	\ee^\Phi &= \mu_0^{-5/6} \Delta^{1/4} (X_1 X_2)^{-5/8} \,,
\end{align}
but now we have also a non-vanishing two-form field
\begin{equation}
	B_{(2)} = \frac{1}{2} \mu_0^{2/3} B \,.
\end{equation}
Here $\dd s_6^2$ stands for the line element of the six-dimensional space~$M_6$ and the one-forms $\sigma_i \equiv \dd \phi_i -g A_i$ are built up from the six-dimensional gauge fields. The angular coordinates~$\phi_1$, $\phi_2$ have canonical $2\pi$ periodicities, whereas the coordinates $\mu_a$, with $a=0,1,2$, satisfy the constraint $\sum \mu_a ^2=1$. Lastly, the warp factor reads
\begin{equation}
  \Delta = \sum_{a=0}^2 X_a \mu_a^2 \,.
\end{equation}
A possible parameterization of the coordinates~$\mu_a$ is
\begin{equation}
  \mu_0 = \sin\xi \,,  \qquad  \mu_1 = \cos\xi \sin\eta \,,  \qquad  \mu_2 = \cos\xi \cos\eta \,,
\end{equation}
with $\eta\in[0,\pi/2]$ and $\xi\in(0,\pi/2]$, where the range of $\xi$ is fixed by the necessity of having $\mu_0 >0$. In this way the ten-dimensional metric~\eqref{uplift_metric} parameterizes, at each point of~$M_6$, a four-dimensional hemisphere, denoted as $\hemi$ from now on. The metric on $\hemi$ is in general squashed, and becomes the metric on ``half the round four-sphere'' when $X_1 =X_2 =1$.
The RR sector of massive type~IIA supergravity comprises the ten-dimensional Romans mass and the two-form flux
\begin{equation}
	F_{(0)} = m = \frac{2g}{3} \,,  \qquad  F_{(2)} = \frac{g}{3} \mu_0^{2/3} B \,,
\end{equation}
along with the four-form flux, conveniently written in terms of its Hodge dual as\footnote{The four-flux differs in a global sign from~(2.16) of~\cite{Faedo:2021nub}. Although this fact does not affect the equations of motion therein, which, in that case, are invariant under $F_{(4)} \mapsto -F_{(4)}$, the expression~\eqref{hodge star F4 ansatz} correctly reduces to~\cite{Cvetic:1999un} when only one scalar field is retained.}
\begin{align}
	\star_{10} F_{(4)} &= -g U \vol{M_6} + \frac{1}{g^2} \sum_i X_i^{-2} \mu_i (\star_6 F_i) \wedge \dd\mu_i \wedge \sigma_i - \frac{1}{g} \sum_a X_a^{-1} \mu_a (\star_6 \dd X_a) \wedge \dd\mu_a \nonumber \\
  \label{hodge star F4 ansatz}
	& + \frac{\mu_1 \mu_2}{3g^3} \, \Delta^{-1} X_0 \, B \wedge \dd\mu_1 \wedge \dd\mu_2 \wedge \sigma_1 \wedge \sigma_2 \\[0.25em]
	& - \frac{\mu_1 \mu_2}{2g^3} \, \Delta^{-1} \, H \wedge (X_2 \mu_2 \, \dd\mu_1 - X_1 \mu_1 \, \dd\mu_2) \wedge \sigma_1 \wedge \sigma_2 \,, \nonumber
\end{align}
where we defined
\begin{equation}
  U = 2 \sum_{a=0}^2 X_a^2 \mu_a^2- \biggl[\frac{4}{3}X_0 + 2(X_1+X_2) \biggr] \Delta \,.
\end{equation}
The Hodge star operators $\star_{10}$ and $\star_{6}$ are computed using the string frame metric~\eqref{uplift_metric} and $\dd s_6^2$, respectively. Even though we did not prove that this truncation is consistent, we tested it successfully against our $\AdS_2 \times \riemann \ltimes \spindle_2$~\eqref{riem_solution} and $\AdS_2 \times \spindle_1 \ltimes \spindle_2$~\eqref{2spin_solution} solutions.

Provided the equations of motion of the six-dimensional gauged supergravity are satisfied, the above ansatz should solve the equations of motion of massive type~IIA supergravity, which can be derived from the string frame action\footnote{Here we use the shortcut $B_{(2)}^n$ to denote the wedge product of $B_{(2)}$ with itself $n$ times, divided by~$n!$.}
\begin{align}
	S_\mathrm{mIIA} = \frac{1}{16\pi G_{(10)}} \biggl\{ &\int d^{10}x \sqrt{-g} \Bigl[ \ee^{-2\Phi} \bigl( R + 4|\dd\Phi|^2 - \frac12|H_{(3)}|^2 \bigr) -\frac12\bigl( F_{(0)}^2 + |F_{(2)}|^2 + |F_{(4)}|^2 \bigl) \Bigr] \nonumber \\
	- \frac12 &\int \bigl( B_{(2)} \wedge \dd C_{(3)} \wedge \dd C_{(3)} + 2F_{(0)} B_{(2)}^3 \wedge \dd C_{(3)} + 6F_{(0)}^2 B_{(2)}^5 \bigr) \biggr\} ,
\end{align}
where, in these conventions, the field strengths are
\begin{equation}
	H_{(3)} = \dd B_{(2)} \,,  \quad  F_{(2)} = \dd C_{(1)} + F_{(0)} B_{(2)} \,,  \quad  F_{(4)} = \dd C_{(3)} - H_{(3)} \wedge C_{(1)} + \frac12 F_{(0)} B_{(2)} \wedge B_{(2)} \,.
\end{equation}

As noticed in \cite{Faedo:2021nub}, the ten-dimensional equations of motion are invariant under a scaling symmetry, which defines an ``improved uplift'' and whose action is
\begin{equation} \label{scaling symmetry}
	\begin{aligned}
		\dd\hat{s}_\mathrm{s.f.}^2 &= \lambda^ 2 \dd s^{2}_\mathrm{s.f.} \,,  \qquad  & \ee^{\hat{\Phi}} &= \lambda^{2}\ee^{\Phi} \,,  \qquad  & \hat{B}_{(2)} &= \lambda^2 B_{(2)} \,,  \\
    \hat{F}_{(0)} &= \lambda^{-3} F_{(0)} \,,  \qquad  & \hat{C}_{(1)} &= \lambda^{-1} C_{(1)} \,,  \qquad  & \hat{C}_{(3)} &= \lambda \, C_{(3)} \,,
	\end{aligned}
\end{equation}
where $\lambda$ is any strictly positive constant. This additional parameter~$\lambda$ enters in the regularity analysis of the uplifted ten-dimensional solutions. In particular, its presence will be necessary for the fluxes to be correctly quantized \cite{Faedo:2021nub}.

In the next subsections we will specify our proposed truncation ansatz to the solutions presented in~\eqref{riem_solution} and~\eqref{2spin_solution}. After having applied the scaling symmetry~\eqref{scaling symmetry}, we will quantize the fluxes, thus ensuring the global regularity of the ten-dimensional solutions.

\subsection{AdS$_2\times \riemann \ltimes \spindle_2$ solutions}

We begin this section by presenting the key ingredients for the flux quantization and the  computation of the entropy. The quantization conditions are
\begin{equation} \label{10d_quantization}
	(2\pi\ell_s) F_{(0)} = n_0 \in \NN \,,  \quad  \frac{1}{(2\pi\ell_s)^3} \int_{\hemi} F_{(4)} = N \in \NN \,,  \quad  \frac{1}{(2\pi \ell_s)^3} \int_{\Morb_4} F_{(4)} = K \in \NN \,,
\end{equation}
where $\Morb_4 = \riemann\ltimes\spindle_2$. At $\xi=0$, where $\mu_0=0$ and the warp factor is singular, there are an O8-plane and $N_f =8-n_0 $ coincident D8-branes, while $N$ D4-branes are wrapped over $\riemann\ltimes\spindle_2$. These integers get contributions from the following fluxes
\begin{align}
	\label{Romans_mass}
	F_{(0)} &= \frac{2g}{3\lambda^3} \,, \\
	F_{(4)}&= -\frac{\lambda \mu_0^{1/3} h_1 h_2}{g^3 \Delta_h} \bigg\{ \frac{U_h}{\Delta_h} \frac{\mu_1 \mu_2}{\mu_0} \, \dd\mu_1 \wedge \dd\mu_2 \wedge \sigma_1 \wedge \sigma_2 \nonumber \\
	& - \sum_{i\neq j} \Bigl[ g\,F_i \wedge \dd\phi_j - \frac{y^3 (h_i' - 3y^{-1} h_i) h_j}{\Delta_h h_i} \mu_i^2 \, \dd y \wedge \sigma_i \wedge \sigma_j \Bigr] \wedge \bigl( \mu_0 \mu_j \, \dd\mu_j - y^3 h_j^{-1} \mu_j^2 \, \dd\mu_0 \bigr) \bigg\} \nonumber \\
	\label{F4_riem}
	& + \frac{\lambda y}{2} \mu_0^{4/3} \, \vol{\riemann} \wedge \dd y \wedge \dd z - \frac{\lambda y^2}{3} \mu_0^{1/3} \, \vol{\riemann} \wedge \dd z \wedge \dd\mu_0 \,,
\end{align}
while the metric and the dilaton enter in the computation of the entropy
\begin{align}
	\begin{split} \label{explicit uplifted metric riemann}
		\dd s_\mathrm{s.f.}^2 &= \lambda^2 \mu_0^{-1/3} y^{-1} \Delta_h^{1/2} \bigg\{ \frac{1}{4} \, \dd s_{\AdS_2}^2 + \dd s_{\riemann}^2 + \frac{y^2}{F} \, \dd y^2 + \frac{F}{h_1 h_2} \Bigl( \dd z - \frac{1}{2m} \, \omega_\riemann \Bigr)^2 \\
		& + g^{-2} y \, \Delta_h^{-1} \bigl[ y^3 \, \dd\mu_0^2 + h_1 \bigl(\dd\mu_1^2 + \mu_1^2 \sigma_1^2\bigr) + h_2 \bigl(\dd\mu_2^2 + \mu_2^2 \sigma_2^2\bigr) \bigr] \bigg\} \,,
	\end{split} \\
  \ee^{\Phi} &= \lambda^2 \mu_0^{-5/6} y^{-3/2} \Delta_h^{1/4} \,.
\end{align}
For convenience, we defined the functions
\begin{equation} \label{DUh}
  \begin{split}
    \Delta_h &= h_1 h_2 \, \mu_0^2 + y^3 h_2 \, \mu_1^2 + y^3 h_1 \, \mu_2^2 \,, \\
    U_h &= 2 \bigl[ (y^3 - h_1)(y^3 - h_2) \mu_0^2 - y^6 \bigr] - \frac43 \Delta_h \,.
  \end{split}
\end{equation}
Additionally, we report for completeness also the other ten-dimensional fields, which however do not play an important role in our analysis
\begin{equation}
  B_{(2)} = \frac{\lambda^2 y}{4m} \mu_0^{2/3} \, \vol{\AdS_2} \,,  \qquad  F_{(2)} = \frac{gy}{6\lambda m} \mu_0^{2/3} \, \vol{\AdS_2} \,.
\end{equation}

Plugging our solution in the first two conditions of~\eqref{10d_quantization} gives
\begin{equation} \label{10d_quant_cond_global}
  g^8 = \frac{1}{(2\pi\ell_s)^8} \frac{18\pi^6}{N^3 n_0} \,,  \qquad  \quad \lambda^8 = \frac{8\pi^2}{9N n_0^3} \,.
\end{equation}
It is worth noting that, being $N$ and $n_0$ integers, the second equation would be inconsistent with $\lambda=1$. Such a problem arises from the fact that, without introducing~$\lambda$, the first two constraints in~\eqref{10d_quantization} would have to be imposed on an unique dimensionless parameter, namely $(g\ell_s)$. This makes the scaling symmetry~\eqref{scaling symmetry} crucial to establish the regularity of the ten-dimensional uplifted solution.
The integration in the third condition of~\eqref{10d_quantization} is performed along a representative of~$\Morb_4$, which we take to be at the pole of the hemisphere~$\hemi$, \ie\ in $\xi=\pi/2$. The result can be implicitly expressed in terms of the integer parameters $(\mathrm{g},n_\pm,\mathtt{z},N)$ by means of equations~(3.31) and~(3.38) of~\cite{Faedo:2021nub}, and reads
\begin{equation} \label{F4 flux riemann}
  K = N \chi_2 (\mathrm{g} - 1) \, \frac{3 [3\mu(\x^2 + 1) - \x(\x^2 + 5)]}{8\x (\x^2 + 3)} \,.
\end{equation}
The analysis of this quantization condition is left to appendix~\ref{app:quantization.exe}.

We now move to the computation of the entropy. The first step is to write the ten-dimensional metric in the form
\begin{equation}
  \dd s_\mathrm{s.f.}^2 = \ee^{2A} \bigl( \dd s^2_{\AdS_2} + \dd s^2_{M_8} \bigr) \,,
\end{equation}
where $\dd s^2_{M_8}$ is the metric on the internal space~$M_8$, that is the total space of an $\hemi$ bundle over~$\Morb_4$, namely $\hemi \hookrightarrow M_8 \to \Morb_4$.
There is a $U(1)^2$ symmetry acting on $\hemi$, with $g A_i$ connections on the associated circle bundles, and the correct quantization of the corresponding magnetic fluxes~$\flt_i$ and~$\fls_i^\pm$, as discussed in section~\ref{subsec:6d_riem}, ensures that the ten-dimensional solution is a well-defined orbifold. The entropy of our solution can be read from the two-dimensional effective Newton constant $G_{(2)}$ as
\begin{equation} \label{entropy_def}
S = \frac{1}{4G_{(2)}} = \frac{8\pi^2}{(2\pi\ell_s)^8} \int \ee^{8A-2\Phi} \, \vol{M_8} \,.
\end{equation}
Applying this to metric~\eqref{explicit uplifted metric riemann}, we obtain
\begin{equation} \label{riem_entropy}
  \begin{split}
    S_{\riemann\ltimes\spindle_2} &= \frac{8\pi^2}{(2\pi\ell_s)^8} \frac{3\pi^2 \lambda^4}{20g^4} \, 4\pi (\mathrm{g} - 1) \, (\YB^3 - \YA^3) \Delta z \\
    &= (\mathrm{g} - 1) \, F_{S^3\times\spindle_2} \,,
  \end{split}
\end{equation}
where $F_{S^3\times\spindle_2}$ is the free energy of $d=3$, $\mathcal{N}=2$ SCFTs that arise from a system of $N$ D4-branes and $N_f$ D8-branes wrapped on a spindle~\cite{Faedo:2021nub}
\begin{equation} \label{free_energy_faedo}
  F_{S^3\times\spindle_2} = \chi_2 \, \frac{\sqrt3 \pi N^{5/2}}{5 \sqrt{8 - N_f}} \frac{[3\mu (\x^2+1) - \x (\x^2+5)]^{3/2}}{\x (\x^2+3) (\mu-\x)^{1/2}} \,.
\end{equation}

\subsection{AdS$_2\times \spindle_1 \ltimes \spindle_2$ solutions}

Similarly to the previous section we present the explicit expressions of the fluxes, which are then quantized. Apart from the Romans mass, which is the same as in \eqref{Romans_mass}, the complete uplifted solution includes the four-flux
\begin{align}
	F_{(4)}&= -\frac{\lambda \mu_0^{1/3} h_1 h_2}{g^3 \Delta_h} \bigg\{ \frac{U_h}{\Delta_h} \frac{\mu_1 \mu_2}{\mu_0} \, \dd\mu_1 \wedge \dd\mu_2 \wedge \sigma_1 \wedge \sigma_2 \nonumber \\
	& - \sum_{i\neq j} \Bigl[ g\,F_i \wedge \dd\phi_j - \frac{y^3 (h_i' - 3y^{-1} h_i) h_j}{\Delta_h h_i} \mu_i^2 \, \dd y \wedge \sigma_i \wedge \sigma_j \Bigr] \wedge \bigl( \mu_0 \mu_j \, \dd\mu_j - y^3 h_j^{-1} \mu_j^2 \, \dd\mu_0 \bigr) \bigg\} \nonumber \\
	& + \frac{\lambda \mathtt{a} y}{2x^2} \mu_0^{4/3} \, \dd x \wedge \dd\psi \wedge \dd y \wedge \dd z - \frac{\lambda \mathtt{a} y^2}{3x^2} \mu_0^{1/3} \, \dd x \wedge \dd\psi \wedge \dd z \wedge \dd\mu_0 \,,
\end{align}
and the metric and dilaton
\begin{align}
	\begin{split}
		\dd s_\mathrm{s.f.}^2 &= \lambda^2 \mu_0^{-1/3} y^{-1} \Delta_h^{1/2} \bigg\{ \frac{x^{2}}{4} \, \dd s_{\AdS_2}^2 +\frac{x^{2}}{q}\dd x^{2}+\frac{q}{4x^{2}}\dd \psi^{2} + \frac{y^2}{F} \, \dd y^2 \\
		& + \frac{F}{h_1 h_2} \Bigl( \dd z - \frac{1}{2m} \Bigl(1 - \frac{\mathtt{a}}{x}\Bigr) \, \dd\psi \Bigr)^2 \\
		& + g^{-2} y \, \Delta_h^{-1} \bigl[ y^3 \, \dd\mu_0^2 + h_1 \bigl(\dd\mu_1^2 + \mu_1^2 \sigma_1^2\bigr) + h_2 \bigl(\dd\mu_2^2 + \mu_2^2 \sigma_2^2\bigr) \bigr] \bigg\} \,,
	\end{split} \\
	\ee^{\Phi} &= \lambda^ 2 \mu_0^{-5/6} y^{-3/2} \Delta_h^{1/4} \,,
\end{align}
where $\Delta_h$ and $U_h$ were defined in~\eqref{DUh}. Again, even if they are not fundamental for the computations, we also have
\begin{equation}
	B_{(2)} = \frac{\lambda^2 \mathtt{a} y}{4m} \mu_0^{2/3} \,\vol{\AdS_2} \,,  \qquad  F_{(2)} = \frac{\mathtt{a} g y}{6\lambda m} \mu_0^{2/3} \, \vol{\AdS_2} \,.
\end{equation}

The quantization conditions are the same as in~\eqref{10d_quantization}, where now $\Morb_4 = \spindle_1\ltimes\spindle_2$, and the first two give again the relations~\eqref{10d_quant_cond_global}. The last one, expressed in terms of $(m_\pm,n_\pm,\mathtt{z},N)$, reads
\begin{equation} \label{F4 flux spindle}
  K = N \chi_2 \frac{m_+ - m_-}{m_+ m_-} \, \frac{3 [3\mu(\x^2 + 1) - \x(\x^2 + 5)]}{8\x (\x^2 + 3)} \in \NN \,.
\end{equation}
We refer to appendix~\ref{app:quantization.exe} for a detailed analysis.

The ten-dimensional geometry is similar to the solution of the previous subsection, with the internal space~$M_8$ having a fibration structure $\hemi \hookrightarrow M_8 \to \Morb_4$. Again, there is a $U(1)^2$ symmetry acting on $\hemi$, with $g A_i$ connections on the associated circle bundles. Therefore, the uplifted solution can be considered a well-defined orbifold thanks to the correct quantization of the magnetic fluxes of~$g A_i$, as discussed in section~\ref{subsec:6d_2spin}.

The entropy of this solution can be computed following the same steps of the previous case and the final result is
\begin{equation} \label{2spin_entropy}
\begin{split}
	 S_{\spindle_1\ltimes\spindle_2} &= \frac{8\pi^2}{(2\pi\ell_s)^8} \frac{3\pi^2\lambda^4}{20g^4} \, (\XB - \XA) \Delta\psi \, (\YB^3 - \YA^3) \Delta z \\
   & = \frac{1}{2\pi} \, A_{\spindle_1} F_{S^3\times\spindle_2}  \,,
  \end{split}
\end{equation}
where $F_{S^3\times\spindle_2}$ is given in~\eqref{free_energy_faedo} and $A_{\spindle_1}$ is the area of the  horizon of the  four-dimensional supersymmetric black hole solutions with $\AdS_2\times\spindle_1$ near-horizon studied in~\cite{Ferrero:2020twa}, namely
\begin{equation}
  A_{\spindle_1} = \frac{\sqrt{2} \sqrt{m_+^2 + m^2_-} - (m_+ + m_-)}{m_+ m_-} \, \pi \,.
\end{equation}

\section{Toric geometry}
\label{sec:toric_geometry}

In this section we will provide a description  of the orbifold $\spindle_1\ltimes\spindle_2$ in terms of certain combinatorial data, that will be used in the next section to construct the entropy function. Since $\spindle_1\ltimes\spindle_2$ admits a $U(1)^2$ action, corresponding to the $U(1)^2$ isometry of the metric~\eqref{2spin_solution}, it is natural to seek a description in terms of toric geometry. Below we will discuss a method for extracting a set of toric data that is sufficient for the purposes of this paper, leaving a more detailed study of the geometry for future work. As a warm-up, we will begin with an $S^2\ltimes\spindle_2$ geometry, namely a spindle fibred over a smooth two-sphere.
This can be obtained formally as an analytic continuation of the $\riemann\ltimes\spindle_2$ family and although it is not an actual supersymmetric background, it solves the equations of motion. This case can be analysed in the framework of symplectic toric geometry and we will derive a \emph{labelled polytope}~\cite{Lerman:1995aaa} from the image of an associated moment map.
From the algebraic geometry point of view, the corresponding fan of labelled normals to the polytope does not lead to a toric variety, but to a generalization known as a ``stack''. The reason is that in ordinary toric varieties orbifold singularities along divisors cannot appear, as all the singularities are in co-dimension higher than one. In the broader context of stacks, the key combinatorial gadget is called a \emph{stacky fan}, which roughly speaking is a simplicial fan with a distinguished lattice point on each ray of the fan (see \eg~\cite{Borisov:2005,Lerman:2012,Sakai:2013,Hochenegger:2012}).
For the more general $\spindle_1\ltimes\spindle_2$ family we will not be able to find a symplectic structure and therefore we will have to use an indirect method to extract the toric data. These will again take the form of a stacky fan and will be taken as inputs for constructing an entropy function in terms of gravitational blocks in the next section. As further illustration of our approach, in appendix~\ref{app:toric-theory} we present the derivation of the Delzant polytope of the first Hirzebruch surface~$\mathbb{F}_1$, starting from the explicit metric found in \cite{Martelli:2004wu}, as well as the toric data relevant for the seven-dimensional solutions of~\cite{Cheung:2022ilc}.

\subsection{AdS$_2\times S^2 \ltimes \spindle_2$ solutions}
\label{subsec:riem_toric_geometry}

Before studying the $\spindle_1\ltimes\spindle_2$ orbifold, in which case the analysis is more involved, we begin with a geometry characterized by a smooth two-dimensional base.
Since a Riemann surface with $g>0$  is not toric, we need to consider its analytic continuation to a two-sphere. Although non-supersymmetric, the $\AdS_2\times S^2 \ltimes \spindle_2$ backgrounds form a family of solutions to the equations of motion very similar to~\eqref{riem_solution}, but with a different value of the radius of the Riemann surface, now a sphere. Specifically, the space we consider is the four-dimensional toric orbifold $S^2\ltimes\spindle_2$ with metric
\begin{equation} \label{toric_S2xspindle}
  \dd s_{S^2\ltimes\spindle_2}^2 = \frac16 \bigl( \dd\theta^2 + \sin^2\!\theta \, \dd\psi^2 \bigr) + \frac{y^2}{F} \, \dd y^2 + \frac{F}{h_1 h_2} \Bigl( \dd z + \frac{1}{2m} \cos\theta \, \dd\psi \Bigr)^2 \,.
\end{equation}
In this case, the condition on~$t$~\eqref{riem_t} in order to have a well-defined fibration reads
\begin{equation}
  t = -\frac{1}{m} \frac{2\pi}{\Delta z} \in \ZZ \,,
\end{equation}
where $t$ results to be negative. For fixed values of~$y$, metric~\eqref{toric_S2xspindle} describes an $S^3/\ZZ_{-t}$, where $S^3$ is a squashed three-sphere written as a Hopf fibration, therefore a natural basis of an effective two-torus action is~\cite{Martelli:2004wu}\footnote{We exchanged $e_1$ and $e_2$ with respect to~\cite{Martelli:2004wu} in order to have the vectors~$\vec{n}_\ell$ ordered counter-clockwise.}
\begin{equation} \label{riem_killing-basis}
  e_1 = \partial_{\spang2} \,,  \qquad  e_2 = \partial_\psi - \frac{t}{2} \, \partial_{\spang2} \,,
\end{equation}
where we introduced the $2\pi$-periodic coordinate $\spang2 = \frac{2\pi}{\Delta z} z$. In total we have four fixed points~$\fp_\ell$ ($\ell=1,\ldots,4$) under the action of the two-torus, which are all the possible combinations obtained by pairing the poles of the sphere ($\theta=0,\pi$) and the poles of the spindle ($y=\YAB$)
\begin{equation} \label{riem_fixed-points}
  \begin{aligned}
    \fp_1 &= \{\theta = 0, \, y = \YA\} \,,    \qquad  & \fp_2 &= \{\theta = 0, \, y = \YB\} \,, \\
    \fp_3 &= \{\theta = \pi, \, y = \YB\} \,,  \qquad  & \fp_4 &= \{\theta = \pi, \, y = \YA\} \,.
  \end{aligned}
\end{equation}

We now consider the conformally rescaled metric $\dd s^2 = \Gamma(y) \, \dd s_{S^2\ltimes\spindle_2}^2$, with $\Gamma(y)>0$, with compatible symplectic two-form
\begin{equation} \label{symplectic}
  \omega = \Gamma(y) \biggl[ \frac16 \sin\theta \, \dd\theta \wedge \dd\psi + \frac{y}{(h_1 h_2)^{1/2}} \, \dd y \wedge \Bigl( \dd z + \frac{1}{2m} \cos\theta \, \dd\psi \Bigr) \biggr] \,.
\end{equation}
When $\Gamma'(y) = -\frac{3y}{m (h_1 h_2)^{1/2}}\Gamma(y)$~\footnote{A real and positive solution to this differential equation exists and is unique, up to an irrelevant overall constant.}, the two-form~\eqref{symplectic} is closed and can be written as
\begin{equation}
  \omega = \dd\psi \wedge \dd\biggl[ \frac16 \Gamma(y) \cos\theta \biggr] + \dd\spang2 \wedge \dd\biggl[ -\frac{1}{3t} \Gamma(y) \biggr] \,.
\end{equation}
From this expression we can derive the moment maps with respect to the basis~\eqref{riem_killing-basis}
\begin{equation}
  \vec{\mu} = -\frac{1}{3t} \Gamma(y) \, \left( 1, -\frac{t (1 + \cos\theta)}{2} \right) \,,
\end{equation}
and compute the image of the fixed points~\eqref{riem_fixed-points}
\begin{equation}
  \begin{aligned}
    \vec{\mu}(\fp_1) &= -\frac{1}{3t} \Gamma(\YA) \, (1, -t) \,,  \qquad  & \vec{\mu}(\fp_2) &= -\frac{1}{3t} \Gamma(\YB) \, (1, -t) \,, \\
    \vec{\mu}(\fp_3) &= -\frac{1}{3t} \Gamma(\YB) \, (1, 0)  \,,  \qquad  & \vec{\mu}(\fp_4) &= -\frac{1}{3t} \Gamma(\YA) \, (1, 0) \,,
  \end{aligned}
\end{equation}
which are vertices of the moment polytope. Vertices are connected by facets and we define the facet~$D_\ell$ to be the one that joins~$\fp_{\ell-1}$ and~$\fp_\ell$, where $\ell$ is modulo 4\footnote{With a slight abuse of notation, we denote as $\fp_\ell$ both the fixed points and the vertices. Moreover, we denoted as~$D_\ell$ both the facets and the corresponding divisors, \ie\ their preimages under~$\vec{\mu}$.}.
In order to correctly draw the polytope we notice that, since $\Gamma'(y)<0$, we have $\Gamma(\YA)>\Gamma(\YB)$.

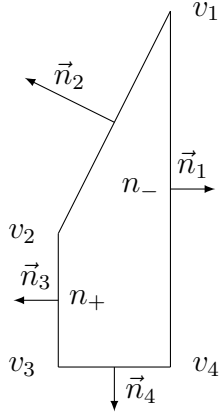
\begin{figure}[ht]
	\centering

	\begin{tikzpicture}[scale=0.59pt]
		\node [label=right : {$\fp_1$}] (x1) at (4,8)   {};
		\node [label=left  : {$\fp_2$}] (x2) at (1.5,3) {};
		\node [label=left  : {$\fp_3$}] (x3) at (1.5,0) {};
		\node [label=right : {$\fp_4$}] (x4) at (4,0)   {};

		\draw (x4.center) -- node[left]{$n_-$}  (x1.center) node[midway,anchor=center] (d1) {};
		\draw (x1.center) --                    (x2.center) node[midway,anchor=center] (d2) {};
		\draw (x2.center) -- node[right]{$n_+$} (x3.center) node[midway,anchor=center] (d3) {};
		\draw (x3.center) --                    (x4.center) node[midway,anchor=center] (d4) {};

		\draw [-latex] (d1.center) -- node[above] {$\vec{n}_1$} ($(d1)+(1,0)$);
		\draw [-latex] (d2.center) -- node[above] {$\vec{n}_2$} ($(d2)+(-2,1)$);
		\draw [-latex] (d3.center) -- node[above] {$\vec{n}_3$} ($(d3)+(-1,0)$);
		\draw [-latex] (d4.center) -- node[right] {$\vec{n}_4$} ($(d4)+(0,-1)$);
	\end{tikzpicture}

	\caption{Labelled polytope corresponding to the $S^2\ltimes\spindle_2$ orbifold. The different facets are labelled with the integers $m_\ell$. In the figure $t=-2$.}
	\label{fig:riem-polytope}
\end{figure}

Due to the orbifold nature of the spindle, we can construct an associated labelled polytope following~\cite{Lerman:1995aaa}, which extends the classic Delzant construction.
In particular, if the preimage under moment maps of a facet has an orbifold singularity locally modelled on $\CC/\ZZ_{m_\ell}$, we label the facet with the corresponding integer~$m_\ell$, otherwise there is no label (formally, we have $m_\ell=1$). Specifically, the $S^2\ltimes\spindle_2$ orbifold is characterized by
\begin{equation} \label{riem_toric-m}
  \begin{aligned}
    D_1 &= \{y = \YA\}:  & m_1 &= n_- \,,  \qquad  & D_2 &= \{\theta = 0\}:    & m_2 &= 1 \,, \\
    D_3 &= \{y = \YB\}:  & m_3 &= n_+ \,,  \qquad  & D_4 &= \{\theta = \pi\}:  & m_4 &= 1 \,.
  \end{aligned}
\end{equation}
The resulting labelled polytope is depicted in figure~\ref{fig:riem-polytope}, and if the labels are stripped off
this is exactly the polytope corresponding to the Hirzebruch surfaces
$\mathbb{F}_{-t}$. This polytope can also be described uniquely by a set of inequalities
\begin{equation}
  \langle \vec{\mu}, m_\ell \vec{n}_\ell \rangle \leq \lambda_\ell \,,
\end{equation}
where $\langle\cdot,\cdot\rangle$ is the standard inner product between vectors in $\RR^2$, $\vec{\mu}$ is a generic point in the $(\mu_1,\mu_2)$-plane, $\lambda_\ell$ is a real number and each $\vec{n}_\ell$ is a primitive element of~$\ZZ^2$, which represents the outward-pointing normal vector to the facet~$D_\ell$. In particular, we have
\begin{equation} \label{riem_vec-n}
  \vec{n}_1 = (1, 0) \,,  \qquad  \vec{n}_2 = (t, 1) \,,  \qquad  \vec{n}_3 = (-1, 0) \,,  \qquad  \vec{n}_4 = (0, -1) \,.
\end{equation}

The method we employed to construct the labelled polytope above is rigorous and the result fits in the  theory of symplectic toric orbifolds~\cite{Lerman:1995aaa}.
However, we will now study this toric orbifold from a different point of view, in order to derive a method that can be applied also to orbifolds which do not admit a conformally rescaled symplectic form. A direct computation using the metric~\eqref{toric_S2xspindle}  shows that there  are  four distinct loci $D_\ell$ where a linear combination of the two Killing vectors~$\partial_\psi$ and~$\partial_z$ degenerates. These are precisely  the toric divisors which  in the symplectic context  correspond  to the preimages of the four facets of the polytope and intersect at the four fixed points. Explicitly, denoting with $\xi_{(\ell)}$ the Killing vector that degenerates at~$D_\ell$, they are
\begin{equation} \label{killing-vectors}
	\xi_{(1)} = n_- \, \partial_{\spang2} \,,  \qquad  \xi_{(2)} = \partial_\psi + \frac{t}{2} \, \partial_{\spang2} \,,  \qquad
	\xi_{(3)} = n_+ \, \partial_{\spang2} \,,  \qquad  \xi_{(4)} = \partial_\psi - \frac{t}{2} \, \partial_{\spang2} \,,
\end{equation}
normalized so to have unitary surface gravity
\begin{equation}
  \kappa_\mathrm{grav}^2 = \frac{\partial_\mu|\xi|^2 \, \partial^\mu|\xi|^2}{4|\xi|^2} = 1  \qquad  \text{where}  \qquad  |\xi|^2 = 0 \,.
\end{equation}
Expanded in the basis~\eqref{riem_killing-basis}, the degenerating Killing vectors read
\begin{equation}
	\xi_{(1)} = n_- \, e_1 \,,  \qquad  \xi_{(2)} = t \, e_1 + e_2 \,,  \qquad
	\xi_{(3)} = n_+ \, e_1 \,,  \qquad  \xi_{(4)} = e_2 \,.
\end{equation}
We notice that
\begin{equation} \label{killing-and-toric}
	\xi_{(\ell)} = m_\ell \, \vec{n}_\ell \cdot (e_1, e_2) \,,
\end{equation}
where we packed the basis vectors~\eqref{riem_killing-basis} into a two-dimensional vector $(e_1, e_2)$ and $(\vec{n}_\ell,m_\ell)$ are the toric data of the corresponding labelled polytope.

We can use relation~\eqref{killing-and-toric} to infer the toric data starting from the information about the degenerate Killing vectors. In particular, given a toric divisor $D_\ell$ where the Killing vector~$\xi_{(\ell)}$ is degenerate, studying the (possible) orbifold singularity on~$D_\ell$ we can obtain the corresponding integer~$m_\ell$. Expanding $\xi_{(\ell)}$ on a suitable basis---in which the two-torus act effectively---according to~\eqref{killing-and-toric} we can deduce the normal vectors~$\vec{n}_\ell$, up to a global sign depending on the conventions. This sign can be chosen arbitrarily for one of the vectors and for the others it follows requiring the vectors~$\vec{n}_\ell$ to be ordered counter-clockwise, in our conventions.

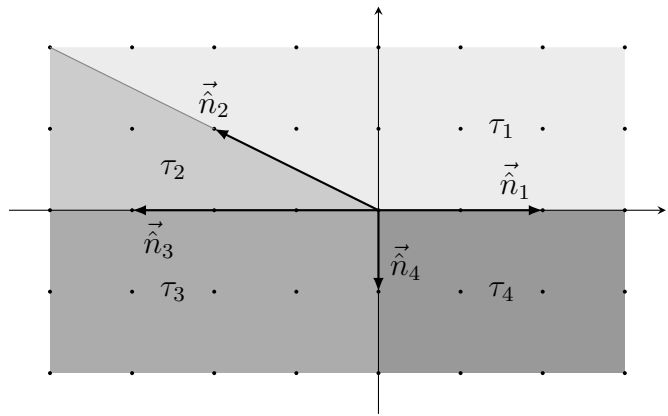
\begin{figure}[ht]
	\centering

	\begin{tikzpicture}[x=1.08cm, y=1.08cm]
		\coordinate (00)  at (0,0);
		\coordinate (w1)  at (2,0);
		\coordinate (w2)  at (-2,1);
		\coordinate (w3)  at (-3,0);
		\coordinate (w4)  at (0,-1);
		\coordinate (w2x) at (-4,2);

		\path[fill=gray!15] (00) -- (3,0)  -- (3,2)   -- (w2x)  -- cycle;
		\path[fill=gray!40] (00) -- (w2x)  -- (-4,0)            -- cycle;
		\path[fill=gray!65] (00) -- (-4,0) -- (-4,-2) -- (0,-2) -- cycle;
		\path[fill=gray!80] (00) -- (0,-2) -- (3,-2)  -- (3,0)  -- cycle;

		\draw [-stealth] (-4.5,0) -- (3.5,0);
		\draw [-stealth] (0,-2.5) -- (0,2.5);

		\foreach \x in {-4,...,3}
		\foreach \y in {-2,...,2} {
			\fill (\x,\y) circle (0.75pt);
		}

		\draw [gray] (w2) -- (w2x);

		\draw [thick, -latex] (00) -- (w1) node[above left]  {$\vec{\hat{n}}_1$};
		\draw [thick, -latex] (00) -- (w2) node[above]       {$\vec{\hat{n}}_2$};
		\draw [thick, -latex] (00) -- (w3) node[below right] {$\vec{\hat{n}}_3$};
		\draw [thick, -latex] (00) -- (w4) node[above right] {$\vec{\hat{n}}_4$};

		\node (c1) at (1.5,1)    {$\tau_1$};
		\node (c2) at (-2.5,0.5) {$\tau_2$};
		\node (c3) at (-2.5,-1)  {$\tau_3$};
		\node (c4) at (1.5,-1)   {$\tau_4$};
	\end{tikzpicture}

	\caption{Stacky fan corresponding to the vectors~\eqref{riem_vec-n-stack}. The cones generated by~$\vec{\hat{n}}_\ell$ and $\vec{\hat{n}}_{\ell+1}$ are denoted as~$\tau_\ell$. In this example we took $n_+=3$, $n_-=2$ and $t=-2$. }
	\label{fig:riem_fan}

\end{figure}

For an un-labelled, convex, rational polytope (not necessarily a Delzant polytope) one can pass to the algebraic point of view, where the primitive normals to the polytope represent the fan of a toric variety. In this context, the one-dimensional cones (rays) of the fan are in a one-to-one correspondence with the toric divisors and the order of the orbifold singularities at the intersection between two divisors can be read off from the determinant of two adjacent generators, namely $\det(\vec{n}_\ell,\vec{n}_{\ell+1})$. Orbifold singularities along divisors cannot arise in toric varieties, so the information of a labelled polytope cannot be accommodated in an ordinary fan.
An equivalent way of thinking about the labels $m_\ell$ is by dropping the primitive requirement and considering $\vec{\hat{n}}_\ell = m_\ell\,\vec{n}_\ell$ instead of $\vec{n}_\ell$ \cite{abreu2009}. We will refer to the $\vec{n}_\ell$ as ``short'' vectors and to the $\vec{\hat{n}}_\ell$ as ``long'' vectors.
In algebraic geometry the fan obtained from such ``long'' (non-primitive) normals can be understood in the broader context of stacks and it is referred to as a stacky fan. This turns out to be the correct combinatorial data needed in the construction of the entropy function in the next section.
The ``long'' vectors characterizing the $S^2\times\spindle_2$ orbifold then read
\begin{equation} \label{riem_vec-n-stack}
  \vec{\hat{n}}_1 = (n_-, 0) \,,  \qquad  \vec{\hat{n}}_2 = (t, 1) \,,  \qquad  \vec{\hat{n}}_3 = (-n_+, 0) \,,  \qquad  \vec{\hat{n}}_4 = (0, -1) \, ,
\end{equation}
and the corresponding stacky fan, dual to the labelled polytope of figure~\ref{fig:riem-polytope}, is presented in figure~\ref{fig:riem_fan}.

Let us make some comments on the orbifold  $\riemann\ltimes\spindle_2$, with~$\mathrm{g}>1$. This cannot be toric in any sense, therefore none of the previous considerations can be applied.
Nevertheless, for our purposes, we can think about extending the results obtained for the $S^2\ltimes\spindle_2$ toric orbifold to this case, and adopt as ``toric data'' the same vectors~$\vec{n}_\ell$~\eqref{riem_vec-n} and the same integers~$m_\ell$~\eqref{riem_toric-m}. The information on the genus of the Riemann surface does not enter in any way in the ``toric data'', but it will appear in the twisting procedure of section~\ref{sec:entropy-function}, as discussed in the examples.
Clearly, for $\riemann\ltimes\spindle_2$ the divisors~$D_2$ and~$D_4$ have no physical meaning, and likewise the four fixed points~$\fp_\ell$, but they are all artefacts needed for the application of the conjectural entropy function we will present in section~\ref{sec:entropy-function}.

\subsection{AdS$_2\times \spindle_1 \ltimes \spindle_2$ solutions}
\label{subsec:toric_2spin_6d}

After the warm-up with the $S^2\ltimes\spindle_2$ case, we are now ready to study the $\spindle_1\ltimes\spindle_2$ orbifold. In this case we can not construct a conformal symplectic structure, therefore we do not have a moment map to derive the labelled polytope.
Nevertheless, we are still able to obtain the toric data that we need by analysing where the Killing vectors of $\spindle_1\ltimes\spindle_2$ degenerate.
Recall that the metric on $\spindle_1\ltimes\spindle_2$ is
\begin{equation} \label{toric_spindlexspindle}
  \dd s_{\spindle_1\ltimes\spindle_2}^2 = \frac{x^2}{q} \, \dd x^2 + \frac{q}{4x^2} \, \dd\psi^2 + \frac{y^2}{F} \, \dd y^2 + \frac{F}{h_1 h_2} \Bigl( \dd z - \frac{1}{2m} \Bigl(1 - \frac{\mathtt{a}}{x}\Bigr) \, \dd\psi \Bigr)^2 \,.
\end{equation}
A simple direct calculation shows that it has four Killing vectors degenerating at the poles of the two spindles, defining the four loci
\begin{equation}
D_1 = \{y = \YA\} \,,  \qquad  D_2 = \{x = \XA\} \,,  \qquad  D_3 = \{y = \YB\} \,,  \qquad  D_4 = \{x = \XB\} \,,
\end{equation}
intersecting at the the four fixed points
\begin{equation}
  \begin{aligned}
    \fp_1 &= \{x = \XA, \, y = \YA\} \,,  \qquad  & \fp_2 &= \{x = \XA, \, y = \YB\} \,, \\
    \fp_3 &= \{x = \XB, \, y = \YB\} \,,  \qquad  & \fp_4 &= \{x = \XB, \, y = \YA\} \,.
  \end{aligned}
\end{equation}
normalizing the degenerating Killing vectors so that they have unitary surface gravity, these read
\begin{equation} \label{2spin_killing}
  \begin{aligned}
    \xi_{(1)} &= n_- \, \partial_{\spang2} \,,  \qquad  & \xi_{(2)} &= m_- \Bigl( \partial_{\spang1} - \frac{\mathtt{a} - \XA}{2m \XA} \frac{\Delta\psi}{\Delta z} \, \partial_{\spang2} \Bigr) \,, \\
    \xi_{(3)} &= n_+ \, \partial_{\spang2} \,,  \qquad  & \xi_{(4)} &= m_+ \Bigl( \partial_{\spang1} - \frac{\mathtt{a} - \XB}{2m \XB} \frac{\Delta\psi}{\Delta z} \, \partial_{\spang2} \Bigr) \,,
  \end{aligned}
\end{equation}
where, in addition to $\spang2 = \frac{2\pi}{\Delta z}z$, we introduced also the $2\pi$-periodic coordinate $\spang1 = \frac{2\pi}{\Delta\psi}\psi$.

Reading off the labels associated with the $D_\ell$ turns out to be a bit tricky. We will first present a naive argument leading to slightly wrong conclusions, and then we will explain how to amend this.
Each of the  Killing vectors in~\eqref{2spin_killing} is in a one-to-one correspondence with a specific pole of either~$\spindle_1$ or~$\spindle_2$, whose rotation around the axis is generated by the $2\pi$-periodic
Killing vector~$\partial_{\spang1}$ or~$\partial_{\spang2}$, respectively. This suggests that the integer coefficient in front of this rotational Killing vector corresponds to a $\mathbb{C}/\mathbb{Z}_{m_\ell}$ quotient singularity, with $m_\ell=(n_-,m_-,n_+,m_+)$.
While for $D_1$ and $D_3$ this is correct, for $D_2$ and $D_4$ this conclusion is not precise, as we will show below.

In order to derive the normal vectors~$\vec{n}_\ell$ we need to find an appropriate basis in which to expand the~$\xi_{(\ell)}$. Inspired by~\eqref{riem_killing-basis}, where $\partial_{\spang2}$ is the first basis element and $\xi_{(4)}$ plays the role of $e_2$, we consider the following basis
\begin{equation} \label{2spin_killing-basis}
  e_1 = \partial_{\spang2} \,,  \qquad  e_2 = \partial_{\spang1} - \frac{\mathtt{a} - \XB}{2m \XB} \frac{\Delta\psi}{\Delta z} \, \partial_{\spang2} \,,
\end{equation}
in which the degenerating  Killing vectors are decomposed as
\begin{equation} \label{2spin_killing2}
  \begin{aligned}
    \xi_{(1)} &= n_- \, e_1 \,,  \qquad  & \xi_{(2)} &= \frac{t}{m_+} \, e_1 + m_- \, e_2 \,, \\
    \xi_{(3)} &= n_+ \, e_1 \,,  \qquad  & \xi_{(4)} &= m_+ \, e_2 \,,
  \end{aligned}
\end{equation}
where we made use of the quantization condition on~$t$ in~\eqref{2spin_t} in the form
\begin{equation}
	\frac{t}{m_+ m_-} = -\frac{\mathtt{a} (\XB - \XA)}{2m \XA \XB} \frac{\Delta\psi}{\Delta z} \,.
\end{equation}
Note that, even though $\XAB$ and $\mathtt{a}$ are real numbers, all the coefficients in~\eqref{2spin_killing2} are rational.
Taking the  labels as above, for the ``short'' vectors  
we obtain
\begin{equation} \label{2spin_vec-n}
  \vec{n}_1 = (1, 0) \,,   \qquad  \vec{n}_2 = \Bigl(\frac{t}{m_+ m_-}, 1\Bigr) \,,  \qquad  \vec{n}_3 = (-1, 0) \,,  \qquad  \vec{n}_4 = (0, -1) \,,
\end{equation}
where we picked the sign of~$\vec{n}_1$ as in~\eqref{riem_vec-n} and the others followed accordingly.
However, since these vectors  are not primitive elements of~$\ZZ^2$, they cannot be regarded as the normal vectors to the facets of a polytope, in the sense of~\cite{Lerman:1995aaa}.
Even adopting the alternative description in which the labels are included in the normal vectors, the ``long'' vectors associated with~\eqref{2spin_vec-n} are not valued in~$\ZZ^2$:
\begin{equation}
	\vec{\hat{n}}_1 = (n_-, 0) \,,  \qquad  \vec{\hat{n}}_2 = \Bigl(\frac{t}{m_+}, m_-\Bigr) \,,  \qquad  \vec{\hat{n}}_3 = (-n_+, 0) \,,  \qquad  \vec{\hat{n}}_4 = (0, -m_+) \,.
\end{equation}
Although in the mathematical literature there are examples where the normal vectors are allowed to take rational \cite{abreu2009,legendre2009} as well as real values~\cite{Battaglia:2018ffm}, the fact that the Killing vectors in~\eqref{2spin_killing2} are not linear combinations of the basis $\{e_1,e_2\}$ with integer coefficients, indicates that the two-torus is not acting effectively.
A better basis can be obtained rotating the vectors~$\vec{\hat{n}}_\ell$ through the following  $SL(2,\QQ)$ matrix
\begin{equation} \label{S-matrix}
	S = \begin{pmatrix} 1 & -r_-/m_+ \\ 0 & 1 \end{pmatrix} \,,
\end{equation}
where $r_-$ is an integer prime to~$m_+$. Defining $\vec{\hat{w}}_\ell \equiv S\,\vec{\hat{n}}_\ell$, we have
\begin{equation} \label{2spin_vec-n-stack}
	\vec{\hat{w}}_1 = (n_-, 0) \,,  \qquad  \vec{\hat{w}}_2 = (r_+, m_-) \,,  \qquad  \vec{\hat{w}}_3 = (-n_+, 0) \,,  \qquad  \vec{\hat{w}}_4 = (r_-, -m_+) \,,
\end{equation}
where $r_\pm\in\ZZ$ are chosen such that $t = r_+\,m_+ + r_-\,m_-$, and exist by B\'ezout's lemma\footnote{By B\'ezout's lemma, if $m_+$ and $m_-$ are co-prime, there exist $a_\pm\in \ZZ$ such that $ a_+\,m_+ + a_-\,m_-=1$. Then we take $r_+=t\,a_+$ and $r_-=t\,a_-$.}, for co-prime $m_\pm$.
The vectors $\vec{\hat{w}}_\ell$ now take values in~$\ZZ^2$, with primitive $\vec{\hat{w}}_2$, $\vec{\hat{w}}_4$, hence $m_2 = m_4 = 1$, and non-primitive $\vec{\hat{w}}_1$, $\vec{\hat{w}}_3$, corresponding to the labels $m_1 = n_-$ and $m_3 = n_+$. We can now define a new basis $\{E_1,E_2\}$, obtained as $E_i=S^{-1}_{ji}e_j$, such that
\begin{equation} \label{new_killing_and_toric}
	\xi_{(\ell)} = \vec{\hat{w}}_\ell \cdot (E_1,E_2) \, ,
\end{equation}
and the degenerating Killing vectors take the form
\begin{equation} \label{kill_2spin_6d_new_basis}
\begin{aligned}
	\xi_{(1)} &= n_- \, E_1 \,,  \qquad  & \xi_{(2)} &= m_- \, E_2 + r_+ \, E_1 \,, \\
	\xi_{(3)} &= n_+ \, E_1 \,,  \qquad  & \xi_{(4)} &= m_+ \, E_2 - r_- \, E_1 \,.
\end{aligned}
\end{equation}
All the coefficients of this decomposition are now integers, indicating that in  this basis the torus action is effective. For completeness we also write the basis vectors in terms of the Killing vectors~$\partial_{\spang1}$ and~$\partial_{\spang2}$\footnote{Notice that $E_2$ is finite also when $m_+=m_-=1$. Indeed, in this limit we have $r_++r_-=t$, which is proportional to $m_+-m_-$ (\cf~\eqref{2spin_t}), thus eliminating the vanishing denominator.}
\begin{equation}\label{new_basis_6d}
	E_1 = \partial_{\spang2} \,,  \qquad  E_2 = \partial_{\spang1} + \frac{r_+ + r_-}{m_+ - m_-} \, \partial_{\spang2} \,.
\end{equation}

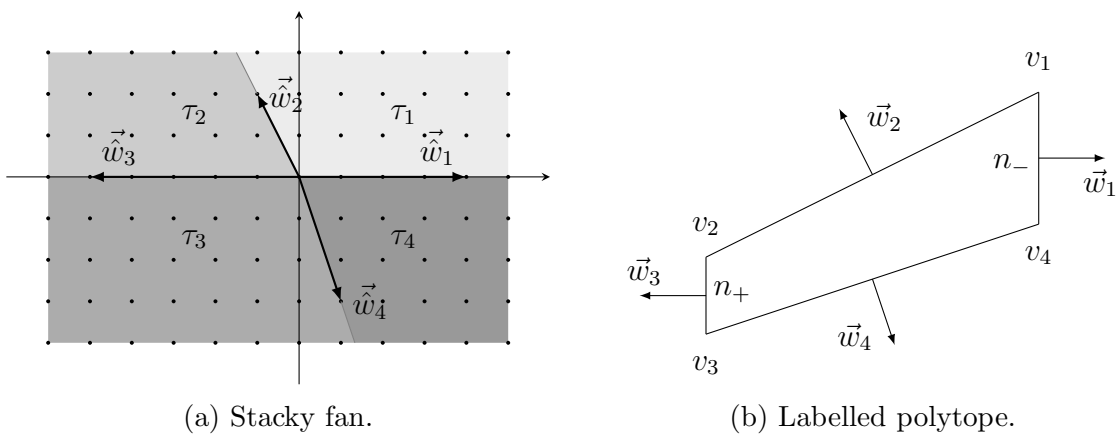
\begin{figure}[ht]
\centering

	\begin{subfigure}[b]{0.48\textwidth} \centering

	\begin{tikzpicture}[x=0.55cm, y=0.55cm]
		\coordinate (00)  at (0,0);
		\coordinate (w1)  at (4,0);
		\coordinate (w2)  at (-1,2);
		\coordinate (w3)  at (-5,0);
		\coordinate (w4)  at (1,-3);
		\coordinate (w2x) at (-1.5,3);
		\coordinate (w4x) at (4/3,-4);

		\path[fill=gray!15] (00) -- (5,0)  -- (5,3)   -- (w2x)  -- cycle;
		\path[fill=gray!40] (00) -- (w2x)  -- (-6,3)  -- (-6,0) -- cycle;
		\path[fill=gray!65] (00) -- (-6,0) -- (-6,-4) -- (w4x)  -- cycle;
		\path[fill=gray!80] (00) -- (w4x)  -- (5,-4)  -- (5,0)  -- cycle;

		\draw [-stealth] (-7,0) -- (6,0);
		\draw [-stealth] (0,-5) -- (0,4);

		\foreach \x in {-6,...,5}
		\foreach \y in {-4,...,3} {
			\fill (\x,\y) circle (0.75pt);
		}

		\draw [gray] (w2) -- (w2x);
		\draw [gray] (w4) -- (w4x);

		\draw [thick, -latex] (00) -- (w1) node[above left]  {$\vec{\hat{w}}_1$};
		\draw [thick, -latex] (00) -- (w2) node[right]       {$\vec{\hat{w}}_2$};
		\draw [thick, -latex] (00) -- (w3) node[above right] {$\vec{\hat{w}}_3$};
		\draw [thick, -latex] (00) -- (w4) node[right]       {$\vec{\hat{w}}_4$};

		\node (c1) at (2.5,1.5)   {$\tau_1$};
		\node (c2) at (-2.5,1.5)  {$\tau_2$};
		\node (c3) at (-2.5,-1.5) {$\tau_3$};
		\node (c4) at (2.5,-1.5)  {$\tau_4$};
	\end{tikzpicture}

	\subcaption{Stacky fan.}
	\label{fig:2spin_fan}
	\end{subfigure}
	\begin{subfigure}[b]{0.48\textwidth} \centering

	\begin{tikzpicture}[scale=1.75]
		\node [label=below : {$\fp_3$}] (x3) at (3.5,7/6)   {};
		\node [label=below : {$\fp_4$}] (x4) at (6,6/3)     {};
		\node [label=above : {$\fp_1$}] (x1) at (6,6/2)     {};
		\node [label=above : {$\fp_2$}] (x2) at (3.5,21/12) {};

		\draw (x2.center) -- (x3.center) node[midway,anchor=center] (d3) {};
		\draw (x3.center) -- (x4.center) node[midway,anchor=center] (d4) {};
		\draw (x4.center) -- (x1.center) node[midway,anchor=center] (d1) {};
		\draw (x1.center) -- (x2.center) node[midway,anchor=center] (d2) {};

		\node (c1) at (3.7,1.49)   {$n_+$};
		\node (c2) at (5.8,2.45)   {$n_-$};

		\draw [-latex] (d3.center) -- node[above left]{$\vec{w}_3$}  ($(d3)+(-1/2,0)$);
		\draw [-latex] (d4.center) -- node[below left]{$\vec{w}_4$}  ($(d4)+(1/6,-1/2)$);
		\draw [-latex] (d1.center) -- node[below right]{$\vec{w}_1$} ($(d1)+(1/2,0)$);
		\draw [-latex] (d2.center) -- node[above right]{$\vec{w}_2$} ($(d2)+(-1/4,1/2)$);
	\end{tikzpicture}

	\subcaption{Labelled polytope.}
	\label{fig:2spin_polytope}
	\end{subfigure}

	\caption{Toric data of the  $\spindle_1\ltimes\spindle_2$ orbifold corresponding to the vectors~\eqref{2spin_vec-n-stack}. In this example we took $n_+=5$, $n_-=4$, $m_+=3$, $m_-=2$, $r_+=-1$, $r_-=1$ and $t=-1$.}
	\label{fig:stack_geometry}
\end{figure}

The combinatorial data can be presented in the form of a stacky fan, as depicted in figure~\ref{fig:2spin_fan}. Note that for $n_+=n_-=1$ this reduces to the ``Hirzebruch orbifold'' discussed in \cite{Wang:2020} in the context of toric stacks. We can then read off the order of the quotient singularities at the intersections of the four divisors, corresponding to the four cones~$\tau_\ell$, which are given by $\CC^2/\ZZ_{\det(\vec{\hat{n}}_\ell,\vec{\hat{n}}_{\ell+1})}$ and thus read $\CC^2/\ZZ_{n_-m_-}$, $\CC^2/\ZZ_{n_+m_-}$,
$\CC^2/\ZZ_{n_+m_+}$, $\CC^2/\ZZ_{n_-m_+}$, for $\ell=1,\dots,4$ respectively. From the toric description we see that the total space $\spindle_1\ltimes\spindle_2$ is smooth not only away from the poles of $\spindle_2$, but also at the poles of $\spindle_1$, in agreement with \cite{Cheung:2022ilc}.

Despite the fact that we do not have a moment map, it is of course possible to formally define a labelled polytope, encapsulating the same combinatorial data.
This is obtained from the stacky fan in figure~\ref{fig:2spin_fan}, considering the vectors generating the rays of the fan and interpreting  them as normal vectors to each facet of the polytope. The labels are read off from
the ``multiplicities'' of the non-primitive vectors. The result is the rational
convex polytope depicted in figure~\ref{fig:2spin_polytope}, described by the labels~$m_\ell=(n_-,1,n_+,1)$ and the primitive normal vectors $\vec{w}_\ell = \vec{\hat{w}}_\ell/m_\ell$, which together are packed into what we shall refer to as toric data $(\vec{w}_\ell,m_\ell)$.

\subsection{Toric divisors and their intersection}
\label{subsec:divisors}

In this section we will discuss toric divisors with the aim of computing their intersection matrix, that will be a key ingredient in the construction of the entropy function.
We will begin recalling some basic facts about divisors in the context of toric varieties (thus, ignoring the labels) and then we will indicate the modifications needed to pass to the broader setting in which our toric orbifolds fit. Recall that all the toric varieties defined by fans are normal and therefore all the (orbifold) singularities are in complex co-dimension higher than one.
For toric surfaces $\Morb_4$ (\ie\ in real dimension 4), this implies that only points can be singular, while there cannot be singularities along divisors.
Divisors that are torus-invariant are called \emph{toric divisors} and are in a one-to-one correspondence with the primitive generators of the rays of the fan. These are in general Weil divisors, but
for toric surfaces, since all the fans are necessarily simplicial, all Weil divisors $D$ have an integer multiple $l D$ that is a Cartier divisor---namely they are $\QQ$-Cartier divisors\footnote{Varieties with such a property are called $\QQ$-factorial.}. Below we will consider only such $\QQ$-Cartier toric divisors, associated with the primitive generators of a fan, $\vec{\vv}_\ell$, that will be denoted as $D_\ell$. As in the previous sections, $\ell=1,\dots,\fpN$ runs over the generators of the fan in cyclic order.
Furthermore, associated with any toric divisor~$D_\ell$ there is an equivariant line bundle $L_\ell$, whose first Chern class $c_1(L_\ell) \in H^2(\Morb_4,\QQ)$ is Poincar\'e dual to $D_\ell$.
While the second co-homology group of $\Morb_4$, $H^2(\Morb_4,\QQ)$, is ($\fpN-2$)-dimensional, its equivariant extension, generated by the $c_1(L_\ell)$, has clearly dimension~$\fpN$.

The intersection matrix of the toric divisors is defined as
\begin{equation}
  D_{\ell,\ell'} \equiv D_\ell \cdot D_{\ell'} = \int_{\Morb_4} c_1(L_\ell) \wedge c_1(L_{\ell'}) \,,
\end{equation}
and by a standard calculation in toric geometry, it reads
\begin{equation} \label{inter-num}
	D_{\ell,\ell'} [\vec{\vv}_\ell] = D_{\ell,\ell'} = \left\{
	\begin{aligned}
		& \frac{1}{\det(\vec{v}_{\ell-1}, \vec{v}_\ell)}    \qquad  && \text{if} \ \ell' = \ell - 1 \,, \\
		& \frac{1}{\det(\vec{v}_\ell, \vec{v}_{\ell+1})}    \qquad  && \text{if} \ \ell' = \ell + 1 \,, \\[0.5em]
		& -\frac{\det(\vec{v}_{\ell-1}, \vec{v}_{\ell+1})}{\det(\vec{v}_{\ell-1}, \vec{v}_\ell) \det(\vec{v}_\ell, \vec{v}_{\ell+1})}    \qquad  && \text{if} \ \ell' = \ell \,, \\
		& 0    \qquad  && \text{otherwise} \, ,
	\end{aligned}
	\right.
\end{equation}
where we used the notation $D_{\ell,\ell'}[\vec{\vv}_\ell]$ to emphasize that it is the intersection matrix of the toric divisors associated with a fan with generators $\vec{\vv}_\ell$.
It can be easily proven that this satisfies the relation
\begin{equation} \label{Dn=0}
  \sum_{\ell'}  D_{\ell,\ell'} \, \vec{v}_{\ell'} = 0 \,,
\end{equation}
implying that $D_{\ell,\ell'}$, viewed as an $\fpN\times\fpN$ matrix, has two null eigenvalues, and it is therefore not invertible.
This of course is consistent with the fact that $\dim(H^2(\Morb_4,\QQ))=\fpN-2$ implies that only $\fpN-2$ toric divisors $D_\ell$ are independent in homology.

For the special case of toric manifolds, all $\det(\vec{v}_\ell,\vec{v}_{\ell+1})=1$, and the above discussion reduces to that in \cite{Hosseini:2020vgl}. On the other hand, for our purposes we need to incorporate the effect of the labels.
We will refrain from attempting a rigorous treatment, indicating a natural way to take into account this additional information.
If we have a labelled polytope/stacky fan described by ``long'' vectors $\vec{\hat{\vv}}_\ell = m_\ell\vec{\vv}_\ell$, it is natural to associate a divisor $\hat{D}_\ell$ with $\vec{\hat{\vv}}_\ell$, which is related to $D_\ell$ simply by
\begin{equation}
	\hat{D}_\ell = \frac{1}{m_\ell} D_\ell \,.
\end{equation}
In particular, this implies that for any two-form $\Lambda$ the integrals over the two sets of divisors are related as
\begin{equation}
	m_\ell \int_ {\hat{D}_\ell} \Lambda = \int_{D_\ell} \Lambda \,,
\end{equation}
and the intersection matrix of the divisors ${\hat{D}_\ell}$ is  given by
\begin{equation}
	\hat{D}_{\ell,\ell'} \equiv D_{\ell,\ell'}[\vec{\hat{\vv}}_\ell] = \frac{1}{m_\ell m_{\ell'}} D_{\ell,\ell'}[\vec{\vv}_\ell] \,.
\end{equation}
Correspondingly, the first Chern classes of the dual line bundles are related as $c_1(L_\ell) = m_\ell\,c_1(\hat{L}_\ell)$.
Since we can use either $c_1(L_\ell)$ or $c_1(\hat{L}_\ell)$ as basis for $H^2(\Morb_4,\QQ)$, these two descriptions are essentially interchangeable.

We note that $\hat{D}_{\ell,\ell'}$ is invariant under $SL(2,\ZZ)$ transformations of the basis  and more generally for any transformation in $SL(2,\RR)$. This follows immediately from the fact that given any two vectors $\vec{v}_1,\vec{v}_2 \in \RR^2$, $\det(\vec{v}_1,\vec{v}_2)$ is invariant under $SL(2,\RR)$ transformations of the two vectors.
In particular, taking a generic matrix $S \in SL(2,\RR)$ and transforming the vectors as $\pvec{v}'_a = S\,\vec{v}_a$, we have that
 $S$ acts on the matrix $(\vec{v}_1,\vec{v}_2)$ simply by matrix multiplication, \ie\ $(\pvec{v}'_1,\pvec{v}'_2) = S\,(\vec{v}_1,\vec{v}_2)$, therefore
\begin{equation}
  \det(\pvec{v}'_1,\pvec{v}'_2) = \det(S) \det(\vec{v}_1,\vec{v}_2) = \det(\vec{v}_1,\vec{v}_2) \,.
\end{equation}
This property implies that formally the intersection matrix of the divisors computed in the two bases~$\vec{\hat{w}}_\ell$ and~$\vec{\hat{n}}_\ell$
are identical, being related by the $SL(2,\QQ)$ transformation~\eqref{S-matrix}.

In order to make contact with the computations in supergravity, we need to describe the relation between the toric divisors defined above and the representative of the cycles, that we used to integrate fluxes in our main example of $\spindle_1\ltimes\spindle_2$.
The $S_\pm$ introduced in section~\ref{subsec:6d_2spin} were precisely two copies of the base $\spindle_1$ at the south pole (section at infinity) and north pole (zero section) of the fibre $\spindle_2$ and we can identify these with the toric divisors $D_1$ and $D_3$ as
\begin{equation}
	D_1 = S_- \,,  \qquad  D_3 = S_+ \,.
\end{equation}
On the other hand, $D_2$ and $D_4$ correspond to copies of the fibre~$\spindle_2$ at the poles of the base $\spindle_1$.
In the smooth case ($m_+=m_-=1$), these would be simply homologous to~$\spindle_2$. However, the orbifold singularities of the base imply that instead
\begin{equation}
	D_2 = X_- \,,  \qquad  D_4 = X_+ \,,
\end{equation}
where $X_\pm \equiv \spindle_2/\ZZ_{m_\pm}$, respectively.
Therefore, in order to compare the integral of a two-form $\Lambda$ over $X_\pm$ with the integral over $\spindle_2$ at a generic point of $\spindle_1$ we should use the relation
\begin{equation}
	m_\pm \int_ {X_\pm} \Lambda = \int_{\spindle_2} \Lambda \, .
\end{equation}
Since in section~\ref{subsec:6d_2spin} the fluxes were computed integrating on $\spindle_2$ (see \eg~\eqref{nonstackyintegral}), the integrals computed using $X_\pm$ will have to be multiplied by a factor of $m_\pm$ to be compared with the fluxes computed in section~\ref{subsec:6d_2spin}.

\section{The entropy function}
\label{sec:entropy-function}

We conjecture that the solutions presented in section~\ref{subsec:6d_2spin} are holographically dual to one-dimensional SCQMs obtained compactifying on a $\spindle_1\ltimes\spindle_2$ the five-dimensional SCFTs dual to the solution of~\cite{Brandhuber:1999np}.
In this section we provide supporting evidence for this conjecture, proposing an entropy function, whose extremization reproduces the entropy~\eqref{2spin_entropy} of the (putative) black hole near-horizon $\AdS_2\times\spindle_1\ltimes\spindle_2$. Starting from first principles, this function should  be derived from the localized partition function of the $d=5$ SCFT, placed on the background of $S^1\times\spindle_1\ltimes\spindle_2$, and then taking the large~$N$ limit.
Following the idea of ``gravitational blocks''~\cite{Hosseini:2019iad}, we will propose a large~$N$ entropy function on $S^1\times\spindle_1\ltimes\spindle_2$ obtained by suitably gluing the $S^5$ free energy of the $d=5$ theories.
With some further assumptions, this prescription can be applied also to the $\AdS_2\times\riemann\ltimes\spindle_2$ solutions of section~\ref{subsec:6d_riem}. The prescription that we propose  extends the results of~\cite{Faedo:2021nub} to a broader class of configurations, including solutions arising in $D=7$ supergravity.

\subsection{Twisting data}

In this section we consider the compactification of five- and six-dimensional SCFTs on a four-dimensional toric orbifold~$\Morb_4$. Specifically, we place the theory on $\Morb_4$ and couple it to two background gauge fields $A_i$, with appropriately quantized magnetic fluxes.
All the information about $\Morb_4$ is encoded in the toric data $(\vec{\vv}_\ell,m_\ell)$, which can be thought of as the data defining a labelled polytope.  The vectors~$\vec{\vv}_\ell$ must be primitive, $\ZZ^2$-valued and, in our conventions, ordered counter-clockwise. The recipe we will formulate is in principle applicable to toric orbifolds (including smooth toric manifolds) with an arbitrary number $\fpN \geq 3$ of fixed points.

Consider the two line bundles~$E_i$ on which the one-forms $-A_i$ are the connections. Since the set of $c_1(L_{\ell})$ form a basis for the equivariant extension of $H^2(\Morb_4,\QQ)$, we can decompose the first Chern class of~$E_i$ as
\begin{equation} \label{p_def}
  c_1(E_i)  = -\frac{\dd A_i}{2\pi} = -\sum_{\ell} \flp_i^{(\ell)} c_1(L_\ell) \,,
\end{equation}
where $\flp_i^{(\ell)} \in \QQ$. The ``physical fluxes'' are defined
as\footnote{Here and in the following we shall rename the background gauge fields as $g A_i \mapsto A_i$, which is more natural from the field theory point of view.}
\begin{equation} \label{q_def}
  \flq_i^{(\ell)} \equiv \frac{1}{2\pi} \int_{D_\ell} F_i = -\int_{D_\ell} c_1(E_i) \,,
\end{equation}
and using~\eqref{p_def} we obtain the relation
\begin{equation} \label{physfluxes}
  \flq_i^{(\ell)} = \sum_{\ell'} D_{\ell,\ell'} \, \flp_i^{(\ell')} \,,
\end{equation}
where $D_{\ell,\ell'}$ is the intersection matrix defined in~\eqref{inter-num}.
We recall that $H^2(\Morb_4,\QQ)$ has dimension $\fpN-2$, thus the whole set of $\flp_i^{(\ell)}$ is a redundant parameterization. In particular, for fixed~$i$, only  $\fpN-2$ of the $\flp_i^{(\ell)}$  are linearly independent. The two additional degrees of freedom are the expression of a gauge symmetry that we will discuss in the next subsection.

A further consequence of relation~\eqref{Dn=0} is that, combined with~\eqref{physfluxes}, it gives
\begin{equation} \label{qn=0}
	\sum_\ell \flq_i^{(\ell)} \vec{\vv}_\ell = 0 \,.
\end{equation}
This is a vector equation and, for fixed~$i$, we have $\fpN$ fluxes~$\flq_i^{(\ell)}$ constrained by two equations, one for each component of the two-dimensional vectors~$\vec{\vv}_\ell$. Therefore, only $\fpN-2$ physical fluxes are linearly independent. In what follows, given a generic toric orbifold with arbitrary fluxes, we shall require them to satisfy the constraint~\eqref{qn=0}.
We can now go back to relation~\eqref{physfluxes}. As we explained, it represents a linear system of $\rank(D_{\ell,\ell'}) = \fpN-2$ independent equations for $\fpN-2$ independent unknowns~$\flp_i^{(\ell)}$. Eliminating the redundant equations, we can solve this system and obtain~$\flp_i^{(\ell)}$ in terms of~$\flq_i^{(\ell)}$, up to gauge transformations.

Equation~\eqref{qn=0} imposes a similar condition on the $R$-symmetry fluxes~$\flq_R^{(\ell)} \equiv \flq_1^{(\ell)}+\flq_2^{(\ell)}$, specifically $\sum_\ell \flq_R^{(\ell)} \vec{\vv}_\ell = 0$. This constraint can be solved writing
\begin{equation} \label{physfluxes-relation}
	\flq_R^{(\ell)} = \sum_{\ell'} D_{\ell,\ell'} \, \frac{\sigma^{(\ell')}}{m_{\ell'}} \,,
\end{equation}
where $m_\ell$ are the labels of the labelled polytope associated with~$\Morb_4$ and, \textit{a priori}, $\sigma^{(\ell)}$ are $n$ arbitrary coefficients.
Denoting as $E_R$ the $R$-symmetry (orbifold) line bundle, equation~\eqref{physfluxes-relation} can be rewritten as
\begin{equation} \label{gentwist}
	c_1(E_R) = -\sum_{\ell} \sigma^{(\ell)} c_1(\hat{L}_\ell) \,,
\end{equation}
or, following standard practice, as $c_1(E_R) = -\sum_{\ell}\sigma^{(\ell)} \hat D_{\ell}$.
Note that the standard topological twist corresponds to identifying $E_R$ with the orbifold canonical line bundle $K^\mathrm{orb}_{\Morb_4} = -\sum_\ell \hat D_\ell$ and hence $\sigma^{(\ell)}_\mathrm{top\mbox{-}twist}=(+,\dots,+)$. See \eg~\cite{Martelli:2007pv} for a related discussion of orbifold line bundles.

In analogy with the case of the spindle~\cite{Ferrero:2021etw}, we conjecture that the only possible values of $\sigma^{(\ell)}$ are $\pm1$.
This hypothesis is supported by the different examples analysed, as we shall see.
In other words, we expect that it should be possible to  compactify a SCFT on a toric orbifold $\Morb_4$, turning on a background $R$-symmetry gauge field with magnetic fluxes given in (\ref{physfluxes-relation}), with $\sigma^{(\ell)}$ parameterizing the different supersymmetry-preserving twists.
Of course, we have not proven that all these different twists preserve supersymmetry, nor that there cannot exist more general twists. It would be interesting to carry out such an analysis extending the results of~\cite{Ferrero:2021etw}, where it was demonstrated that, in whole generality, on a spindle the only two supersymmetry-preserving twists are the twist and the anti-twist.

\subsection{The recipe}
\label{subsec:recipe}

We conjecture that given a general class of SCFTs in $d=5,6$ compactified on a four-dimensional toric orbifold~$\Morb_4$, with an arbitrary twist parameterized by a set of signs~$\sigma^{(\ell)}$ as in~\eqref{physfluxes-relation}, the corresponding entropy/central charge, respectively, is determined extremizing the following \emph{off-shell free energy}
\begin{equation} \label{F-extr}
	\Ispindle(\varphi_i, \epsilon_i; \flq_i^{(\ell)}, m_\ell) = k_d \sum_{\ell} \frac{\eta_d^{(\ell)}}{d_{\ell,\ell+1}} \frac{\mathcal{F}_d(\Phi_i^{(\ell)})}{\epsilon_1^{(\ell)} \epsilon_2^{(\ell)}} \,.
\end{equation}
The sum runs over the~$\fpN$ fixed points~$\fp_\ell$ of the toric orbifold~$\Morb_4$, $k_d$ is a numerical constant, which, \textit{a posteriori}, takes the values
\begin{equation}
	k_5 = -1 \,,  \qquad  k_6 = \frac{64}{9} \,,
\end{equation}
$d_{\ell,\ell+1}$ is defined as $d_{\ell,\ell'} \equiv d_{\ell,\ell'}[\vec{\vv}_\ell] = \det(\vec{\vv}_\ell, \vec{\vv}_{\ell'})$ and $\mc{F}_d$  are the usual gravitational blocks  (\cf\ table~2 of~\cite{Faedo:2021nub})
\begin{equation}
	\mathcal{F}_5(\Delta_i) = -\frac{4\sqrt2 \pi \, N^{5/2}}{15 \sqrt{8 - N_f}} (\Delta_1 \Delta_2)^{3/2} \,,  \qquad  \mathcal{F}_6(\Delta_i) = -\frac{9 N^3}{256} (\Delta_1 \Delta_2)^2 \,.
\end{equation}
The variables~$\Phi_i^{(\ell)}$ are defined as
\begin{equation} \label{def_Phi}
  \Phi_i^{(\ell)} = \varphi_i - \flp_i^{(\ell)} \epsilon_1^{(\ell)} - \flp_i^{(\ell+1)} \epsilon_2^{(\ell)} \,,
\end{equation}
and the auxiliary quantities $\epsilon_1^{(\ell)}$ and $\epsilon_2^{(\ell)}$ read
\begin{equation} \label{def_epsilon}
  \epsilon^{(\ell)}_1 = -\frac{\det(\vec{\vv}_{\ell+1}, \vec{\epsilon})}{d_{\ell,\ell+1}} \,,  \qquad
	\epsilon^{(\ell)}_2 = \frac{\det(\vec{\vv}_\ell, \vec{\epsilon})}{d_{\ell,\ell+1}} \,,
\end{equation}
with $\vec{\epsilon}=(\epsilon_1,\epsilon_2)$\footnote{The two variables~$\epsilon_{1,2}$ are different from the ones adopted in~\cite{Faedo:2021nub}. Specifically, $\epsilon_i^\mathrm{here} = 2\epsilon_i^\mathrm{there}$.}. Here, $\epsilon_1$ and $\epsilon_2$ may be interpreted as the fugacities associated with the two $U(1)$ rotational symmetries and parameterize their mixing with the $R$-symmetry.
$\eta_d^{(\ell)}=\pm$ are signs that at this stage must be tuned by hand: in particular we believe that in general\footnote{Due to different conventions they could also be all~$-$, as explained at the end of subsection~\ref{subsec:applicatin-ads3}.} in $d=6$ they are all~$+$, while for $d=5$ they are related to the type of twist and we speculate that $\eta_5^{(\ell)} = \sigma^{(\ell)}\sigma^{(\ell+1)}$.
In the next section we will see some explicit examples applying this prescription to the $\AdS_2 \times \spindle_1 \ltimes \spindle_2$ family.
In this construction $\varphi_i$ and $\epsilon_i$ are variables with respect to which one has to extremize~$\Ispindle(\varphi_i,\epsilon_i)$, $\vec{\vv}_\ell$ and $m_\ell$ are data describing the toric orbifold~$\Morb_4$ and $\flp_i^{(\ell)}$ are related to the physical fluxes~$\flq_i^{(\ell)}$ through~\eqref{physfluxes}.

The variables $\varphi_i$ and $\epsilon_i$ are subject to the  constraint
\begin{equation} \label{constraint}
  \varphi_1 + \varphi_2 - \det(\vec{W}, \vec{\epsilon}) = 2 \,,
\end{equation}
where $\vec{W}$ is a two-dimensional constant vector parameterizing a ``gauge invariance'' of the problem, that we discuss below. This condition is inherited from the $R$-symmetry constraint $\Delta_1+\Delta_2=2$, where $\Delta_i$ are the fugacities parameterizing the $R$-symmetry within the Cartan subgroup of the global symmetries of the $d$-dimensional parent theory. The vector $\vec{W}$ can be determined imposing
\begin{equation} \label{relation-p-W}
  \flp_1^{(\ell)} + \flp_2^{(\ell)} = \frac{\sigma^{(\ell)}}{m_\ell} + \det(\vec{W}, \vec{\vv}_\ell) \, ,
\end{equation}
which  is a system of $\fpN$ equations -- one for each value of~$\ell$ -- for the two components of~$\vec{W}$, and it is thus overdetermined. Yet, a solution always exists. By means of equations~\eqref{physfluxes} and~\eqref{relation-p-W} we can write the following chain of equalities
\begin{equation}
  \flq_R^{(\ell)} = \sum_{\ell'} D_{\ell,\ell'} \bigl( \flp_1^{(\ell')} + \flp_2^{(\ell')} \bigr) = \sum_{\ell'} D_{\ell,\ell'} \, \frac{\sigma^{(\ell')}}{m_{\ell'}} + \det\Bigl( \vec{W}, \sum_{\ell'} D_{\ell,\ell'} \vec{\vv}_{\ell'} \Bigr) = \sum_{\ell'} D_{\ell,\ell'} \, \frac{\sigma^{(\ell')}}{m_{\ell'}} \,,
\end{equation}
where, in the last step, we made use of~\eqref{Dn=0}. First of all, this relation proves the consistency of~\eqref{relation-p-W}, since, as it should be, the physical quantities $\flq_R^{(\ell)}$ do not depend on the auxiliary vector~$\vec{W}$. Moreover, it shows that $\fpN-2$ independent linear combinations of equations~\eqref{relation-p-W}, obtained contracting them with~$D_{\ell,\ell'}$, are already satisfied once~\eqref{physfluxes-relation} is imposed. The result is a linear system of two independent equations with two unknowns, that can always be solved.

As we already mentioned, the quantity $\det(\vec{v}_1,\vec{v}_2)$ is invariant under $SL(2,\RR)$ transformations of the two vectors $\vec{v}_1,\vec{v}_2 \in \RR^2$. This property is fundamental for the consistency of our construction. A given toric orbifold can be described by an infinite number of polytopes equivalent under $SL(2,\ZZ)$ transformations of the vectors $\vec{\vv}_\ell$. These transformations are generated by rotations of the basis vectors $(e_1,e_2)$ that give rise to effective torus actions.
On the other hand, we expect the free energy not to be affected by these transformations and, indeed, this is the case. If we perform an $SL(2,\ZZ)$ rotation of the vectors~$\vec{\vv}_\ell$, the expression of the intersection matrix $D_{\ell,\ell'}$ is retained, and the explicit form of the off-shell free energy is not modified provided we apply exactly the same transformation to $\vec{\epsilon}$ and $\vec{W}$.

The off-shell entropy function we constructed enjoys also another symmetry, the gauge symmetry we mentioned before. Thanks to~\eqref{Dn=0}, relation~\eqref{physfluxes} is left invariant under the ``gauge transformation''
\begin{equation} \label{gauge-trans}
  \tilde{\flp}_i^{(\ell)} = \flp_i^{(\ell)} + \det(\vec{\lambda}_i, \vec{\vv}_\ell) \,,
\end{equation}
for any two-dimensional constant vector~$\vec{\lambda}_i$. This transformation affects the value of the vector~$\vec{W}$ appearing in the constraint. Specifically, since $\sigma^{(\ell)}$ and $m_\ell$ are fixed, we must impose
\begin{equation}
  \tilde{\flp}_1^{(\ell)} + \tilde{\flp}_2^{(\ell)} - \det(\vec{\tilde{W}}, \vec{\vv}_\ell) = \flp_1^{(\ell)} + \flp_2^{(\ell)} - \det(\vec{W}, \vec{\vv}_\ell) \,,
\end{equation}
which returns $\vec{\tilde{W}}=\vec{W}+\vec{\lambda}_1+\vec{\lambda}_2$. Introducing the new variables $\tilde{\varphi}_i = \varphi_i+\det(\vec{\lambda}_i,\vec{\epsilon})$ it is possible to keep the functional expression of $\Phi_i^{(\ell)}$ unmodified, \ie\ $\Phi_i^{(\ell)}(\tilde{\varphi},\tilde{\flp}) = \Phi_i^{(\ell)}(\varphi,\flp)$. Remarkably, the constraint changes accordingly
\begin{equation}
  \tilde{\varphi}_1 + \tilde{\varphi}_2 - \det(\vec{\tilde{W}}, \vec{\epsilon}) = \varphi_1 + \varphi_2 - \det(\vec{W}, \vec{\epsilon}) = 2 \,.
\end{equation}
As a result, both the off-shell free energy and the constraint retain the same functional expression under the gauge transformation~\eqref{gauge-trans}, thus the value of the former at its critical points will be the same.

The proposed prescription takes inspiration from~\cite{Hosseini:2020vgl} and extends their construction to compactifications on generic toric orbifolds.
Moreover, it also applies to $d=5$, generalizing the formulas presented in~\cite{Faedo:2021nub} for compactifications on $\Sigma_{\mathrm{g}_1}\times\Sigma_{\mathrm{g}_2}$ and $\spindle\times\riemann$, to the case of fibred spaces. The constrained extremization of $\Ispindle(\varphi_i,\epsilon_i)$ can be performed defining the function
\begin{equation}\label{entropy_funct}
  \mathcal{S}(\varphi_i, \epsilon_i, \Lambda; \flq_i^{(\ell)}, m_\ell) = \Ispindle(\varphi_i, \epsilon_i; \flq_i^{(\ell)}, m_\ell) + \Lambda \bigl( \varphi_1 + \varphi_2 - \det(\vec{W}, \vec{\epsilon}) - 2 \bigr)
\end{equation}
and extremizing it with respect to~$\varphi_i$, $\epsilon_i$ and the Lagrangian multiplier~$\Lambda$.
In order to solve this problem it might be helpful to notice that $\eqref{F-extr}$ is homogeneous of degree~$h$ in~$\varphi_i$ and $\epsilon_i$ ($h=1$ for $d=5$, $h=2$ for $d=6$). As a consequence, by Euler's theorem, we have $\Ispindle(\varphi_i^*,\epsilon_i^*)=-\frac{2}{h}\Lambda^*$.

\subsection{Application to the AdS$_2\times \spindle_1 \ltimes \spindle_2$ solutions}

The first example to which we can apply the prescription presented in the previous section is the $\AdS_2 \times \spindle_1 \ltimes \spindle_2$ solution~\eqref{2spin_solution}. We now briefly collect some of the results needed in the construction. The toric orbifold $\Morb_4 = \spindle_1\ltimes\spindle_2$ can be described by the toric data (see section~\ref{subsec:toric_2spin_6d})
\begin{equation}
  \begin{aligned}
    m_1 &= n_- \,,  \quad  & \vec{w}_1 &= (1, 0) \,,   \qquad\quad  & m_2 &= 1 \,,  \quad  & \vec{w}_2 &= (r_+, m_-) \,, \\
    m_3 &= n_+ \,,  \quad  & \vec{w}_3 &= (-1, 0) \,,  \qquad\quad  & m_4 &= 1 \,,  \quad  & \vec{w}_4 &= (r_-, -m_+) \,,
  \end{aligned}
\end{equation}
with $t = r_+\,m_+ + r_-\,m_-$. The vectors~$\vec{w}_\ell$ are integers, primitive and ordered counter-clockwise, as required by the recipe. Keeping in mind all the observations about divisors made in section~\ref{subsec:divisors}, the physical fluxes defined in~\eqref{q_def} read
\begin{equation}
	\begin{aligned}
		\flq_i^{(1)} &= \frac{g}{2\pi} \int_{S_-} F_i = \fls_i^- \,,  \qquad  & \flq_i^{(2)} &= \frac{g}{2\pi} \int_{X_-} F_i = \frac{g}{2\pi} \frac{1}{m_-} \int_{\spindle_2} F_i = \frac{\flt_i}{m_-} \,, \\
		\flq_i^{(3)} &= \frac{g}{2\pi} \int_{S_+} F_i = \fls_i^+ \,,  \qquad  & \flq_i^{(4)} &= \frac{g}{2\pi} \int_{X_+} F_i = \frac{g}{2\pi} \frac{1}{m_+} \int_{\spindle_2} F_i = \frac{\flt_i}{m_+} \,,
	\end{aligned}
\end{equation}
where $\fls_i^+$, $\fls_i^-$ and $\flt_i$ are given, respectively, in~\eqref{2spin_quantS}, \eqref{2spin_quantN} and~\eqref{quant_F-spin2}. The intersection matrix $D_{\ell,\ell'}$, computed straightforwardly  using~\eqref{inter-num}, reads
\begin{equation}
	D_{\ell,\ell'} = \left( \begin{array}{cccc}
		-\frac{t}{m_+ m_-} & \frac{1}{m_-} & 0 & \frac{1}{m_+} \\
		\frac{1}{m_-} & 0 & \frac{1}{m_-} & 0 \\
		0 & \frac{1}{m_-} & \frac{t}{m_+ m_-} & \frac{1}{m_+} \\
		\frac{1}{m_+} & 0 & \frac{1}{m_+} & 0
	\end{array} \right)\, ,
\end{equation}
and imposing~\eqref{physfluxes-relation} we obtain the vector of twists
\begin{equation} \label{free_sigma}
  \sigma^{(\ell)} = (+, +, +, -) \,.
\end{equation}

Taking advantage of the gauge symmetry, we impose $\flp_i^{(1)}=\flp_i^{(3)}$ and $\flp_i^{(2)}/m_-=\flp_i^{(4)}/m_+$. Therefore, the relation~\eqref{physfluxes} yields the identifications
\begin{equation}
  \flp_i^{(1)} = \flp_i^{(3)} = \frac{\flt_i}{2} \,,  \qquad  \flp_i^{(2)} = \frac{m_- \, \fls_i^*}{2} \,,  \qquad  \flp_i^{(4)} = \frac{m_+ \, \fls_i^*}{2} \,,
\end{equation}
where we defined $\fls_i^* \equiv \frac{\fls_i^+ + \fls_i^-}{2}$. In particular $\fls_1^*$ explicitly reads
\begin{equation}
  \fls_1^* = \frac{t \chi_2}{m_+ m_-} \, \frac{\x^3 - 3\mu \x^2 + (3 - 2\mathtt{z}) \x - 3\mu}{8\x^2} \,,
\end{equation}
and $\fls_2^*$ can be obtained flipping the sign of~$\mathtt{z}$. With this definition, $\fls_i^\pm$ can be written as
\begin{equation}
  \fls_i^\pm = \fls_i^* \pm \frac{t}{2m_+ m_-} \, \flt_i \,.
\end{equation}
System~\eqref{relation-p-W} can be solved to get~$\vec{W}$, which gives the constraint~\eqref{constraint}
\begin{equation} \label{free_constraint}
  \varphi_1 + \varphi_2 + \frac{n_+ - n_-}{2n_+ n_-} \epsilon_1 + \Bigl( \frac{m_+ + m_-}{2m_+ m_-} - \frac{r_+\,m_+ - r_-\,m_-}{m_+ m_-} \frac{n_+ - n_-}{4n_+ n_-} \Bigr) \epsilon_2 = 2 \,.
\end{equation}
The ingredients $\epsilon_{1,2}^{(\ell)}$ and $\Phi_i^{(\ell)}$ can be constructed by means of~\eqref{def_epsilon} and~\eqref{def_Phi}, whereas the off-shell free energy~\eqref{F-extr} reads
\begin{equation} \label{free_2spin_F-extr}
  \Ispindle(\varphi_i, \epsilon_i; \fls_i^\pm, \flt_i, m_\ell) =
  - \frac{\mathcal{F}_5(\Phi_i^{(1)})}{d_{1,2} \, \epsilon_1^{(1)} \epsilon_2^{(1)}}
  - \frac{\mathcal{F}_5(\Phi_i^{(2)})}{d_{2,3} \, \epsilon_1^{(2)} \epsilon_2^{(2)}}
  + \frac{\mathcal{F}_5(\Phi_i^{(3)})}{d_{3,4} \, \epsilon_1^{(3)} \epsilon_2^{(3)}}
  + \frac{\mathcal{F}_5(\Phi_i^{(4)})}{d_{4,1} \, \epsilon_1^{(4)} \epsilon_2^{(4)}} \,,
\end{equation}
where the relative signs have been fixed using the prescription  $\eta_5^{(\ell)} = \sigma^{(\ell)}\sigma^{(\ell+1)}$.

We can now proceed to the extremization of the off-shell free energy and, after some work, we get the critical values
\begin{equation}
  \begin{aligned}
    \varphi_i^* &= \frac{2m_+ m_-}{m_+ - m_-} \Bigl( 1 + \eta \, \frac{m_+ + m_-}{2m_+ m_-} \frac{2\pi}{\Delta\psi} \Bigr) \, \fls_i^* \,, \\
    \epsilon_1^* &= -\frac{1}{m} \frac{2\pi}{\Delta z} + \eta \, \frac{2(r_+ + r_-)}{m_+ - m_-} \frac{2\pi}{\Delta\psi} \,,  \qquad
		\epsilon_2^* = -2\eta \, \frac{2\pi}{\Delta\psi} \,,
  \end{aligned}
\end{equation}
where the sign ambiguity $\eta=\pm1$ arises by solving the equations over the complex numbers\footnote{Generically, in presence of rotation we have to work with the complex numbers~\cite{Cassani:2021dwa}, therefore we continue to do so also in the static case.}. Inserting these values back into~\eqref{free_2spin_F-extr} we obtain
\begin{equation}
  \Ispindle(\varphi_i^*, \epsilon_i^*; \fls_i^\pm, \flt_i, m_\ell) = \frac{-\eta \sqrt2 \sqrt{m_+^2 + m_-^2} - (m_+ + m_-)}{2m_+ m_-} \, F_{S^3\times\spindle_2} \,,
\end{equation}
where $F_{S^3\times\spindle_2}$ is given in~\eqref{free_energy_faedo}. In order to get a positive entropy we need to pick $\eta=-1$ and in this case the result agrees with the gravitational entropy~\eqref{2spin_entropy}.

We propose that the procedure we formulated can be extended also to non-toric four-dimensional orbifolds, \eg\ in presence of Riemann surfaces as submanifold. In this case we can adopt the same toric data obtained replacing $\riemann$ ($\mathrm{g}>1$) with~$S^2$, setting $\mathrm{g}=0$ whenever it occurs in~\eqref{relation-p-W} and~\eqref{physfluxes-relation}. The genus is preserved in the explicit expression of the fluxes.
Specifically, we will test our conjecture against the $\AdS_2 \times \riemann \ltimes \spindle_2$ system~\eqref{riem_solution}. The ``toric data'' are given in~\eqref{riem_vec-n} and~\eqref{riem_toric-m}, the physical fluxes are the same as in the previous example, with $m_\pm=1$, and now $\fls_i^\pm$ can be read in~\eqref{riem_quantS} and~\eqref{riem_quantN}.
The fluxes~$\flp_i^{(\ell)}$ are identified as before and $\fls_i^*$ retains the same form, but, again, with $m_\pm=1$ and $t$ as in~\eqref{riem_t}.
The constraint on the variables~$\varphi_i$ and~$\epsilon_i$ is
\begin{equation} \label{constr_riemm_6d}
  \varphi_1 + \varphi_2 + \frac{n_+ - n_-}{2n_+ n_-} \epsilon_1 - t \, \frac{n_+ - n_-}{4n_+ n_-} \epsilon_2 = 2 \,,
\end{equation}
whereas the off-shell free energy reads
\begin{equation}
  \Ispindle(\varphi_i, \epsilon_i; \fls_i^\pm, \flt_i, m_\ell) =
  \frac{\mathcal{F}_5(\Phi_i^{(1)})}{d_{1,2} \, \epsilon_1^{(1)} \epsilon_2^{(1)}}
  + \frac{\mathcal{F}_5(\Phi_i^{(2)})}{d_{2,3} \, \epsilon_1^{(2)} \epsilon_2^{(2)}}
  + \frac{\mathcal{F}_5(\Phi_i^{(3)})}{d_{3,4} \, \epsilon_1^{(3)} \epsilon_2^{(3)}}
  + \frac{\mathcal{F}_5(\Phi_i^{(4)})}{d_{4,1} \, \epsilon_1^{(4)} \epsilon_2^{(4)}} \,,
\end{equation}
where again the relative signs were chosen to be $\eta_5^{(\ell)} = \sigma^{(\ell)}\sigma^{(\ell+1)}$.
These different gluing signs are due to the different types of twist that we have. In particular, on both $\riemann$ and~$\spindle_2$ the twist is realized, in contrast to the previous case, in which we had anti-twist on~$\spindle_1$. The off-shell free energy is extremized by
\begin{equation}
  \varphi_i^* = \frac{\fls_i^*}{\mathrm{g} - 1} \,,  \qquad  \epsilon_1^* = -\frac{1}{m} \frac{2\pi}{\Delta z} \,,  \qquad  \epsilon_2^* = 0 \,,
\end{equation}
to which corresponds the critical value
\begin{equation}
  \Ispindle(\varphi_i^*, \epsilon_i^*; \fls_i^\pm, \flt_i, m_\ell) = (\mathrm{g} - 1) \, F_{S^3\times\spindle_2} \,.
\end{equation}
Also in this case, the result returned by the extremization agrees with the entropy computed from the ten-dimensional supergravity solution~\eqref{riem_entropy}.

\subsection{Application to the AdS$_3\times \spindle_1 \ltimes \spindle_2$ solutions}
\label{subsec:applicatin-ads3}

The conjectured prescription presented at the beginning of this section can also be applied to another interesting family, the $\AdS_3 \times \spindle_1 \ltimes \spindle_2$ solutions constructed in~\cite{Cheung:2022ilc}\footnote{In order to make the analogy with the $\AdS_2 \times \spindle_1 \ltimes \spindle_2$ systems more manifest, we exchanged $n_+ \leftrightarrow n_-$ and $m_+ \leftrightarrow m_-$ with respect to~\cite{Cheung:2022ilc}.}. The toric data of the $\spindle_1\ltimes\spindle_2$ toric orbifold are derived in appendix~\ref{app:toric_2spin_7d} and we recall them here for the reader's convenience
\begin{equation}
  \begin{aligned}
    m_1 &= n_- \,,  \quad  & \vec{w}_1 &= (1, 0) \,,   \qquad\quad  & m_2 &= 1 \,,  \quad  & \vec{w}_2 &= (r_+, m_-) \,, \\
    m_3 &= n_+ \,,  \quad  & \vec{w}_3 &= (-1, 0) \,,  \qquad\quad  & m_4 &= 1 \,,  \quad  & \vec{w}_4 &= (r_-, -m_+) \,,
  \end{aligned}
\end{equation}
with $t = r_+\,m_+ + r_-\,m_-$. Global conditions lead to
\begin{equation} \label{global_cond_7d}
  t = \frac{-6 (m_+ - m_-) n_+ n_- (n_+ - n_-)}{[\mathtt{s} - (p_1 + p_2)] [\mathtt{s} + 2(p_1 + p_2)]} \,,  \qquad  \mathtt{s} = \sqrt{7 (p_1^2 + p_2^2) + 2p_1 p_2 - 6(n_+^2 + n_-^2)} \,,
\end{equation}
where $p_1$ and $p_2$ are two integers related to the magnetic fluxes as in~(2.11) of~\cite{Cheung:2022ilc}.
The physical fluxes can be computed as in the previous example and, compactly, they read
\begin{equation} \label{physical_fluxes_7d}
  \flq_i^{(\ell)} = \Bigl( \fls_i^-, \frac{\flt_i}{m_-}, \fls_i^+, \frac{\flt_i}{m_+} \Bigr)  \,,
\end{equation}
where $\fls_1^-$ corresponds to the flux computed in~(3.15) of~\cite{Cheung:2022ilc}, $\fls_2^-$ and $\fls_i^+$ can be derived from the former and $\flt_i=\frac{p_i}{n_+ n_-}$. The gauge symmetry allows us to impose $\flp_i^{(1)}=\flp_i^{(3)}$ and $\flp_i^{(2)}/m_-=\flp_i^{(4)}/m_+$, which yields, through~\eqref{physfluxes}, the identifications
\begin{equation}\label{identif_7d}
  \flp_i^{(1)} = \flp_i^{(3)} = \frac{\flt_i}{2} \,,  \qquad  \flp_i^{(2)} = \frac{m_- \, \fls_i^*}{2} \,,  \qquad  \flp_i^{(4)} = \frac{m_+ \, \fls_i^*}{2} \,.
\end{equation}
For later convenience we defined $\fls_1^* \equiv \frac{\fls_i^+ + \fls_i^-}{2}$, whose expression is
\begin{equation}
  \fls_i^* = \frac{t p_i [-6p_i + 4(n_+ + n_-) - \mathtt{s}]}{6m_+ m_- n_+ n_- (n_+ - n_-)} \,.
\end{equation}
The fluxes $\fls_i^\pm$ are related to~$\fls_i^*$ through
\begin{equation}
  \fls_i^\pm = \fls_i^* \pm \frac{t}{2m_+ m_-} \, \flt_i \,.
\end{equation}
The constraint can be derived following the path traced in the previous case and reads
\begin{equation}
  \varphi_1 + \varphi_2 + \frac{n_+ - n_-}{2n_+ n_-} \epsilon_1 + \Bigl( \frac{m_+ + m_-}{2m_+ m_-} - \frac{r_+\,m_+ - r_-\,m_-}{m_+ m_-} \frac{n_+ - n_-}{4n_+ n_-} \Bigr) \epsilon_2 = 2 \,.
\end{equation}
Notice that this expression is identical to~\eqref{free_constraint}.
As we discussed in section \ref{subsec:recipe}, in $d=6$ we set $\eta_6^{(\ell)}=+$ for all~$\ell$. Remarkably, this implies that $F$ is \emph{quadratic} in the $\varphi_i$, as expected for the off-shell central charge of a two-dimensional SCFT \cite{Benini:2012cz}. Explicitly, this reads
\begin{align}
	 \Ispindle(\varphi_i, \epsilon_i; \fls_i^\pm, \flt_i, m_\ell) = &- \biggl[ \frac{\flt_2 \fls_1^* + \flt_1 \fls_2^*}{8} \biggl( \flt_1 \flt_2 \epsilon_1^2 + \frac{(r_+^2 m_+^2 + r_-^2 m_-^2) \flt_1 \flt_2 + 2m_+^2 m_-^2 \fls_1^* \fls_2^*}{2m_+^2 m_-^2} \, \epsilon_2^2 \nonumber \\
	& - \frac{r_+ m_+ - r_- m_-}{m_+ m_-} \, \flt_1 \flt_2 \epsilon_1 \epsilon_2 \biggr) + \frac{\flt_2 \fls_2^* \varphi_1^2 + \flt_1 \fls_1^* \varphi_2^2}{2} + (\flt_2 \fls_1^* + \flt_1 \fls_2^*) \varphi_1 \varphi_2 \nonumber \\
	& + \frac{2m_+ m_- \epsilon_1 - (r_+ m_+ - r_- m_-) \epsilon_2}{8m_+^2 m_-^2} \, t \, \flt_1 \flt_2 (\flt_2 \varphi_1 + \flt_1 \varphi_2) \biggr] N^3\, .
\end{align}

The extremization procedure, realized by means of Lagrangian multipliers, gives the critical values\footnote{We notice that the expression of~$\varphi_i^*$, written in this seemingly cumbersome way, is identical to the corresponding quantity of the $\AdS_2 \times \spindle_1 \ltimes \spindle_2$ extremization.}
\begin{equation}
  \begin{aligned}
    \varphi_i^* &= \frac{2m_+ m_-}{m_+ - m_-} \Bigl( 1 - \frac{m_+ + m_-}{2m_+ m_-} \frac{2\pi}{\Delta\psi} \Bigr) \, \fls_i^* \,, \\
    		\epsilon_1^* &= -\frac43 \frac{2\pi}{\Delta z} - \frac{2 (r_+ + r_-)}{m_+ - m_-} \frac{2\pi}{\Delta\psi} \,,  \qquad
		\epsilon_2^* = 2 \, \frac{2\pi}{\Delta\psi} \,,
  \end{aligned}
\end{equation}
which, plugged into the off-shell central charge, gives
\begin{equation} \label{2spin_central-charge}
  \Ispindle(\varphi_i^*, \epsilon_i^*; \fls_i^\pm, \flt_i, m_\ell) = \frac{4 (m_+ - m_-)^3}{3m_+ m_- (m_+^2 + m_+ m_- + m_-^2)} \, a_{4\mathrm{d}} \,,
\end{equation}
where $a_{4\mathrm{d}}$ is the central charge of $d=4$, $\mathcal{N}=1$ SCFTs that arise from $N$ M5-branes wrapped on a spindle~\cite{Ferrero:2021wvk}
\begin{equation} \label{cent_charge_4d}
  a_{4\mathrm{d}} = \frac{3p_1^2 p_2^2 (\mathtt{s} + p_1 + p_2)}{8n_+ n_- (n_+ - p_1) (p_2 - n_+) [\mathtt{s} + 2(p_1 + p_2)]^2} \, N^3 \,.
\end{equation}
This matches exactly the central charge of the $\AdS_3\times \spindle_1 \ltimes \spindle_2$ system computed in~\cite{Cheung:2022ilc}.
In general, we expect that the anomaly polynomial computation of~\cite{Cheung:2022ilc} and the prescription presented in this paper should be equivalent and connected by a suitable gauge choice and a possible redefinition of $\epsilon_{1,2}$, mixing the two related $U(1)$ isometries.

As explained in the previous section, our recipe can be applied also to non-toric four-dimensional orbifolds. In this class we consider $\AdS_3 \times \riemann\ltimes\spindle_2$, whose ``toric data'' are deduced in appendix~\ref{app:toric_riem_7d} for $\AdS_3 \times S^2 \ltimes \spindle_2$ and read
\begin{equation}
	\begin{aligned}
		m_1 &= n_- \,,  \quad  & \vec{n}_1 &= (-1, 0) \,,  \qquad\quad  & m_2 &= 1 \,,  \quad  & \vec{n}_2 &= (0,-1) \,, \\
		m_3 &= n_+ \,,  \quad  & \vec{n}_3 &= (1, 0) \,,  \qquad\quad  & m_4 &= 1 \,,  \quad  & \vec{n}_4 &= (-t, 1) \,.
	\end{aligned}
\end{equation}
Global regularity leads to
\begin{equation}
	t = \frac{12(\mathrm{g}-1) n_+ n_- (n_+ - n_-)}{[\mathtt{s} - (p_1 + p_2)] [\mathtt{s} + 2(p_1 + p_2)]} \,,
\end{equation}
with $\mathtt{s}$ given in equation~\eqref{global_cond_7d}. We take the physical fluxes to be as in~\eqref{physical_fluxes_7d}, where $\flt_i=\frac{p_i}{n_+ n_-}$ is unchanged and $\fls_i^\pm$ can be read from (4.7) of \cite{Cheung:2022ilc}.
The gauge symmetry allows us to identify the $\flp_i^{(\ell)}$ as in equation~\eqref{identif_7d}, with $\fls_i^\pm = \fls_i^*\pm\frac{t}{2}\flt_i$ and
\begin{equation}
\fls_i^* = \frac{t p_i [-6p_i + 4(n_+ + n_-) - \mathtt{s}]}{6 n_+ n_- (n_+ - n_-)} \,.
\end{equation}
We can then solve equation~\eqref{physfluxes-relation} to obtain $\sigma^{(\ell)}=(+,-,+,-)$, use it to identify the auxiliary vector $\vec{W}$ through~\eqref{relation-p-W} and write down the constraint~\eqref{constraint}
\begin{equation}
	\varphi_1 + \varphi_2 - \frac{n_+ - n_-}{2n_+ n_-} \epsilon_1 - t \, \frac{n_+ - n_-}{4n_+ n_-} \epsilon_2 = 2 \,.
\end{equation}
This is the same as in~\eqref{constr_riemm_6d}, with $t \mapsto -t$ and $\epsilon_{1,2} \mapsto -\epsilon_{1,2}$, due to the different conventions used in~\cite{Cheung:2022ilc}.

Taking all the $\eta_6^{(\ell)}=-$ and extremizing equation~\eqref{entropy_funct} with respect to $\varphi_i, \,\epsilon_i$ and $\Lambda$, we obtain the critical values
\begin{equation}
	\varphi_i^* = -\frac{\fls_i^*}{\mathrm{g} - 1} \,,  \qquad  \epsilon_1^* = \frac43 \frac{2\pi}{\Delta z} \,,  \qquad  \epsilon_2^* = 0 \,,
\end{equation}
as well as the off-shell central charge at the extremum
\begin{equation} \label{riem_central-charge}
	\Ispindle(\varphi_i^*, \epsilon_i^*; \fls_i^\pm, \flt_i, m_\ell) = \frac{32}{3} (\mathrm{g} - 1) \, a_{4\mathrm{d}} \,,
\end{equation}
with $a_{4\dd}$ given in~\eqref{cent_charge_4d}. This result agrees with the central charge of the $\AdS_3 \times \riemann \ltimes \spindle_2$ system studied in~\cite{Cheung:2022ilc}.
The origin of the choice of the $\eta_6^{(\ell)}$ can be traced to the sign of the gauge potential of the five-dimensional solution that the authors of~\cite{Cheung:2022ilc} uplifted to obtain the $\AdS_3 \times \riemann \ltimes \spindle_2$ backgrounds.
The component along $\AdS_3$ of their Killing spinors satisfy the corresponding Killing spinor equations, but with a minus, namely ${\hat{\nabla}_{\hat{\mu}} \vartheta = -\frac12 \alpha_{\hat{\mu}} \vartheta}$, which implies that in the two-dimensional dual SCFTs $(2,0)$ supersymmetries are preserved, instead of $(0,2)$ as in our case. This fact leads to a different sign in the extraction of the central charge from the anomaly polynomial.

\section{Discussion}

In this paper we have presented two new families of supersymmetric $\AdS_2$ solutions of massive type~IIA supergravity, associated with D4-branes wrapped on four-dimensional orbifolds $\Morb_4$, which may be viewed as two different generalizations of the Hirzebruch surfaces.
In one case $\Morb_4 = \riemann\ltimes\spindle$ is a spindle $\spindle$ fibred over a smooth Riemann surface~$\riemann$ of genus $\mathrm{g}>1$, while in the other case $\Morb_4 = \spindle\ltimes\spindle $ is a spindle $\spindle$ fibred over another spindle~$\spindle$. We have argued that these are dual to ${\cal N}=2$ SCQMs arising from different twisted compactifications of the $d=5$, ${\cal N}=1$ $USp(2N)$ supersymmetric gauge theories on~$\Morb_4$. The structure of these solutions is analogous to that of the $\AdS_3$ solutions constructed in~\cite{Cheung:2022ilc} and indeed for the $\Morb_4 = \spindle\ltimes\spindle$ family we have provided a toric geometry description and formulated a conjectural extremal problem that applies to both classes.

As in \cite{Cheung:2022ilc}, the solutions that we found have less parameters than the data specifying the orbifolds $\Morb_4$. In particular, the twisting parameter $t$ is not an arbitrary integer, but it is unnaturally related  to the spindle data $n_\pm$, $m_\pm$. We therefore expect more general solutions to exist, although the ansatz describing them is likely to be quite different from those employed here and in \cite{Cheung:2022ilc}. Furthermore, the toric description of the $\Morb_4 = \spindle\ltimes\spindle$ solutions strongly suggests that there exist solutions corresponding to toric orbifolds with an arbitrary number of fixed points, and could include the case of toric manifolds.

While in this paper we provided a basic description of the orbifolds, which was adequate for characterizing the present solutions, it would be desirable to study more systematically their underlying structure.
For example, it would be useful  to perform a  global analysis of the gauge fields and  Killing spinors on~$\Morb_4$, determining the bundles of which these are sections.
Relatedly, given a generic compact four-dimensional toric orbifold $\Morb_4$, it would be nice to determine what are the allowed supersymmetry-preserving twists.
It could also be useful to cast our solutions in terms of classifications of supersymmetric solutions of massive type~IIA supergravity, or to characterize them directly in the context of $D=6$ and $D=7$ gauged supergravities. For example, we expect that generic~$\Morb_4$'s will be endowed with a canonical complex
structure\footnote{We note that the $\spindle\ltimes\spindle$ orbifolds have been discussed in the mathematical literature in the context of toric K\"ahler geometry and
were referred to as ``Hirzebruch orbifold surfaces'' \cite{Apostolov:2013}. In particular, the authors of \cite{legendre2009} classified four-dimensional toric K\"ahler orbifolds associated with labelled polytopes with four edges (quadrilaterals) and showed that they admit an extremal K\"ahler metric of Calabi type if and only if the base spindle is actually a smooth two-sphere (\ie\ $m_+ = m_- = 1$).}.
Another natural extension of our results is to investigate whether supersymmetric $\AdS_6$ black holes with~$\Morb_4$ horizons can be constructed.

In this paper we have proposed an entropy function constructed assembling \emph{gravitational blocks}, extending the proposal of~\cite{Faedo:2021nub} from the spindle to a very general class of four-dimensional orbifolds. We believe that our approach will be applicable to orbifolds more general than those for which we constructed explicit solutions here. However, finding additional explicit examples may lead to a more refined formulation of our recipe.
Similarly to the spindle off-shell free energies  \cite{Faedo:2021nub}, it is striking that the structure we introduced applies to both $D=6$ and $D=7$ supergravities, with minor tweaks.
In the context of $D=7$ supergravity, our proposal reproduces the results obtained by integrating the M5-branes anomaly polynomial on compact  toric four-manifolds~\cite{Hosseini:2020vgl} and on the $\Morb_4 = \spindle\ltimes\spindle$ orbifolds~\cite{Cheung:2022ilc}.
On the other hand, in the context of $D=6$ supergravity, at present we do not have access to any field theory computation and therefore our proposal gives a prediction for the large $N$ limit of the localized partition function of the $d=5$, ${\cal N}=1$ $USp(2N)$ on $S^1\times\Morb_4$.

\subsection*{Acknowledgments}

We thank Ugo Bruzzo, Michele Rossi and Alberto Zaffaroni for helpful discussions. We also thank the authors of~\cite{Couzens:2022lvg} for informing us about their upcoming paper.
The research of FF is supported by the project ``Nonlinear Differential Equations'' of Universit\`a degli Studi di Torino.


\appendix

\section{Some solutions to the Diophantine equation}
\label{app:quantization.exe}

In this appendix we address the quantization conditions we stated through the paper. Although we were not able to prove that these constraints can be satisfied appropriately tuning the parameters of the solutions, nevertheless we could find some examples in which the quantization can be realized.

\textit{A priori}, there are no constraints on~$\x$ apart from being solution of the quartic equation~\eqref{quartic} and lying in the range $(0,1)$. Nevertheless, in order to construct the integer quantities we are searching for, it is easier to work with rational values of~$\x$, if they exist. Indeed they exist, and in table~\ref{tab:x-solutions} we present some of them, along with the corresponding values of the integers $n_+$, $n_-$ and $p_1$ (recall that $p_2 = n_++n_--p_1$).
\begin{table}[ht]
\centering
\begin{tabular}[t]{ | c | c | c | c | }
\hline
$n_+$ & $n_-$ & $p_1$ & $\x$ \\
\hline\hline
11 & 1  & $-$22 & 1/6 \\
11 & 4  & $-$6  & 1/5 \\
14 & 3  & $-$19 & 3/17 \\
17 & 8  & $-$98 & 1/25 \\
17 & 12 & $-$52 & 1/29 \\
24 & 1  & $-$81 & 3/25 \\
31 & 3  & $-$98 & 2/17 \\
\hline
\end{tabular}
\qquad
\begin{tabular}[t]{ | c | c | c | c | }
\hline
$n_+$ & $n_-$ & $p_1$ & $\x$ \\
\hline\hline
32 & 25 & $-$21 & 1/19 \\
36 & 11 & $-$85 & 5/47 \\
41 & 19 & $-$19 & 1/6 \\
43 & 3  & $-$8  & 12/23 \\
43 & 17 & $-$87 & 1/10 \\
53 & 13 & $-$6  & 4/11 \\
59 & 13 & $-$39 & 1/4 \\
\hline
\end{tabular}
\qquad
\begin{tabular}[t]{ | c | c | c | c | }
\hline
$n_+$ & $n_-$ & $p_1$ & $\x$ \\
\hline\hline
67 & 50 & $-$15 & 1/13 \\
73 & 42 & $-$27 & 3/23 \\
80 & 19 & $-$18 & 1/3 \\
82 & 5  & $-$42 & 11/29 \\
83 & 7  & $-$33 & 2/5 \\
83 & 47 & $-$12 & 2/13 \\
90 & 53 & $-$93 & 1/11 \\
\hline
\end{tabular}
\caption{Examples of solutions to the quartic equation~\eqref{quartic} for $n_+ \leq 100$ and $|p_1| \leq 100$.}
\label{tab:x-solutions}
\end{table}

Inspired by~\cite{Cheung:2022ilc} we propose an algorithm to construct a family of solution to the quantization conditions. For the combinations of parameters presented in table~\ref{tab:x-solutions}, $\x$ is rational and positive, thus we can write it as $\x = \mathtt{p}/\mathtt{q}$, with $\mathtt{p},\mathtt{q}\in\NN$. Then, we define~$t$ as
\begin{equation} \label{t_ansatz}
  t = -8n_+ n_- \mathtt{p}^2 \mathtt{q} \, u \,,
\end{equation}
with $u \in \NN$, which makes $t$ integer by construction and, moreover, multiple of both~$n_+$ and~$n_-$. We now focus on the $\AdS_2 \times \spindle_1 \ltimes \spindle_2$ system, the case with Riemann surface proceeds similarly. Substituting the expression of~$t$ in the constraint~\eqref{2spin_t} we obtain
\begin{equation}
  m_+ - m_- = 2 (\mathtt{p}^2 + 3\mathtt{q}^2) [(n_+ - n_-)\mathtt{q} - (n_+ + n_-)\mathtt{p}] \, u \,,
\end{equation}
which can be solved to find $m_\pm$ in terms of~$n_\pm$, $\mathtt{z}$ and~$u$. The quantization condition that follows from~\eqref{2spin_quantN} is automatically satisfied given $t$ as in~\eqref{t_ansatz}, indeed it becomes
\begin{equation}
  (m_+ m_-) \, \fls_1^- = -\bigl[ (n_+ + n_-) \mathtt{p}^3 - (p_1 - 3p_2 + 6n_+) \mathtt{p}^2 \mathtt{q} + (p_1 + 5p_2) \mathtt{p} \mathtt{q}^2 - 3(n_+ - n_-) \mathtt{q}^3 \bigr] u \,,
\end{equation}
which is evidently an integer. The last requirement~\eqref{F4 flux spindle} can be met considering that $3\mu(\x^2 + 1) - \x(\x^2 + 5) > 0$ (\cf~\cite{Faedo:2021nub}) and that all the quantities appearing are rational, hence $N$ can be tuned appropriately so that $K\in\NN$.

\section{Riemann surfaces from spindles}
\label{app:spindle-2-riemann}

In presence of twist, it was explicitly shown in~\cite{Ferrero:2021etw} that the spindle factor of the known $\AdS_{D-2} \times \spindle$ solutions, with $D=4,5,7$, can be turned into a round sphere performing a particular limit of the coordinates and parameters of the system. In the same spirit, as we will see in this appendix, a similar mechanism applies to the $\AdS_{D-2} \times \spindle$ solutions with anti-twist, so far constructed only in~$D=4$ and~$D=5$. In this case the spindle becomes the two-dimensional hyperbolic space, which can later be quotiented to give a constant curvature Riemann surface with genus $\mathrm{g}>1$.

\subsection{AdS$_2\times\Morb_4$ solutions}

Let us begin considering the local spindle metric~\eqref{spin1_metric}
\begin{equation}
  \dd s_{\spindle_1}^2 = \frac{x^2}{q} \, \dd x^2 + \frac{q}{4x^2} \, \dd\psi^2 \,,
\end{equation}
where, we recall, $q(x) = x^4 - 4x^2 + 4\mathtt{a} x - \mathtt{a}^2$. Performing the change of coordinates
\begin{equation} \label{hyper-coord}
  x = 1 + \epsilon \, \xi \,,  \qquad  \psi = \frac{\phi}{\epsilon} \,,  \qquad  \text{with}  \qquad  \mathtt{a} = 1 - \epsilon^2 \,,
\end{equation}
and computing the limit $\epsilon \to 0^+$, the spindle metric becomes
\begin{equation} \label{spin1_metric-H2}
  \dd s_{\spindle_1}^2 = \frac12 \biggl( \frac{\dd\xi^2}{\xi^2 -1} + \bigl(\xi^2 -1\bigr) \, \dd\phi^2 \biggr) = \frac12 \, \dd s_{H^2}^2 \,,
\end{equation}
which is the local metric of the hyperbolic space~$H^2$, with radius squared equal to 1/2. This space can then be quotiented to obtain a Riemann surface with $\mathrm{g}>1$. In order for metric~\eqref{spin1_metric-H2} to genuinely describe~$H^2$, the coordinates~$\xi$ and~$\phi$ must have the correct range of definition, \ie\ $|\xi|\in(1,+\infty)$ and $\phi\in[0,2\pi]$. To address this issue, we begin considering that the coordinate~$x$ is restricted between~$\XA$ and~$\XB$, where~\cite{Ferrero:2020twa}
\begin{equation}
  \XA = -1 + \sqrt{1 + \mathtt{a}} \,,  \qquad  \XB = 1 - \sqrt{1 - \mathtt{a}} \,.
\end{equation}
By means of~\eqref{hyper-coord}, these bounds are mapped to
\begin{equation}
  \xi_2 = \frac{-2 + \sqrt{2 - \epsilon^2}}{\epsilon} \to -\infty \,,  \qquad  \xi_3 = -\frac{\sqrt{\epsilon^2}}{\epsilon} = -1\,,
\end{equation}
in the limit $\epsilon \to 0^+$, thus correctly giving $\xi\in(-\infty,-1)$. The periodicity of~$\phi$ can be studied analysing the conditions under which~$\spindle_1$ is proper spindle~\cite{Ferrero:2020twa}
\begin{equation} \label{spindle1-cond}
  \sqrt{1 + \mathtt{a}} \cdot \Delta\psi = \frac{2\pi}{m_-} \,,  \qquad  \sqrt{1 - \mathtt{a}} \cdot \Delta\psi = \frac{2\pi}{m_+} \,,
\end{equation}
with $m_\pm$ two co-prime integers such that $m_-<m_+$. By means of the change of coordinates~\eqref{hyper-coord}, these conditions yield
\begin{equation}
  \sqrt{2 - \epsilon^2} \cdot \frac{\Delta\phi}{\epsilon} = \frac{2\pi}{m_-} \,,  \qquad  \sqrt{\epsilon^2} \cdot \frac{\Delta\phi}{\epsilon} = \frac{2\pi}{m_+} \,,
\end{equation}
hence we have $\Delta\phi = 2\pi$ if and only if, at first order,
\begin{equation} \label{hyper-ms}
  m_- = \frac{\epsilon}{\sqrt2} \to 0 \,,  \qquad  m_+ = 1 \,.
\end{equation}
Consistently, plugging these values of~$m_\pm$ into equation~\eqref{spin1_a+psi} gives $\mathtt{a} = 1 - \epsilon^2$, as in~\eqref{hyper-coord}.
The global quantities characterizing the spindle~$\spindle_1$ are not mapped directly to the corresponding ones for a Riemann surface. Indeed, if we take as an examples the Euler characteristic of~$\spindle_1$~\cite{Ferrero:2020laf}
\begin{equation}
  \chi_1 = \frac{1}{4\pi} \int_{\spindle_1} R_{\spindle_1} \vol{\spindle_1} = \frac{m_+ + m_-}{m_+ m_-} \,,
\end{equation}
we see that it diverges when $m_-\to0$ and $m_+=1$. The reason for this behaviour can be found in the definition of~$\riemann$ as a quotient of the hyperbolic space, whose Ricci curvature, integrated over the whole space, diverges, since it is not compact. Therefore, we need a way to ``regularize'' this limit. To this end, we notice that, adopting the substitution~\eqref{hyper-ms},
\begin{equation}
  2(1 - \mathrm{g}) \cdot \chi_1 = \frac{2\sqrt2 (1 - \mathrm{g})}{\epsilon} + 2(1 - \mathrm{g}) \,,
\end{equation}
that is, we obtain a simple pole at $\epsilon=0$ plus the Euler characteristic of a Riemann surface. We propose the following prescription, that we will test against different examples: given a system comprising a $\riemann$~factor that can be obtained from~$\spindle_1$ through limit~\eqref{hyper-ms}, the global quantities of the former can be computed multiplying by $2(1-\mathrm{g})$ the corresponding quantities of the latter and taking the finite term in the Laurent series.

We can now transfer this technology to higher-dimensional spacetimes. Applying the change of coordinates~\eqref{hyper-coord}---driven by the underlying limit~\eqref{hyper-ms}---to the $\AdS_2\times\spindle_1$ system described by~\eqref{4d_spindle} we obtain the $\AdS_2\times\riemann$ background in~\eqref{4d_riemann}. Likewise are related the respective Killing spinors. Indeed, in this limit $Q_1$ vanishes and $Q_2$ becomes constant, thus \eqref{4d_spindle-killing} turns into~\eqref{4d_riemann-killing}, apart from an irrelevant overall factor.
For what concerns the total charge of the two systems, applying the recipe illustrated for the Euler characteristic~$\chi_1$ we obtain (\cf~\eqref{4d_flux})
\begin{equation}
  2(1 - \mathrm{g}) \cdot \frac{1}{2\pi} \int_{\spindle_1} F_{4\mathrm{d}}^{\spindle_1} = \frac{a_{-1}}{\epsilon} + \frac{1}{2\pi} \int_{\riemann} F_{4\mathrm{d}}^{\riemann} \,,
\end{equation}
where $a_{-1}$ is the residue.

The change of coordinates~\eqref{hyper-coord} also maps the $\AdS_2 \times \spindle_1 \ltimes \spindle_2$ solution and Killing spinors to the $\AdS_2 \times \riemann \ltimes \spindle_2$ corresponding quantities. This relation reflects in a formal connection between the respective entropies by means of~\eqref{hyper-ms}:
\begin{equation}
	2(1 - \mathrm{g}) \cdot S_{\spindle_1\ltimes\spindle_2} = \frac{a_{-1}}{\epsilon} + S_{\riemann\ltimes\spindle_2} + \mathcal{O}(\epsilon) \,,
\end{equation}
where $S_{\spindle_1\ltimes\spindle_2}$ is given in~\eqref{2spin_entropy} and $S_{\riemann\ltimes\spindle_2}$ in~\eqref{riem_entropy}.

\subsection{AdS$_3\times\Morb_4$ solutions}

Moving to the local spindle metric presented and studied in~\cite{Ferrero:2020laf}\footnote{To stick to our notation we relabelled the coordinates $(y,z)$ as~$(x,\psi)$ and the integers~$n_\pm$ as~$m_\mp$.}
\begin{equation}
  \dd s_{\spindle_1}^2 = \frac{x}{q} \, \dd x^2 + \frac{q}{36x^2} \, \dd\psi^2 \,,
\end{equation}
where $q(x) = 4x^3 - 9x^2 + 6\mathtt{a} x - \mathtt{a}^2$, we perform the change of coordinates
\begin{equation} \label{hyper-coord-7}
  x = 1 + 2\epsilon \, \xi \,,  \qquad  \psi = \frac{\phi}{\epsilon} \,,  \qquad  \text{with}  \qquad  \mathtt{a} = 1 - 3\epsilon^2 \,.
\end{equation}
In the $\epsilon \to 0^+$ limit, the spindle metric becomes
\begin{equation}
  \dd s_{\spindle_1}^2 = \frac13 \biggl( \frac{\dd\xi^2}{\xi^2 -1} + \bigl(\xi^2 -1\bigr) \, \dd\phi^2 \biggr) = \frac13 \, \dd s_{H^2}^2 \,.
\end{equation}
Once again, we can quotient the hyperbolic space to give a constant curvature Riemann surface with genus $\mathrm{g}>1$. It can be shown that the metric obtained in this limit describes a genuine hyperbolic space if
\begin{equation} \label{hyper-ms-7}
  m_- = \frac{2\epsilon}{3} \to 0 \,,  \qquad  m_+ = 1 \,.
\end{equation}
Applying this recipe to the $\AdS_3\times\spindle_1$ solution of~\cite{Ferrero:2020laf} we have
\begin{equation}
  \dd s_{5\mathrm{d}}^2 = \frac49 \, \dd s_{\AdS_3}^2 + \frac13 \, \dd s_{H^2}^2 \,,  \qquad  A_{5\mathrm{d}} = \frac12 \, \omega_{H^2} \,,
\end{equation}
where $\omega_{H^2}$ is such that $\dd\omega_{H^2} = \vol{H^2}$. This is the $\AdS_3 \times H^2$ near-horizon of the black string solution of $D=5$, $\mc{N}=2$ minimal gauged supergravity constructed in~\cite{Klemm:2000nj}\footnote{The gauge potential has a different sign with respect to~\cite{Klemm:2000nj}, but the solution is equivalent and still supersymmetric.}.
Like in the six-dimensional case, applying the change of coordinates~\eqref{hyper-coord-7}, followed by the $\epsilon\to0$ limit, to the $\AdS_3 \times \spindle_1 \ltimes \spindle_2$ solution of~\cite{Cheung:2022ilc} gives the $\AdS_3 \times \riemann \ltimes \spindle_2$ background therein.
Moreover, it is possible to retrieve the central charge of the latter, presented in~\eqref{riem_central-charge}, multiplying the central charge of the former, given in~\eqref{2spin_central-charge}, by $2(1-\mathrm{g})$, substituting $m_\pm$ as in~\eqref{hyper-ms-7} and taking the finite part of the Laurent series. Explicitly,
\begin{equation}
  2(1 - \mathrm{g}) \cdot c_{\spindle_1\ltimes\spindle_2} = \frac{a_{-1}}{\epsilon} + c_{\riemann\ltimes\spindle_2} + \mathcal{O}(\epsilon) \,.
\end{equation}

\section{More toric data from explicit metrics}
\label{app:toric-theory}

In this appendix we will provide further examples of how to extract toric data from explicit metrics, as discussed in the main body of the paper.
The first example we consider is the Hirzebruch surface $\mathbb{F}_1$, that is a well known symplectic toric manifold. Its associated polytope is a Delzant polytope (see \eg~\cite{dasilvabook}) and below we show how to obtain it starting from an explicit metric on $\mathbb{F}_1$ \cite{Martelli:2004wu}, using two methods. Namely, computing the image of the moment map associated with a symplectic structure compatible with a conformally rescaled metric and studying the Killing vectors generating the $U(1)^2$ isometry of the metric.
We will then employ these methods to extract the toric data associated with the metrics presented in~\cite{Cheung:2022ilc}. These are used in section~\ref{sec:entropy-function} to show that our recipe reproduces the central charge computed in~\cite{Cheung:2022ilc}.

\subsection{Hirzebruch $\mathbb{F}_1$}

We now illustrate an example of Delzant polytope and its construction studying the first Hirzebruch surface~$\mathbb{F}_1$, whose metric is given by (\cf\ equation~(7.11) of~\cite{Martelli:2004wu})\footnote{Here, we replaced~$\ell$ with~$l$ in order to avoid confusion with the index of the fixed points.}
\begin{equation} \label{F1_metric}
	\dd s_{\mathbb{F}_1}^2 = \frac{1-y}{6} \bigl( \dd\theta^2 + \sin^2\!\theta \, \dd\phi^2 \bigr) + \frac{\dd y^2}{w \, q} + \frac{l^2 w \, q}{36 F} (\dd\Omega + \cos\theta \, \dd\phi)^2 \,,
\end{equation}
where the expression of the functions $w(y)$, $q(y)$ and $F(y)$ can be found in~\cite{Martelli:2004wu}. The angular coordinates lie in the ranges $\theta\in[0,\pi]$, $\phi\in[0,2\pi]$, $\Omega\in[0,4\pi]$, while $y$ spans the range $[y_1,y_2]$, where $y_1$ and $y_2$ are, respectively, the negative and the smallest positive roots of~$q(y)$. To ensure the correct signature of the metric we require $1-y>0$.
After having rescaled the metric as $\dd s^2 = \Gamma(y)\,\dd s_{\mathbb{F}_1}^2$, we consider the symplectic two-form
\begin{equation} \label{F1_symp-2form}
	\omega = \Gamma(y) \biggl[ \frac{1-y}{6} \sin\theta \, \dd\theta \wedge \dd\phi + \frac{l}{6 F^{1/2}} \, \dd y \wedge (\dd\Omega + \cos\theta \, \dd\phi) \biggr] \,.
\end{equation}
When $\Gamma(y)$ is such that $\Gamma'(y) = \frac{F^{1/2}-l}{(1-y) F^{1/2}} \Gamma(y)$, the two-form~\eqref{F1_symp-2form} is closed and can be written as
\begin{equation}
	\omega = \dd\phi \wedge \dd \biggl[ \frac{1-y}{6} \Gamma(y) \cos\theta \biggr] + \dd\Omega \wedge \dd \biggl[ \frac{1-y}{6} \Gamma(y) \biggr] \,.
\end{equation}

\begin{figure}[ht]
	\centering

	\begin{subfigure}[b]{0.4\textwidth}
		\centering

		\begin{tikzpicture}[scale=1.12]
				\node [label=right : {$\fp_{1}$}] (NN) at (4,4) {};
				\node [label=left  : {$\fp_{2}$}] (NS) at (1.5,1.5) {};
				\node [label=left  : {$\fp_{3}$}] (SS) at (1.5,0) {};
				\node [label=right : {$\fp_{4}$}] (SN) at (4,0) {};

				\draw (SN.center) -- node[below]{}      (NN.center) node[midway,anchor=center] (d1) {};
				\draw (NN.center) -- node[above left]{} (NS.center) node[midway,anchor=center] (d2) {};
				\draw (NS.center) -- node[above]{}      (SS.center) node[midway,anchor=center] (d3) {};
				\draw (SS.center) -- node[right]{}      (SN.center) node[midway,anchor=center] (d4) {};

				\draw [-latex] (d1.center) -- node[above] {$\vec{n}_1$} ($(d1)+(1,0)$);
				\draw [-latex] (d2.center) -- node[right] {$\vec{n}_2$} ($(d2)+(-1,1)$);
				\draw [-latex] (d3.center) -- node[above] {$\vec{n}_3$} ($(d3)+(-1,0)$);
				\draw [-latex] (d4.center) -- node[right] {$\vec{n}_4$} ($(d4)+(0,-1)$);
			\end{tikzpicture}

		\subcaption{Delzant polytope of~$\mathbb{F}_1$.}
		\label{fig:F1_toric-poly}
	\end{subfigure}
\qquad\qquad
	\begin{subfigure}[b]{0.4\textwidth}
		\centering

		\begin{tikzpicture}[x=1.05cm, y=1.05cm]
			\coordinate (00)  at (0,0);
			\coordinate (w1)  at (1,0);
			\coordinate (w2)  at (-1,1);
			\coordinate (w3)  at (-1,0);
			\coordinate (w4)  at (0,-1);
			\coordinate (w2x) at (-2,2);

			\path[fill=gray!15] (00) -- (2,0)  -- (2,2)   -- (w2x)  -- cycle;
			\path[fill=gray!40] (00) -- (w2x)  -- (-2,0)  -- cycle;
			\path[fill=gray!65] (00) -- (-2,0) -- (-2,-2) -- (0,-2) -- cycle;
			\path[fill=gray!80] (00) -- (0,-2) -- (2,-2)  -- (2,0)  -- cycle;

			\draw [-stealth] (0,-2.5) -- (0,2.5);
			\draw [-stealth] (-2.5,0) -- (2.5,0);

			\foreach \x in {-2,...,2}
			\foreach \y in {-2,...,2} {
				\fill (\x,\y) circle (0.75pt);
			}

			\draw [gray] (w2) -- (w2x);

			\draw [thick, -latex] (00) -- (w1) node[above left]  {$\vec{{n}}_1$};
			\draw [thick, -latex] (00) -- (w2) node[right]       {$\vec{{n}}_2$};
			\draw [thick, -latex] (00) -- (w3) node[below right] {$\vec{{n}}_3$};
			\draw [thick, -latex] (00) -- (w4) node[above right] {$\vec{{n}}_4$};

			\node (c1) at (1.5,1.5)   {$\tau_1$};
			\node (c2) at (-1.5,0.5)  {$\tau_2$};
			\node (c3) at (-1.5,-1.5) {$\tau_3$};
			\node (c4) at (1.5,-1.5)  {$\tau_4$};
		\end{tikzpicture}

		\subcaption{Fan of~$\mathbb{F}_1$.}
		\label{fig:F1_toric-fan}
	\end{subfigure}

	\caption{Toric data of the first Hirzebruch surface~$\mathbb{F}_1$.}
	\label{fig:F1_toric}
\end{figure}
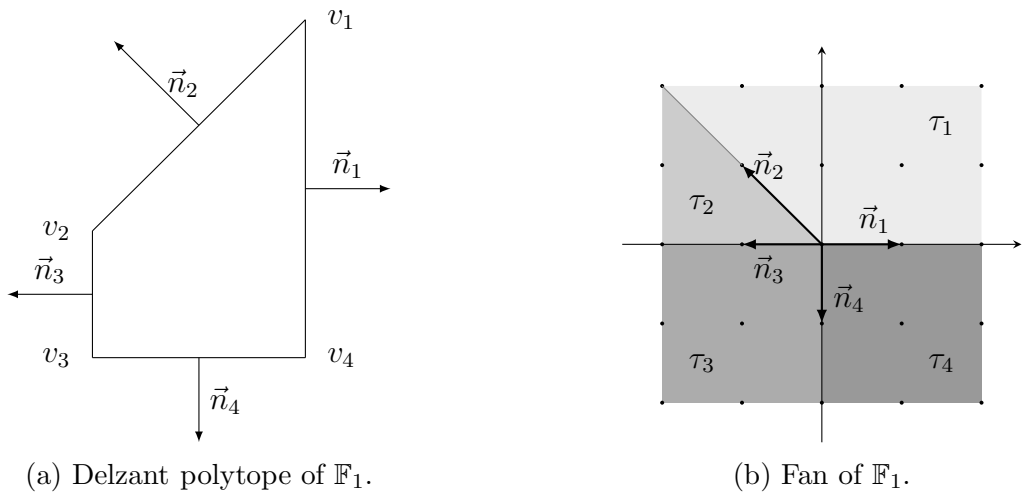

For fixed values of~$y$, metric~\eqref{F1_metric} describes a squashed, smooth three-sphere, corresponding to the system studied in section~4 of~\cite{Martelli:2004wu} for $m=1$. Following their arguments, we take as a basis of an effective torus action\footnote{We exchanged $e_1$ and $e_2$ with respect to~\cite{Martelli:2004wu} in order to stick to our notation of counter-clockwise ordered normal vectors.}
\begin{equation} \label{F1_vector-basis}
	e_1 = V_2 \,,  \qquad  e_2 = V_1 + \frac12 \, V_2 \,,
\end{equation}
where $V_1$ generates the rotations of the two-sphere about the equator with weight one, hence $V_1=\partial_\phi$, and $V_2$ the rotations of the $S^1$ fibre, also with weight one, hence $V_2=\partial_\nu$, where we introduced the new radial coordinate $\nu=\Omega/2$, with period~$2\pi$. Once we have chosen a basis for the torus action, we can derive the related moment maps
\begin{equation}
	\vec{\mu} = \frac{1-y}{3} \Gamma(y) \, \biggl( 1, \frac{\cos\theta + 1}{2} \biggr) \,.
\end{equation}
In total we have four fixed points, corresponding to all the possible combinations pairing the poles of the base two-sphere ($\theta=0,\pi$) and the poles of the fibre two-sphere ($y=y_{1,2}$)
\begin{equation}
	\begin{aligned}
		\fp_{1} &= \{ \theta = 0, \, y = y_1 \} \,,    \qquad  & \fp_{2} &= \{ \theta = 0, \, y = y_2 \} \,, \\
		\fp_{3} &= \{ \theta = \pi, \, y = y_2 \} \,,  \qquad  & \fp_{4} &= \{ \theta = \pi, \, y = y_1 \} \,.
	\end{aligned}
\end{equation}
The image of these points under the moment maps are
\begin{equation}
	\begin{aligned}
		\vec{\mu}(\fp_{1}) &= \Xi_1 (1, 1) \,,  \qquad  & \vec{\mu}(\fp_{2}) &= \Xi_2 (1, 1) \,, \\
		\vec{\mu}(\fp_{3}) &= \Xi_2 (1, 0) \,,  \qquad  & \vec{\mu}(\fp_{4}) &= \Xi_1 (1, 0) \,,
	\end{aligned}
\end{equation}
where we defined, for convenience, the positive constants $\Xi_i = \frac{1-y_i}{3}\Gamma(y_i)$. A numerical analysis shows that $\Xi_1>\Xi_2$, therefore the moment polytope of $\mathbb{F}_1$ can be drawn as in figure~\ref{fig:F1_toric-poly}. A quick investigation proves that it is a Delzant polytope in~$\RR^2$.

The polytope is uniquely defined by the relations
\begin{equation}
	\langle \vec{\mu}, \vec{n}_1 \rangle \leq \Xi_1 \,,  \qquad  \langle \vec{\mu}, \vec{n}_2 \rangle \leq 0 \,,  \qquad  \langle \vec{\mu}, \vec{n}_3 \rangle \leq -\Xi_2 \,,  \qquad  \langle \vec{\mu}, \vec{n}_4 \rangle \leq 0 \,,
\end{equation}
where the $\ZZ^2$ primitive normal vectors~$\vec{n}_\ell$ are
\begin{equation}\label{norm_vect_F1}
	\vec{n}_1 = (1, 0) \,,  \qquad  \vec{n}_2 = (-1, 1) \,,  \qquad  \vec{n}_3 = (-1, 0) \,,  \qquad  \vec{n}_4 = (0, -1) \,.
\end{equation}
The fan generated by these vectors is depicted in figure~\ref{fig:F1_toric-fan} and is dual to the Delzant polytope of figure~\ref{fig:F1_toric-poly}.

A different approach to the construction of the toric data of a given toric manifold goes through the analysis of the degenerating Killing vectors of the corresponding metric. In total we have four Killing vectors~$\xi_{(\ell)}$ degenerating at the loci~$D_\ell$, namely
\begin{equation}
	\begin{aligned}
		D_1 &= \{y = y_1\}:  \quad  & \xi_{(1)} &= 2\partial_\Omega \,,  \qquad  & D_2 &= \{\theta = 0\}:  \quad  & \xi_{(2)} &= \partial_\phi - \partial_\Omega \,, \\
		D_3 &= \{y = y_2\}:  \quad  & \xi_{(3)} &= 2\partial_\Omega \,,  \qquad  & D_4 &= \{\theta = \pi\}:  \quad  & \xi_{(4)} &= \partial_\phi + \partial_\Omega \,.
	\end{aligned}
\end{equation}
These vectors have been normalized such that they have unitary surface gravity when approaching the degeneracy hypersurface. Expanding on the basis~\eqref{F1_vector-basis}, the four Killing vectors read
\begin{equation}
	\xi_{(1)} = \xi_{(3)} = e_1 \,,  \qquad  \xi_{(2)} = e_2 - e_1 \,,  \qquad  \xi_{(4)} = e_2 \,.
\end{equation}
We notice that the four loci~$D_\ell$ are in a one-to-one correspondence with the facets of the polytope in figure~\ref{fig:F1_toric-poly} and that the Killing vectors~$\xi_{(\ell)}$ can be written in terms of the related normal vectors~$\vec{n}_\ell$ as $\xi_{(\ell)} = \vec{n}_\ell \cdot (e_1, e_2)$.

\subsection{AdS$_3\times S^2 \ltimes \spindle_2$ solutions}
\label{app:toric_riem_7d}

We consider here the  $\AdS_3 \times \riemann \ltimes \spindle_2$ geometry presented in section 4 of \cite{Cheung:2022ilc} subjected, as before, to the analytic continuation $\riemann \mapsto S^2$.
The metric is given by\footnote{In order to make the comparison with~\cite{Cheung:2022ilc} easier, we relabelled $\phi$ and $a$ therein as $z$ and $\mathtt{a}$ and exchanged $n_+ \leftrightarrow n_-$.}
\begin{equation}
  \dd s^2 = \frac{4 (y P)^{1/5}}{9} \biggl[ \dd s_{\AdS_3}^2 + \frac34 \, \dd s_{S^2}^2 + \frac{9y}{16Q} \, \dd y^2 + \frac{9Q}{4P} \Bigl( \dd z - \frac23 \, \omega_{S^2} \Bigr)^2 \biggr] \,,
\end{equation}
where $\dd\omega_{S^2} = -\vol{S^2}$ and
\begin{equation} \label{func-hPQ}
	h_i(y) = y^2 + q_i \,,  \qquad  P(y) = h_1(y) \, h_2(y) \,,  \qquad  Q(y) = -y^3 + \frac14 P(y) \,.
\end{equation}
We can take explicitly
\begin{equation}
	\dd s_{S^2}^2 = \dd\theta^2 + \sin^2\!\theta \, \dd\psi^2 \,,  \qquad  \omega_{S^2} = \cos\theta \, \dd\psi \,.
\end{equation}
Since the requirement of having a globally well-defined fibration
\begin{equation}
	\frac{1}{2\pi} \int_{S^2} \dd\eta = t \in \ZZ \,,\qquad\eta \equiv \frac{2\pi}{\Delta z} \Bigl( \dd z - \frac23 \, \omega_{S^2} \Bigr) \,,
\end{equation}
yields the relation\footnote{Notice that here $t$ is a positive integer, due to the different convention of  \cite{Cheung:2022ilc} in defining $\omega$.}
\begin{equation}
	\frac{4}{3t} = \frac{\Delta z}{2\pi} \,,
\end{equation}
we can define a $2\pi$-periodic coordinate $\spang2 = \frac{2\pi}{\Delta z}z$ in terms of which the four-dimensional toric orbifold $S^2\ltimes\spindle_2$ of interest is
\begin{align} \label{7dtoric_riem}
	\dd s_{S^2\ltimes\spindle_2}^2 = \frac13 \, \bigl( d\theta^2 + \sin^2\!\theta \, d\psi^2 \bigr) + \frac{y}{4Q} \, \dd y^2 + \frac{Q}{P} \Bigl( \frac{2}{t} \, \dd\spang2 - \cos\theta \, \dd\psi \Bigr)^2  \,.
\end{align}
For fixed~$y$ this metric describes an $S^3/\ZZ_t$, thus we consider as a basis of an effective torus action
\begin{equation} \label{7dkilling-basis}
	e_1 = \partial_{\spang2} \,,  \qquad  e_2 = \partial_\psi + \frac{t}{2} \, \partial_{\spang2} \,.
\end{equation}
The four fixed points are
\begin{equation} \label{7dfixed-points}
	\begin{aligned}
		\fp_1 &= \{\theta = 0, \, y = y_2\} \,,    \qquad  & \fp_2 &= \{\theta = 0, \, y = y_3\} \,, \\
		\fp_3 &= \{\theta = \pi, \, y = y_3\} \,,  \qquad  & \fp_4 &= \{\theta = \pi, \, y = y_2\} \,,
	\end{aligned}
\end{equation}
where $y_{2,3}$ are the two middle roots of $Q(y)$, with $y_2<y_3$.
These points define the four divisors
\begin{equation}
	D_1 = \{y=y_2\} \,,  \qquad  D_2 = \{\theta=0\} \,,  \qquad  D_3 = \{y=y_3\} \,,  \qquad  D_4 =\{\theta=\pi\} \,,
\end{equation}
and, with respect to the basis~\eqref{7dkilling-basis}, the degenerate normalized Killing vectors and the orbifold labels are
\begin{equation}
	\begin{aligned}
		D_1 &:  & \xi_{(1)} &= n_- \, e_1 \,,  \quad  & m_1 &= n_- \,,  \qquad  & D_2 &:  & \xi_{(2)} &= e_2 \,,  \quad  & m_2 &= 1 \,, \\
		D_3 &:  & \xi_{(3)} &= n_+ \, e_1 \,,  \quad  & m_3 &= n_+ \,,  \qquad  & D_4 &:  & \xi_{(4)} &= e_2 - t \, e_1 \,,  \quad  & m_4 &= 1 \,.
	\end{aligned}
\end{equation}

We can now construct the polytope, considering the four-dimensional conformally-rescaled metric $\dd s^2 = \Gamma(y) \, \dd s_{S^2\ltimes\spindle_2}^2$ and the related symplectic two-form
\begin{equation} \label{7dsymplectic}
	\omega = \Gamma(y) \biggl[ \frac13 \sin\theta \, \dd\theta \wedge \dd\psi + \frac{y^{1/2}}{2 P^{1/2}} \, \dd y \wedge \Bigl( \dd z - \frac23 \cos\theta \, \dd\psi \Bigr) \biggr] \,.
\end{equation}
When $\Gamma'(y) = \frac{y^{1/2}}{P^{1/2}}\Gamma(y)$, the two-form~\eqref{7dsymplectic} is closed and can be written as
\begin{equation}
	\begin{split}
		\omega = \dd\psi \wedge \dd \biggl[ \frac13 \Gamma(y) \cos\theta \biggr] + \dd\spang2 \wedge \dd \biggl[ -\frac{2}{3t} \Gamma(y) \biggr] \,.
	\end{split}
\end{equation}
From this expression we can derive the moment maps with respect to the basis~\eqref{7dkilling-basis}
\begin{equation}
	\vec{\mu} = -\frac{2}{3t} \Gamma(y) \, \Bigl( 1, \frac{t (1 - \cos\theta)}{2} \Bigr) \,,
\end{equation}
whose action on the fixed points~\eqref{7dfixed-points} is
\begin{equation}
	\begin{aligned}
		\vec{\mu}(\fp_1) &= -\frac{2}{3t} \Gamma(y_2) \, (1, 0) \,,  \qquad  & \vec{\mu}(\fp_2) &= -\frac{2}{3t} \Gamma(y_3) \, (1, 0) \,, \\
		\vec{\mu}(\fp_3) &= -\frac{2}{3t} \Gamma(y_3) \, (1, t) \,,  \qquad  & \vec{\mu}(\fp_4) &= -\frac{2}{3t} \Gamma(y_2) \, (1, t) \,.
	\end{aligned}
\end{equation}
Since we have chosen the same basis as in~\eqref{riem_killing-basis}, the resultant polytope is similar to the one constructed in section~\ref{subsec:riem_toric_geometry}. Indeed, it can be read off from figure~\ref{fig:riem-polytope} exchanging $\fp_1 \leftrightarrow \fp_3$, $\fp_2 \leftrightarrow\fp_4$ and $n_+ \leftrightarrow n_-$.
The normal vectors $\vec{n}_\ell$ change accordingly with respect to equation~\eqref{riem_vec-n}, \ie\ $\vec{n}_1 \leftrightarrow \vec{n}_3$ and $\vec{n}_2 \leftrightarrow \vec{n}_4$, together with $t \mapsto -t$. We can quickly verify that the quantities we derived satisfy relation~\eqref{killing-and-toric}.

\subsection{AdS$_3\times\spindle_1\ltimes\spindle_2$ solutions}
\label{app:toric_2spin_7d}

We now move to the $\AdS_3 \times \spindle_1 \ltimes \spindle_2$ solutions presented in section~3 of~\cite{Cheung:2022ilc}\footnote{In addition to $n_\pm$, we also exchanged $m_+ \leftrightarrow m_-$.}. The metric on the toric orbifold is given by
\begin{equation} \label{2spindle}
	\dd s_{\spindle_1\ltimes\spindle_2}^2 = \frac{x}{f} \, \dd x^2 + \frac{f}{36x^2} \, \dd\psi^2 + \frac{y}{4Q} \, \dd y^2 + \frac{Q}{P} \Bigl( \dd z - \frac13 \, \Bigl( 1 - \frac{\mathtt{a}}{x} \Bigr) \, \dd\psi \Bigr)^2 \,,
\end{equation}
where functions $h_i$, $P$ and $Q$ are the same as in~\eqref{func-hPQ} and
\begin{equation}
	f(x) = 4x^3 - 9x^2 + 6\mathtt{a} x - \mathtt{a}^2 \,.
\end{equation}
The coordinate~$x$ ranges between~$x_1$ and~$x_2$, the two smallest roots of $f(x)$ with $x_1<x_2$, and $y$ lies between~$y_2$ and~$y_3$ as before. The fibration is globally well-defined if
\begin{equation}
	\frac{m_+ m_-}{2\pi} \int_{\spindle_1} \dd\eta = t\in\ZZ\,,\qquad \eta \equiv \frac{2\pi}{\Delta z} \Bigl( \dd z - \frac13 \, \Bigl( 1 - \frac{\mathtt{a}}{x} \Bigr) \, \dd\psi \Bigr) \,,
\end{equation}
which gives the relation
\begin{equation}
	\frac{t}{m_+ m_-} = -\frac{\mathtt{a} (x_2 - x_1)}{3x_1 x_2} \frac{\Delta\psi}{\Delta z} \,.
\end{equation}

Taking inspiration from section~\ref{subsec:toric_2spin_6d}, we introduce the following basis $\{E_1,E_2\}$ for an effective torus action:
\begin{equation} \label{new_basis_7d}
	E_1 = \partial_{\spang2} \,,  \qquad  E_2 = \partial_{\spang1} + \frac{r_+ + r_-}{m_+ - m_-} \, \partial_{\spang2} \,,
\end{equation}
where $\spang1 = \frac{2\pi}{\Delta\psi}\psi$ and $r_\pm$ are two integers such that $t = r_+\,m_+ + r_-\,m_-$.
In total there are four Killing vectors degenerating at the divisors $D_1=\{y=y_2\}$, $D_2=\{x=x_1\}$, $D_3=\{y=y_3\}$, $D_4=\{x=x_2\}$, which, in the basis~\eqref{new_basis_7d}, read
\begin{equation}
	\begin{aligned}
		\xi_{(1)} &= n_- \, E_1 \,,  \qquad  & \xi_{(2)} &= m_- \, E_2 + r_+ \, E_1 \,, \\
		\xi_{(3)} &= n_+ \, E_1 \,,  \qquad  & \xi_{(4)} &= m_+ \, E_2 - r_- \, E_1 \,.
	\end{aligned}
\end{equation}
These are formally the same Killing vectors given in~\eqref{kill_2spin_6d_new_basis}.
The ``long'' normal vectors can be obtained by means of~\eqref{new_killing_and_toric}
\begin{equation}
	\vec{\hat{w}}_1 = (n_-, 0) \,,  \qquad  \vec{\hat{w}}_2 = (r_+, m_-) \,,  \qquad
	\vec{\hat{w}}_3 = (-n_+, 0) \,,  \qquad  \vec{\hat{w}}_4 = (r_-, -m_+) \,.
\end{equation}
Following the same arguments of section~\ref{subsec:toric_2spin_6d}, we derive the labels $m_\ell = (n_-,1,n_+,1)$ and the ``short'' vectors $\vec{w}=\vec{\hat{w}}_\ell/m_\ell$, which together constitute the toric data $(\vec{w}_\ell,m_\ell)$ describing the $\spindle_1\ltimes\spindle_2$ toric orbifold.
This orbifold is characterized by the same stacky fan and labelled polytope of figure~\ref{fig:stack_geometry}.


\bibliographystyle{JHEP}
\bibliography{biblio}

\end{document}